\documentclass[11pt]{article}
\pdfoutput=1
\pdfoutput=1
\pdfoutput=1
\pdfoutput=1
\pdfoutput=1
\usepackage[pdftex]{graphics}%   <------- for PDFLatTeX
\usepackage{epsfig}
\usepackage{amsfonts,amssymb,amsmath}
\numberwithin{equation}{section}
\def\co#1{}
\newcounter{qqq}
\begin{document}\thispagestyle{empty}
\DeclareGraphicsExtensions{.jpg,.pdf,.mps,.png} %   <------- for PDFLatTeX
\centerline{\Large\bf Finsler geometrization of classic theory for
fields }
\centerline{\Large\bf on the interphase boundary
including monomolecular 2D-system}

\vspace*{3mm}

\centerline{\bf V. Balan, H.V. Grushevskaya, N.G. Krylova, A. Oana}
\begin{quotation}
{\small
\noindent {\bf Abstract.}

We prove that the Ricci scalar curvature and the Berwald scalar
    curvature of a two-dimensional Finsler space, considered over a
    vector field on the 3-dimensional flat space, are naturally
    related to 2-dimensional electro-capillary phenomena effects
 observed  for a compressed monolayer.
The Cartan tensor and the nonlinear Barthel connection of the
Finsler model are determined, and the geometric objects which
depend on compression speed and on the characteristics of the
electrically charged double layer are used in order to reveal
%point out
several classes of structure formation within the phase transition
of the first order. }
\end{quotation}
\section{Introduction}
The nano-structures, which are formed at the boundary which
separates two distinct phases, may exhibit unique electro-physical
and optical properties \cite{K2008,Suzdal,Drapeza,Anishchik}. A
type of such structures is represented by the Langmuir-Blodgett
(LB) monolayers, which are obtained by compressing a
mono-molecular layer (monolayer) of amphiphilic molecules on the
surface  of a liquid subphase
\cite{Blodgett-1934,Blodgett-1937,Woll,Schwartz}, in case that the
compression is  accompanied by the
    first order phase transition \cite{NPCS-Vol-7,Robazzi,Andelman,Ramos}.
In \cite{Grushevskaya2010Cluj,Grushevskaya2011} a Finsler
geometrization of the contribution of the electro-capillary
interactions to an action $S$, which describes the motion of
particles in the monolayer, was proposed. This
    (pseudo-)Finsler space has the specific feature that a part of the tangent vectors lie on the hypercone similar
    to light cones of Relativity Theory (\cite{Ras,Rund1981}).
    The Finsler structure relies on associated geometric vector bundles over the slit tangent space, which
    infers that the fundamental Finsler function (pseudonorm) may be considered as an action for physical systems.
    In particular, one has the 3-dimensional case where the Finsler function is associated to the physical system
    associated to a 2d-monolayer. In its essence, a bundle is a geometric structure whose associated projection
    mapping transforms a path which lies inside a fiber to a point (a set of null measure) of the base
manifold. There exist similar examples in which a system within
the physical space is given by means of a projection from a space
with larger dimension than the base space of the physical system.
This can be illustrated by the  reducing of the hydrogen-like atom
problem with its symmetry group $SO(4)$ to a consideration of the
harmonic oscillator in the space with the symmetry group $SO(6)$.
In such problems  the    projection of states from the
6-dimensional space to the physical (3,1)-dimensional space is
constructed \cite{gg}.\par
For the determined Finsler function of the monolayer we numerically produced a large number of indicatrix
    classes \cite{Grushevskaya2011}. However, in the absence of the stability of such solutions, the link between
    these solutions and the real observed monolayer structures was not completely sustained. The principle of
    minimal action $S$ needs the fulfillment of the Euler-Lagrange equations for the Lagrange function
    $L$, which provides the action $\int L\, dt = S $. These are differential equations of second-order, hence
    the need to consider the specific structural (Jacobi) stability, which reveals the
    robustness/fragility of the second order system \cite{ABBR}.
To this aim, one has to take into account that the obtained
Finsler metric provides a specific class of invariant    geometric
objects of the Finsler bundle of tangent spaces, which naturally
relate to the surface electrocapillary  physical phenomena
observed for the monolayer. We should note that the 2d-motion of
particles within this in this monolayer is approximated by
geodesics of the geometric structure.

The anomalous behavior of geometric objects like the Ricci and
Berwald scalar curvatures $R_c$ and $B_c$, the Cartan tensor, and
the nonlinear Barthel connection play a special role while
developing the KCC theory of the model and constructing the
associated invariants. We shall determine the relation between the
anomalous behavior of these geometric objects and the behavior of
compression isotherms $\tilde \pi - s$. For $LB$-structures, these
isotherms exhibit a plateau, which is characteristic for the phase
transition of the first order.

%paragraf 153. Van der Vaals theory of phase transition. (p.577, Ландау т.5)

It is known \cite{Landau5,Bellac} that the necessary condition of
phase state stability is a positiveness of Hessian in an expansion
of energy increment over thermodynamic variables.
% Как известно,
%необходимое условие устойчивости фазового состояния -
%положительность гессиана в разложении по термодинамическим
%переменным для приращения энергии.
By analogy one can study the stability of phase state by analyzing
a Hessian for an expansion of Gibbs free  energy
% По аналогии, можно исследовать устойчивость фазового
% состояния, анализируя гессиан свободной энергии Гиббса
 \cite{Bazarov}.
With such an approach one can introduce a metric of a
thermodynamic variables space  to utilize Riemann geometry methods
\cite{Weinhold1,Weinhold2}.
%В таком походе - можно ввести метрику в пространстве
%термодинамических величин и использовать методы римановой
%геометрии \cite{Weinhold1,Weinhold2}.
The scalar curvature in such thermodynamic geometrization can be
expressed in terms of such matter  thermodynamics quantities as
the compressibility $\kappa $, the thermal expansion, the heat
capacity.
% Скалярная кривизна в такой геометризации термодинамики
%выражается через такие термодиначеские параметры вещества, как
%сжимаемость $\kappa _S$, isentropic термическое expansion,
%теплоемкость.
%Преимущество такого подхода заключается в анализе
%одной геометрической структуры -- кривизны в качестве обобщенного
%термодинамического параметра вместо большого числа термодинамических неравенств.
It was expected that a priority of such approach would be the
analysis of one geometric structure -- the curvature, as
generalized thermodynamic parameter, instead of a large number of
thermodynamic inequalities.  The heat capacity and compressibility
have divergences at phase transitions. Divergences of the
thermodynamic quantities   should  involve divergences of the
curvature.
% При фазовом переходе расходимость
%одного из термодинамических параметров влечет расходимость всех
%остальных, в том числе должна расходиться кривизна.
However, for well-known Ruppeiner's \cite{Ruppeiner} and
Weinhold's \cite{Weinhold1,Weinhold2} metrics limited values of
scalar curvatures in a critical point remain finite  when trending
one of
 above mentioned thermodynamic quantities to infinity at fixed other
ones
% Однако для известных Ruppeiner's \cite{Ruppeiner} and
%Weinhold’s \cite{Weinhold1,Weinhold2} метрик значение скалярной
%кривизны остается конечным в критической точке при стремлении
%какого-либо одного параметра к бесконечности при фиксировании
%остальных термодинамических параметров
\cite{Bravetti}. In \cite{Quevedo} the thermodynamics was
geometrized with a metric, scalar curvature of which diverges in a
critical point.
%В \cite{Quevedo} была предложена геометризация термодинамики с
%метрикой, скалярная кривизна которой  расходится в критической
%точке.
The Quevedo's metric describes second-order phase transitions
without metastable states satisfactorily \cite{Bravetti}.
%Quevedo's метрика удовлетворительно описывает фазовые переходы
%2-го рода без метастабильных состояний \cite{Bravetti}.
A physical nature of the first order phase transition in a triple
point is analogous to the second order phase transition
\cite{Binder}.
%Физическая природа фазовых переходов 1-го рода в тройной точке
%аналогична фазовым переходам 2-го рода \cite{Binder}.
But, in contrast to the second order phase transitions in the
vicinity of the first order phase transition $\kappa $  and other
thermodynamic quantities of matter change sign to opposite.
%Однако в отличиии от фазовых переходов 2-го рода в окрестности фазового перехода
%1-го рода $\kappa _S$ и другии термодинамические параметры
%вещества меняют знак на противоположный.
In \cite{Bravetti}  it was proposed a metric of thermodynamic
variables space, the scalar curvature of which both divergences
and differs by sign for stable and metastable states in a critical
point.
%В \cite{Bravetti} была предложена метрика пространства
%термодинамических переменных, скалярная кривизна которой не только
%расходится, но и отличиается знаком для стабильных и
%метастабильных состояний в критической точке.
But, unfortunately this metric is not invariant in respect to
Legendre transformations.
%Однако, к сожалению, эта метрика не инвариантна относительно преобразований Лежандра.
Thus, nowadays a problem of description of the first order phase
transitions is still actual.
%Таким образом, в настоящее время отсутствует удовлетворительная
%теория фазовых переходов 1-го рода.

%
The goal of the present paper is to describe by means of the invariants of the 2D-Finsler space the acceleration
    of the molecules of the monolayer in the field of electro-capillary forces, whose contribution to the action is
    represented as a Finsler-type geometric structure associated to the interactions within the amphiphilic
    monolayer, and to study the structuring of the monolayer which undergoes phase transition.
%
%==================================================================================================================
\section{The physical model}
In \cite{Grushevskaya2011}, there is proposed a simple model of a
monolayer of amphiphilic molecules. Hydrophilic parts of
amphiphilic molecules are in an electrolyte and hydrophobic parts
are outside. In water the hydrophilic parts
 of molecules dissociate on positively charged hydrogen
ions and negatively charged hydrophilic groups ("heads"). The
monolayer and the double electrically  charged layer are regarded
as a plane-parallel capacitor with the
    capacity $C$, and a charge $\pm Q$  on its plates. The potential difference ${\Phi} $ on the interphase
    boundary is determined by the formula
    \begin{equation}\label{eq5} \Phi= \frac{Q}{C} = \frac{QD}{\varepsilon\varepsilon _0 S} =
        \frac{q\rho (\vec r,t)D}{\varepsilon\varepsilon _0 },\end{equation}
where $q$ is the charge of one ionized molecule,  $\rho(\vec r,t)$ is the surface density of molecules
    in the monolayer, $\varepsilon _0 $ is the dielectric constant, $\varepsilon $ is the dielectric permittivity,
    and $S$ is the area of the capacitor plates.
The motion of ionized particles of the monolayer takes place in the potential of electro-capillary forces
    (\cite{Lippman, Frumkin, LandauVol8, Gibbs}). This is why - as shown in \cite{Grushevskaya2011} - the particles
    possess, besides the kinetic energy and the energy of interparticle interactions $E_c$,
    a potential energy $U_{s}(r, t)$ due to the capillary interaction:
\begin{equation}
\label{eq21}
U_s (r,t) = {\sigma }'_r + \frac{{\sigma }'_t }{\dot{r}} +
\frac{\sigma }{r}
\end{equation}
at an arbitrary point of the spherically-symmetric monolayer with the co-ordinates $(r, t)$, with
    $\rho$ having the explicit form:
\begin{equation}\label{eq15}\rho (r,t) = \rho _0 e^{ - \int {\frac{\dot {r}}{r}dt} }.\end{equation}
We shall further choose a cylindric orthogonal coordinate system
($r$, $\phi,\ z$), in which the spherically-symmetric  monolayer
is displaced in the plane $XY$ with $z=0$, and the center is
    located at the origin of coordinates; here the prime denotes the derivative with respect to the
    corresponding index, $\sigma $ is the surface tension of the monolayer (the energy of the 2d-membrane),
    determined by the following expression:
\begin{equation}\label{eq17}\sigma (r,t) = - \frac{\pi ^2q^2}{\varepsilon\varepsilon _0 }\rho _0^2
    \int {\dot {r}dt\,} r^3e^{ -\int\limits_{t_0 }^t{\frac{2\dot {r}}{r(t^{\prime})}\,dt^{\prime}} },\end{equation}
and $\rho_{0}$ is the monolayer density at the initial moment $t_{0}$.
%
%==================================================================================================================
\section{The Lagrange and Finsler structures associated to the model}
 One can
neglect the particle interaction energy $E_c$, since the
negatively charged  "heads"\ are surrounded by the
positively-charged "coats"\ of ions and water molecules. Then, for
the general    case of a particle which moves in the potential
$U_{s}(r, t)$ (\ref{eq21}) in a plane, it is possible to write
down the Lagrange function $L$ corresponding to this motion, in
the form:
\begin{equation}\label{eq50} L = m\frac{dr^2 + r^2d\varphi ^2}{2dt^2} - U_s (r,t),\end{equation}
where the first term in the expression (\ref{eq50}) is a kinetic
energy. The non-relativistic metric function $dF$ of the monolayer
in the Euclidean 3d-space reads \cite{Grushevskaya2011}:
\begin{equation}
\label{eq8}
dF = mc^2dt - L_0 (x,y) dt + eA dz/c,
\end{equation}
where $L_{0 }$ is the Lagrange function which is considered
    without taking into account effects of electro-capillarity,
$dz\sim \dot {r} dr/D$, $z$ is the projection of the position
vector $\vec {R}$ of a particle onto the    $Z$-axis,
 $c$ is the speed of light, and we have:
\begin{equation}
    {e\over c} A dz =- {q\over D} \dot {r} dr \int |E'| dt
\end{equation}
due to the relation
    \begin{equation}{1\over c} {\partial\over \partial t} \vec \nabla \times \vec A =
        -\vec \nabla \times \vec E^{\prime},\end{equation}
where $\vec E^{\prime}\ = \vec\nabla\Phi\ $. Since the monolayer
exhibits non-vanishing curvature, the particle will have an
accelerated motion with the acceleration value $a_u$: $\dot{r}\sim
a_u t $. This means that the ranges of variation for the variable
$z$ differ for various directions $(dx,dy)$.
    Taking into account that $dz\sim a_ut\, dr/D$, the space of particle motion is a space of
    linear elements $(x,y,z,t={D\over a_u}{\partial z\over \partial r})$ with the underlying geometry of
    Finsler type, as stated in the Introduction, and we have
    \begin{equation}\label{action-vector-potential}dF  \sim {e}\ dz ={q\ a_u t\over D} dr .\end{equation}
The equation of the figuratrix $H$ can be obtained from the following expression \cite[pp.34,43-44]{Rund1981},
    \cite{Rund1952}:
\begin{eqnarray}\frac{\partial g_{ik}(x, \dot {x})}{\partial x^h }
        \dot {x}^i \dot {x}^k = - \frac{\partial g^{ik}(x, p)}{\partial x^h} p_ip_k,
    \ \dot {x}^i = g^{ij}p_j ,\;p_i= g_{ij} \dot {x}^j,\end{eqnarray}
if we take into account that
\begin{eqnarray}F^2 (x, \dot {x}) = g_{ik} (x, \dot {x})
        \dot {x}^i \dot {x}^k,\ H^2 (x, p) =g^{ik}(x, p)p_i p_k.
\end{eqnarray}
From this, it follows that the metric function $F$ is connected with the Hamiltonian $H$ by the following relation:
\begin{equation}\label{eq27}
\frac{\partial F^2(x,\dot {x})}{\partial x^h} =-
    \frac{\partial H^2(x,p)}{\partial x^h}.
\end{equation}
Now, we differentiate the right hand side of (\ref{eq27}), and
take into consideration the dependence of the kinetic energy on
momentum only. Then, eq.~(\ref{eq27}) gets the form
\begin{equation}\label{eq28}\frac{\partial F^2(x,\dot {x})}
    {\partial x^h} = - 2H\frac{\partial U_s(x,\dot{x})}{\partial x^h}.
\end{equation}
For potential energy $U_s$ in the expression (\ref{eq28}) we shall use the
model potential  \eqref{eq21}--\eqref{eq17}%: $V_s=-U_s$
. Then,
substituting the expressions \eqref{eq21}--\eqref{eq17} into
    (\ref{eq28}), we rewrite the last relation as
\begin{equation}\label{eq32}
\begin{array}{ll}\frac{\partial F^2(x,\dot{x})}{\partial x^h}
&=  -2 E_\kappa \frac{\partial}{\partial x^h}
\frac{\pi ^2q^2}{\varepsilon \varepsilon _0 }
\rho _0^2\int \dot {r}\,dt\,\left(
- 2 e^{ - \int\limits_{t_0} ^t \frac{2}{r}\dot{r}\,dt^{\prime}}
(3 - 2\frac{r\partial }{\dot {r}\partial t}\int\limits_{t_0} ^t\dot {r}\,dt^{\prime} / r) -  e^{ -
        \int\limits_{t_0} ^t\frac{2}{r}\dot {r}\,dt^{\prime}} \right) r^2\\
    &= - 2E_\kappa \frac{\partial}{\partial x^h}\frac{\pi ^2q^2}{\varepsilon \varepsilon _0 }\rho _0^2
        \left(  - 3\int {\dot {r}\,dt}
 e^{ - \int\limits_{t_0} ^t \frac{2}{r}\dot{r} dt^{\prime}} r^2\right),
    \end{array}
\end{equation}
where $E_\kappa $ is an arbitrarily fixed value of the Hamiltonian function ${H}$.
%
%======================================
\section{From metric to field equation}
We shall further examine the state of the monolayer at the beginning of the compression $t\to t_0=0$,
    when the monolayer represents itself a two-dimensional gas. This is the case of such concentrations
    $\rho(r,t)$, $\rho\to \rho_0 $, at which the interaction of the particles of the monolayer may be
    neglected. In this case velocities of particles are very small:
    $\dot{r}=\delta \to 0$ and  $r=r(0)+ \dot{r} t$, as $t\to 0$.
Then, taking into account the condition $\dot r = -a_u t$, $a_u >0$, the dependence
    $\rho(r,t)$ gets the following explicit form:
\begin{equation}
\begin{array}{ll}
\rho(r,\dot{r}\to 0)
    &=\left.\rho_0 e^{-\int_0^t {\dot r\over r}dt^{\prime}}
        \right|_{\dot{r}=\delta} \to\left.\rho_0 e^{-\int_0^t{\dot{r} d t^{\prime}\over r(0)+\dot{r}
        t^{\prime}}}\right|_{\dot{r}=- a_u t} = \medskip\\
    &=\rho_0 e^{-\int_0^t{- a_u {t^{\prime}}
    d t^{\prime}\over r(0)-a_u {t^{\prime}}^2}}=
    \rho_0 e^{-{1/2}\int_0^t
    { d (r(0)- a_u {t^{\prime}}^2)\over r(0)- a_u {t^{\prime}}^2}}=
        \left.\rho_0 \left({r(0)\over r(0)-|\dot{r}| t}\right)^{1/2}
        \right|_{\dot{r}\to 0},
    \end{array}
        \label{density}
\end{equation}
where $r(0)=r(t=0)$.
For small compression time $t\to 0$, by taking into account~(\ref{density}), the quadratic metric (\ref{eq32})
    without the kinetic energy part can be transformed to the following
form:
\begin{equation}
\label{conform-metric-func}
    \begin{array}{ll}
    F^2        &\sim
    %\varpropto
    \;\; -2 E_\kappa\frac{\pi ^2q^2}{\varepsilon
\varepsilon _0 }\rho_0^2 \left(  3\int {a_u t\,dt}
e^{ - \int\limits_{t_0} ^t
        \frac{2}{r}\dot {r} dt^{\prime}} r^2\right)
    =-6E_\kappa \frac{\pi ^2q^2}{\varepsilon \varepsilon _0 }
        a_u  \int t\, dt \,  \rho^2 (r,t)\,  r^2\medskip\\
     &=-6E_\kappa \frac{\pi ^2q^2}{\varepsilon \varepsilon _0 }a_u
     \rho_0^2 \,r(0) \int (r(0)t - a_u t^3)\, dt =
        -3E_\kappa \frac{\pi ^2q^2}{\varepsilon \varepsilon _0 }
        a_u  \rho^2(r,t)\,t^2\, (r^2-{1\over 2} a_u r t^2) \medskip\\
     &=-3E_\kappa \frac{\pi ^2q^2}{\varepsilon \varepsilon _0 } a_u
     \rho^2(r,t) \,t^2\,[r^2+ {\cal{O}}( t^2)].
        \end{array}
\end{equation}
Due to the admissible freedom in choosing the fixed value of the Hamiltonian function $H$, one can assume that
\begin{eqnarray} E_\kappa = - 1  \mbox{ for } \int|E'| dt =
\sqrt{3}\pi \label{normalization} .
\end{eqnarray}
The assumption (\ref{normalization}) allows us to rewrite the part (\ref{conform-metric-func}) of
    the quadratic metric  and the metric function (\ref{action-vector-potential}) without the kinetic
    energy part as
\begin{eqnarray}\label{conform-metric-func1}\left.F^2(x,\dot {x})\right|_{t\to 0} \equiv
 F^2(x,\dot {x}, \dot t)  \sim
 \frac{3\pi ^2q^2}{\varepsilon \varepsilon _0 }
 a_u   \rho^2(r,t)\, {%\dot
 t}^2 \,    {%\dot
 r}^2,\\
F(x,\dot {x}, \dot t)  \sim - \sqrt{3}\pi{q\ a_u \over D} {%\dot
t
} {%\dot
r}
\label{action-vector-potential1}
%\end{split}
\end{eqnarray}
Comparing (\ref{conform-metric-func1}) with (\ref{action-vector-potential1}),
    we get the acceleration $a_u$ in its explicit form
\begin{equation}\label{acceleration}\begin{split}
    {q \sqrt{a_u } \over \sqrt{ \varepsilon \varepsilon _0 }} \sim
    \frac{q \rho(r,t)\, D}{\varepsilon \varepsilon _0 }.
\end{split}
\end{equation}
The comparison of (\ref{acceleration}) and (\ref{eq5}) leads to the approximation
    \begin{equation}\label{acceleration1}\begin{split}a_u   \sim
    {\varepsilon \varepsilon _0 \over q^2}
    \left(\frac{q \rho(r,t)\, D}{\varepsilon \varepsilon _0 }\right)^2=
    {\varepsilon \varepsilon _0 \over q^2}\left(\Phi\right)^2  .
    \end{split}\end{equation}
Now, one can consider the kinetic energy part of the action
(\ref{eq8}). It is easy to see that $m\vec {\dot{r}}$ is a flux
density per one particle which is proportional to a flux density
$\int \rho \vec{\dot{ r}}\, d^2%\vec
r$
normalized to the area %площадь
$S_1=%(
\pi r^2(0)~%/ \rho)
$%, приходящуюся на частицу в этом потоке
:
\begin{equation}
m\dot{r} \sim \int {\rho \dot{r} \over  \pi r^2(0) } \, d^2 %\vec
r.
\label{flux}
\end{equation}
 Substituting (\ref{action-vector-potential}, \ref{normalization},
\ref{acceleration1})   into (\ref{eq8}), rewriting the kinetic
energy term in consideration of (\ref{flux}), and integrating,
 one gets the action $F$ in the form
\begin{equation}
F=\int dF = \int mc^{2} dt
- \int \int {\alpha \over \rho_0} {{\dot{r}}^2 \over 2
    \pi r(0)^2}\rho^2  \, dt d^2%\vec
     r -{q\over D}\int  t  {\varepsilon \varepsilon _0 \over q^2}
    \left(\Phi\right)^2 d^2
    %\vec
    r,
    \label{action}
 \end{equation}
with accuracy up to coefficient $\alpha $. We consider the coefficient given by the following relation:
    \begin{equation}\alpha = {|q|  D\over 2\varepsilon \varepsilon _0} \label{inverse-mass}\end{equation}
and choose the initial value $\rho_0$ as
    \begin{equation}\begin{split}\rho(0)\propto {1\over q^2}. \label{beginning_rho(0)}\end{split}\end{equation}
Assuming that $q=- |q|$ and due to $t = \int dt$, after substituting the
expressions \eqref{inverse-mass} and
    \eqref{beginning_rho(0)} into the (\ref{action}), one gets that
 \begin{equation}
 \int dF =  \int mc^{2} dt - {|q| \varepsilon \varepsilon _0\over D}
 \left[\int\int\left({q D\over \varepsilon \varepsilon _0}\right)^2 \,
     {{\dot{r}}^2 \over 4 \pi r^2(0)}\rho^2  \, dt\, d^2%\vec
     r
    - \int dt \,\int d^2%\vec
    r {1%\varepsilon \varepsilon _0
    \over q^2}\left(\Phi\right)^2\right]\label{action1}.
 \end{equation}
Due to the (\ref{density}), the term $\rho^2 \dot{r}^2/(4\, r^2(0))$ which enters into the second term of
    (\ref{action1}) can be replaced by $ \left( {\partial \rho (r,t)\over \partial t}\right)^2$, since:
\begin{equation}\begin{array}{ll}\left( {\partial \rho (r,t)\over \partial t}\right)^2
    &={1\over 4}\dot{r}^2 \rho_0^2   {r(0) \over (r(0)-|\dot{r}|\, t)^3}
        = {\rho^2 \dot{r}^2 \over   4 r^2(0)( 1- |\dot{r}|t/r(0))^2}\medskip\\
    &\approx{  \dot{r}^2 \rho ^2\over 4 r^2(0)( 1- 2|\dot{r}|t/r(0)) }
        \to {  \dot{r}^2 \rho ^2\over 4 r^2(0) }, \qquad\mbox{ as } t\to 0
        \label{action1-second-term}.\end{array}\end{equation}
By substituting (\ref{action1-second-term}) into (\ref{action1}) and by
taking into account the expression
    (\ref{eq5}), one obtains that
    \begin{equation}\begin{array}{ll}\int dF
    &= \int mc^{2} dt- {|q| \varepsilon \varepsilon _0\over D}
     \left[ \int dt{1 \over \pi }  \int d^2%\vec
     r \left({\partial \over \partial t}
    {q\rho (r,t)D\over \varepsilon \varepsilon _0}\right)^2 -
    \int dt\int d^2%\vec
    r {1 \over q^2}
    \left(\Phi\right)^2\right] \medskip\\
    &= \int mc^{2} dt - {|q| \varepsilon \varepsilon _0\over D}
         \left[\int dt {1 \over \pi }  \int d^2%\vec
         r \left({\partial \over
        \partial t}\Phi\right)^2 - \int dt \int d^2%\vec
        r {1 \over q^2}
        \left(\Phi\right)^2\right] \label{action2}.
        \end{array}
\end{equation}
The quantity $\Pi^{\prime}$, which is conjugated to ${\partial \over \partial t}\Phi$ and equal to
    \begin{equation}\begin{split}\Pi' = {2\over \pi }{\partial \over \partial t}\Phi,
        \end{split}\label{field-momentum}\end{equation}
is called {\em the momentum} of the field $\Phi$.
We shall establish the physical meaning of (\ref{action2}). To this aim, it is necessary to add and subtract
    the gradient of the field $\Phi$, which is multiplied with $\varepsilon \varepsilon _0$:
\begin{equation}\begin{split}{|q| \varepsilon \varepsilon _0\over D}
    \left({\Phi\over D } \right)^2 \sim {|q| \varepsilon \varepsilon _0\over D}
    \left(\vec \nabla \Phi\right)^2 .\end{split}\label{field-gradient}\end{equation}
By substituting (\ref{field-momentum}) and  (\ref{field-gradient})
into (\ref{action2}), one obtains that
\begin{equation}
\begin{split}dF =  mc^{2} dt -  dt{ |q| \varepsilon \varepsilon _0\over D}
    \int d^2%\vec
    r\left\{{1\over 2}\Pi' {\partial \over \partial t}\Phi -
    \left(\vec \nabla \Phi\right)^2- {1 \over q^2}\left(\Phi\right)^2
    \right\}-{|q| \varepsilon \varepsilon _0\over D}dt\int d^2%\vec
    r
        \left({\Phi\over D } \right)^2    \label{action3}.
    \end{split}
\end{equation}
According to field theory \cite{Dirac,LandauVol2}, the term in
braces which enters into (\ref{action3}) is
    the Lagrangian density $\mathcal{L}$ of the field $\Phi$ stipulated by electrocapillary interactions:
    \begin{equation}\begin{split}\mathcal{L} ={1\over 2}\Pi' {\partial \over \partial t}\Phi -
        \left(\vec \nabla\Phi\right)^2 - {1 \over q^2} \left(\Phi\right)^2
        \label{lagrangian-density}.\end{split}\end{equation}
This infers that the quantity ${1\over q^2}$ is a squared mass of the
 electrocapillary field $\Phi$.\par
\co{Выберем начальное значение $r(0)$, удовлетворяющее условию сохранения вещества. При этом условии время $T={R_0\over |V|}$, за которое монослой начального радиуса $R_0$ сжимается в точку, приблизительно равно
времени, которое затрачивает частица, движущаяся со скоростью $\dot{r}$,
%из точки $R_0$
на прохождение расстояния не более, чем $|V|T$, до
барьера и, затем вместе с барьером, расстояния не более, чем $|V|T$ до центра монослоя:}
We shall choose the initial value $r(0)$, such that this satisfies the
law of conservation of matter%condition of substance-preservation
. Under these circumstances, the duration $T={R_0\over |V|}$ for
which the initial radius $R_0$ compresses up to zero (a point), is
approximatively equal to the time consumed by the particle which
moves with the speed $\dot{r}$, %from the point $R_0$
travelling
along a distance which does not exceed $|V|T$, to the barrier, and
further, jointly with the barrier, along a distance less than
$|V|T$ to the center of the monolayer:
    \begin{equation}\begin{split}{R_0\over |V|} < {2|V|T\over |\dot{r}|},
        \label{beginning1_r_0-condition}\end{split}\end{equation}
where $V$ is the compression speed.
\co{Условие (\ref{beginning1_r_0-condition}) для произвольной точки $r(0)$ монослоя переписывается в виде}
The condition (\ref{beginning1_r_0-condition}) for the arbitrary point $r(0)$ of the monolayer
    can be expressed as
    \begin{equation}\begin{split}{r(0)\over |V|} < {2|V|t\over |\dot{r}|} .
        \label{beginning1_r_0-condition1}\end{split}\end{equation}
The velocity  limitation (\ref{beginning1_r_0-condition1}) allows us to choose $r(0)$ as
    \begin{equation}\begin{split}{r(0)\over |V|t} \le {2|V|\over |\dot{r}|} .
        \label{beginning1_r_0}\end{split}\end{equation}
For the initial conditions \eqref{beginning_rho(0)} and \eqref{beginning1_r_0}, the kinetic energy
    term entering into (\ref{action2}) can be rewritten in the following form:
     \begin{equation}\begin{array}{l}{|q| \varepsilon \varepsilon _0\over D}
        \left[\left({q D\over \varepsilon \varepsilon _0}\right)^2 \,\int  {{\dot{r}}^2 \over 4 \pi r^2(0)
        ( 1- 2|\dot{r}|t/r(0))}\rho^2  \, d^2%\vec
     r\right]={|q| \varepsilon \varepsilon _0}
        \left[\left({q \over \varepsilon \varepsilon _0}\right)^2 \,\int  {\rho \over 4 \pi r^2(0)}\rho_0
        {{\dot{r}}^2 \over 1- {\dot{r}^2\over V^2}} \, D\, d^2%\vec
        r\right]\medskip\\
    \qquad={ \varepsilon \varepsilon _0 \over |q|}\left[\left({q \over \varepsilon \varepsilon _0}
\right)^2 \,\int  {\rho \over 2 \pi r^2(0)}
{{\dot{r}}^2 \over 1- {\dot{r}^2\over V^2}} \, d^3
        %\vec
R\right]={ 2\varepsilon \varepsilon _0 \over |q|\, D}\left[\left({ q\, D \over 2\varepsilon
        \varepsilon _0}\right)^2 \,\int  {\rho_V \over  \pi r^2(0)\, D}
        {{\dot{r}}^2 \over 1- {\dot{r}^2\over V^2}} \, d^3%\vec
    R\right]\medskip\\
    \qquad=\int{ P^2\over 2Z_\kappa} \, {\rho_V \over  \pi r^2(0)\, (D/2)}\,
    d^3%\vec
    R,
        \label{kinetic-energy-action}
    \end{array}
\end{equation}
where
    \begin{eqnarray}&&Z_\kappa  ={|q| D\over 2\varepsilon \varepsilon _0}, \label{free-mass}\medskip\\
        %\end{eqnarray}\begin{eqnarray}
        &&{P^2}    = Z^2_\kappa {{\dot{r}}^2 \over 1- {\dot{r}^2\over V^2}}, \label{free-momentum}\medskip\\
        %\end{eqnarray}\begin{eqnarray}
        &&\rho_V =\left\{\begin{array}{lr}\rho & \mbox{if}\ z=0,\\0& \mbox{if}\ z\neq 0.\end{array}\right.
        \label{free-particle-density}\end{eqnarray}
It follows then that the occurrence of $Z_\kappa$ is related with the
existence of %the field $\Phi$    in
the additional third dimension, i.e., this is a mass of the
particle which freely moves in the 3d-space and has the squared
momentum $P^2$ defined by the expression similar to a relativistic
one in the plane    of the monolayer. The number density
describing a particle distribution  in this 3d-space is $\rho_V$.
    The squared three-dimension-mass $Z_\kappa ^2$ determines the quadratic charge $q^2$, as follows:
\begin{equation}
2 (Z_\kappa ^2/\beta) =q^2 %{ D^2\over 2\varepsilon^2 \varepsilon _0^2}
        ,\qquad\beta = {D^2\over 2\varepsilon^2 \varepsilon^2 _0}
        \label{free-mass1}.
\end{equation}
In principle, we may double the dimension of the space from 3 to 6, by means of complexifying the variables,
    and expecting, by analogy,  identical relations between the mass of the particle and the mass of the field,
    which is related to the exceeding complex coordinate. This analogy has already been considered in \cite{gg},
    if this 6-dimensional manifold is regarded as a 2-dimensional spinor space with two time-like coordinates.
In \cite{gg}, the free motion of a particle with mass $Z$ takes place in a 6-dimensional manifold with two
    space-like complex coordinates (a flat 2-dimensional spinor space) and two time-like coordinates.
The projection of the movement of such a particle with the symmetry group $SO(6)$-equations
    \cite[equations (22) and (61)]{gg} - on the 4-dimensional space-time (3,1) is described by the wave
    equation of the hydrogen-like atom with the symmetry group $SO(4)$-equation \cite[equation (49)]{gg} --
    at the non-relativistic limit.
Here the square of the charge $e$ and mass $m_e$ of the electron are proportional to the square
    of the mass $Z$ \cite{gg}:
    \begin{eqnarray}2 Z ^2 = e^2 .\label{complex-free-mass}
    \end{eqnarray}
Comparing \eqref{free-mass1}  with \eqref{complex-free-mass},
    we conclude that the projection of the motion of the free particle from the spinor
    2-dimensional complex space and 2-dimensional time onto the space-time (3,1) is at
    the non-relativistic limit a complex analogue of the motion in the 3-dimensional space
    endowed with Finsler metric.\par
At the end we remark a peculiarity of the obtained action
(\ref{action3}), namely the fact that it contains a re-normalized
field mass $m_f$ due to the adding the term in
(\ref{field-gradient}):
\begin{equation}
\begin{split}
dF =  mc^{2} dt -  dt {|q| \varepsilon \varepsilon _0\over D} \int
        d^2%\vec
        r \left\{ {1\over 2}\Pi' {\partial \over \partial t}\Phi -\left(\vec \nabla \Phi\right)^2 -
        m_f^2 \left(\Phi\right)^2\right\}
\end{split}\label{action4}
\end{equation}
where
    \begin{equation}m_f^2 = {1 \over q^2} - {1\over D ^2}.\label{renorm-mass}\end{equation}
In this way, at the limit of small concentrations, the energy of
molecules in the compressed Langmuir monolayer on the subphase
surface is determined by the rest mass of the particles and the
electro-capillary field
    contribution to action owing  an inner "spinor"\    degree of freedom. The physical meaning of this contribution
    is the jump of the surface tension of the interphase boundary between liquid and air, while dropping amphiphilic
    molecules on the surface of the subphase under the conditions  of the moving barrier.\par\smallskip
Thus, the fundamental function $F$ of our problem is not defined on the whole tangent space, but only on certain
    distributions over $\widetilde{TM}$ -- the slit tangent space. In the general case, one considers the vertical
    sub-bundle $VTM = Ker(d\pi )$ of the vector bundle $(TTM, d\pi, TM)$ provided by the kernel of the linear
    mapping $d\pi$, and the supplementary sub-bundle $HTM = Ker(N)$ is provided by a Barthel connection
    $N : TTM \to  VTM,\ N = (N^j_i (x,y))_{i,j\in \overline{1,n}}$, which satisfies $N \circ i = \left.
    id\right|_{VTM}$ (where $i : VTM \to TTM$ is the canonic inclusion). This leads to the Whitney decomposition
    \cite{Shen,Balan2009SODE}
    \begin{equation}TTM = HTM \oplus VTM,\label{Whitney}\end{equation}
and induces a local adapted bases for the appropriate sections of these sub-bundles,
    \begin{equation}\left\{ \delta_i = {\delta \over \delta x_i}= {\partial \over \partial x_i} +
        N^j_i {\partial \over \partial y_j}\right\}\subset \Gamma (HTM) , \quad
        \left\{ \dot{\partial}_i = {\partial \over \partial y_i}= \right\}\subset \Gamma (VTM) .
        \label{contra-basis}\end{equation}
The related dual splitting $T^*TM = H^*TM \oplus V^*TM$ leads to similar dual bases,
    \begin{equation}\left\{ dx_i \right\} \subset \Gamma (H^*TM) , \quad\left\{ \delta y_i =
        d y_i + N^i_j d x_j \right\} \subset \Gamma (V^*TM) .\label{co-basis}\end{equation}
In the following section we shall consider the  generalized Finsler structure, whose fundamental function $F$
    is a metric function of the monolayer compressed by the barrier, considered at different speeds $V$.
%===========================================================================
\section{Monolayer structuring process in terms of Finsler space invariants}
\co{В этом параграфе мы проанализируем геометрические инварианты движения частицы в монослое which is compressed by the barrier и дадим физическую интепретацию особенностей их поведения.}
In this section we analyze the geometric invariants of the
particle motion within the monolayer, which is compressed by the
barrier and provide physical interpretations of the corresponding
specific behavior.\par
The electro-capillary potential energy $U_{s}(r, t)$ of the
particle at a point $r(t)$ of
    the mono-molecular layer has the following explicit form \cite{Grushevskaya2011}:
\begin{equation}\label{eq49}
\begin{array}{ll}U_s (r,t)
    =& - \frac{\pi ^2q^2}{\varepsilon \varepsilon _0}
    \frac{\rho _0^2}{R_0^2 }\left( {\left( { - \frac{4}{3}r^5 +
        \frac{16}{15}
    (\vert V\vert t)r^4 + \frac{1}{30}(\vert V\vert t)^2r^3 } \right.}
    \right.\medskip\\
    &+\left. { \frac{1}{45}(\vert V\vert t)^3r^2 + \frac{1}{45}(\vert V\vert t)^4r + \frac{2}{45}(\vert V
        \vert t)^5 - r^5\frac{\vert V\vert }{\dot{r}}} \right)\left. { e^{\frac{2
        \vert V\vert t}{r}} -\frac{4}{45}\frac{(\vert V\vert t)^6}{r}  \mbox{Ei} \left[{\frac{2\vert V
        \vert t}{r}} \right]} \right),\end{array}
\end{equation}
where $\mbox{Ei}\left[{\frac{2\vert V\vert t}{r}} \right]$ is the exponential integral and $m$ is the molecular mass.
The non-relativistic action $dl$ is defined by the relation
\begin{equation}
    dl=mc^2 dt - Ldt .\label{(55)}
\end{equation}
 The substitution of the expressions \eqref{eq50} and \eqref{eq49} into
    (\ref{(55)}) gives
 \begin{equation}
 \begin{array}{ll}dl =
    &mc^2\dot {\xi } - m\frac{\dot {r}^2 + r^2\dot {\varphi }^2}{2
        \dot {\xi}} - \frac{\pi ^2q^2}{\varepsilon \varepsilon _0 }\frac{\rho _0^2}{R_0^2 }
        \left( {\left( { - \frac{4}{3}r^5 + \frac{16}{15}(\vert V\vert t)r^4 + \frac{1}{30}(\vert V\vert t)^2
        r^3  } \right.} \right. \medskip\\
    &+\left. { \frac{1}{45}(\vert V\vert t)^3r^2 +\frac{1}{45}(\vert V\vert t)^4r +\frac{2}{45}(\vert V
        \vert t)^5 - r^5\frac{\vert V\vert }{\dot{r}}\dot {\xi }} \right)
        \left. {e^{\frac{2\vert V\vert t}{r}} -\frac{4}{45}\frac{(\vert V\vert t)^6}{r}\mbox{Ei}
        \left[{\frac{2\vert V\vert t}{r}} \right]} \right)\dot {\xi },\end{array}\label{56}\end{equation}
where  $\dot {\xi }$,  $\dot r$, and $\dot {\varphi }$ correspond to the derivatives of $t$, $r$, and  $\phi $
    with respect to the evolution parameter $\tau $, respectively. We rewrite this action $dl$ as
    \begin{equation}\label{eq52}dl = A  {\dot {\xi }^2\over \dot{r}} + B\dot{\xi}-C
        \frac{(\dot {r}^2 + r^2\dot {\phi }^2)}{2c^2\dot {\xi }},\end{equation}
where the parameters $A,\ B,\ C$ are given by
    \begin{equation}\label{eq53}\begin{array}{l}
    A = p\left| V \right|r^5e^{\frac{2\left| V \right|t}{r}},\medskip\\
    B = mc^2 - p\left( {\left( { - \frac{4}{3}r^5 + \frac{16}{15}(\vert V\vert t)r^4 + \frac{1}{30}
        (\vert V\vert t)^2r^3 + \frac{1}{45}(\vert V\vert t)^3r^2 +
        \frac{1}{45}(\vert V\vert t)^4r  }
        \right.} \right.\medskip\\
    \qquad\quad\left. { + \frac{2}{45}(\vert V\vert t)^5} \right)\left. {e^{\frac{2\vert V\vert t}{r}} -
        \frac{4}{45}\frac{(\vert V\vert t)^6}{r}\mbox{Ei}\left[ {\frac{2\vert V\vert t}{r}} \right]}\right),
        \medskip\\
    C = mc^2, \quad \mbox{ where }p = \frac{\pi ^2q^2}{\varepsilon \varepsilon _0}
        \frac{\rho _0^2 }{R_0^2 },\ \dot {r} = \frac{x\dot {x} + y\dot{y}}{r}.\end{array}\end{equation}
The length element $F$ (which provides a Berwald-M\`oor type metric,
\cite{Matsumoto,Garasko}) can be
    defined by means of the volume element $d\widetilde V $, which
    depends on the differentials of the
    Cartesian coordinates.
Due to the stated above relation $\dot{r} \sim a_u t$, it follows that the 4d-volume $V_4$ is equal to
    $d\vec l  \wedge \vec {\dot\xi} \sim (\vec {\dot r} \wedge\vec{\dot\xi}\,) ^2$ at $dt  =\vec {\dot\xi} $,
    and the equation (\ref{conform-metric-func1}) for the square $F^2$ of the metric function $F \sim dl$
    can be written as:
\begin{equation}
\begin{split}
F^2  \sim (\dot t \dot r)^2\sim|(\vec {\dot\xi}
        \cdot d\vec r) |^2 \sim \dot \xi |d\vec l| =\dot \xi\, dl.
\end{split}\label{conform-metric-func2}
\end{equation}
Hence, (\ref{eq52}) determines a quadratic metric $F^2 \sim dl{\dot\xi} $
    (\ref{conform-metric-func2}) of the following form:
    \begin{equation}\label{quadratic-metric}F^2  = A  {\dot {\xi }^3\over \dot
    {r}} + B  \dot {\xi }^2- C\frac{(\dot {r}^2 + r^2\dot {\phi }^2)}{2c^2},\end{equation}
where the parameters $A,\ B,\ C$ are given by the system~(\ref{eq53}).
    Renormalizing the relation (\ref{quadratic-metric}) by using
\begin{equation}
\label{eq54}
\dot{\tilde {r}} = \frac{\dot {r}}{c}, \quad \dot {\tilde x} =
        \frac{\dot {x}}{c}, \quad \dot {\tilde y%\phi
} = \frac{\dot {y%\phi
}}{c}.
\end{equation}
we express the equation of the metric $F$ in terms of these
normalized coordinates:
\begin{equation}\label{normal-quadratic-metric}
        F^2  = \tilde A  {{\dot{  \xi}   }^3\over \dot{\tilde r}} + B  {\dot{  \xi }}^2     - C
        \frac{({\dot{\tilde r } }^2 + r^2{\dot{\tilde \phi } }^2)}{2},
        \quad \tilde A  ={A\over c}.
\end{equation}
For simplifying the notations, we shall further omit in our considerations the symbol "$\sim $".

\begin{figure}%[hbt]
\vspace*{-6cm}

A)\includegraphics[width=4.cm,height=3.5cm,angle=0]{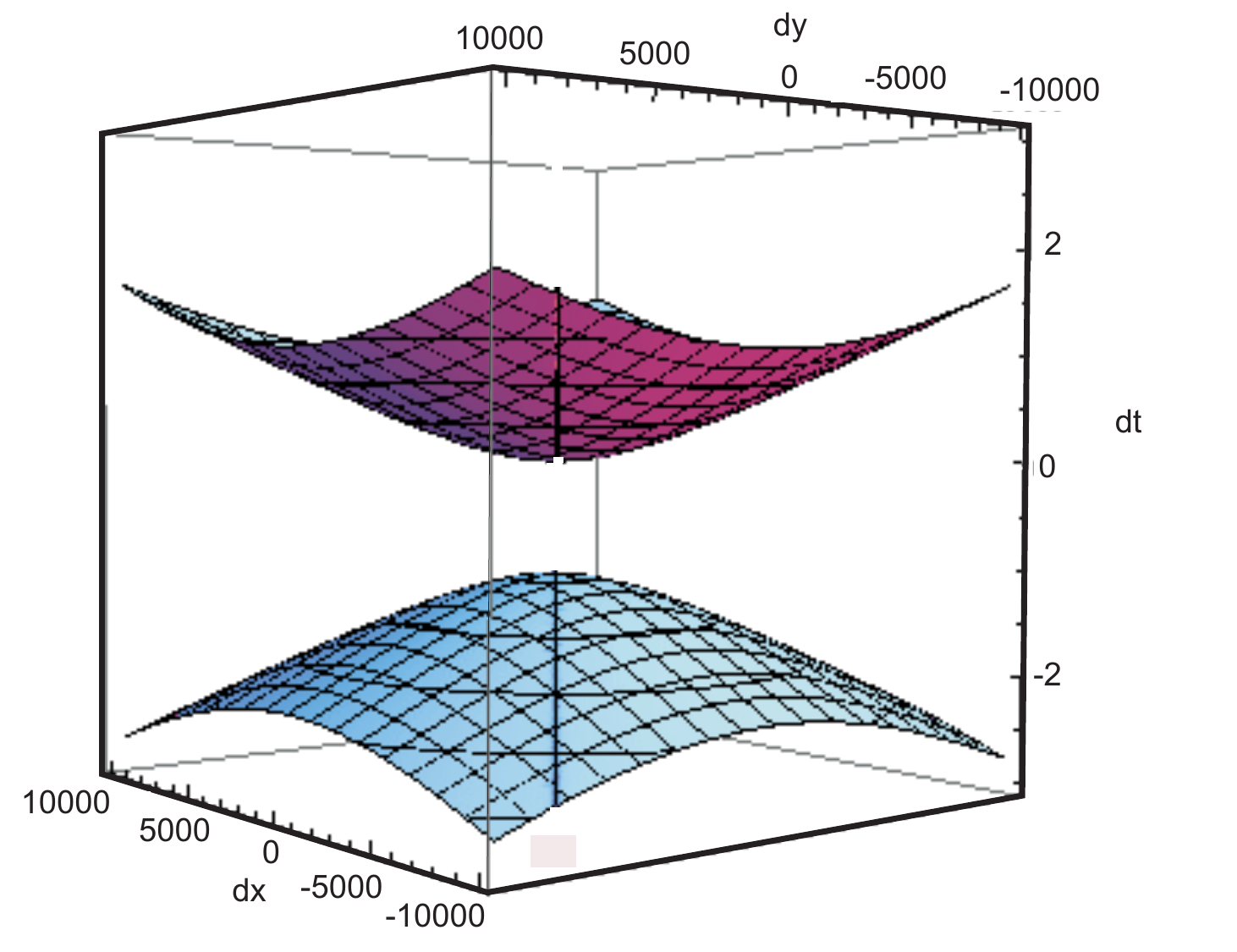}
\includegraphics[width=8.cm,height=8.5cm,angle=0]{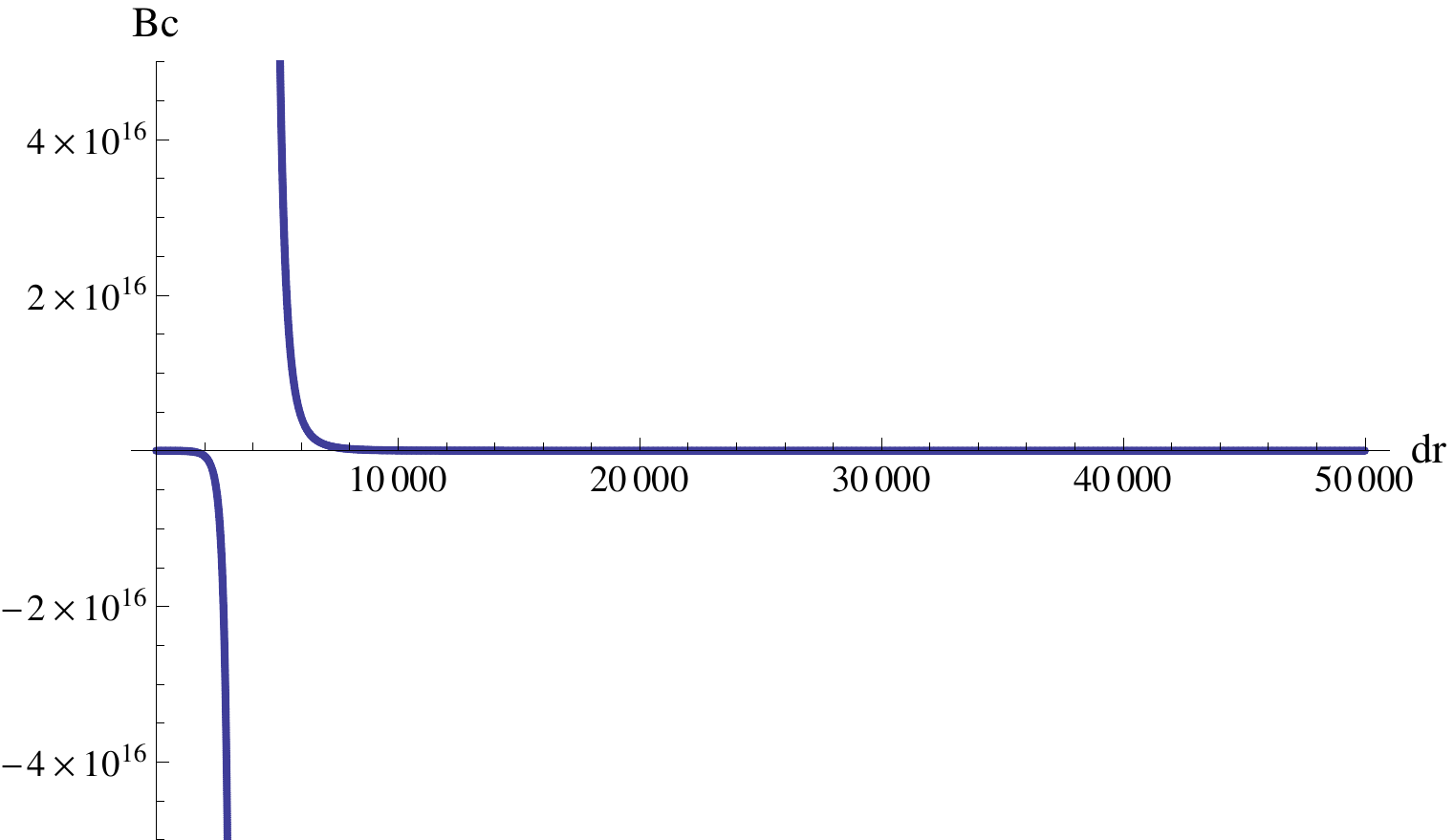}
\includegraphics[width=4.cm,height=3.5cm,angle=0]{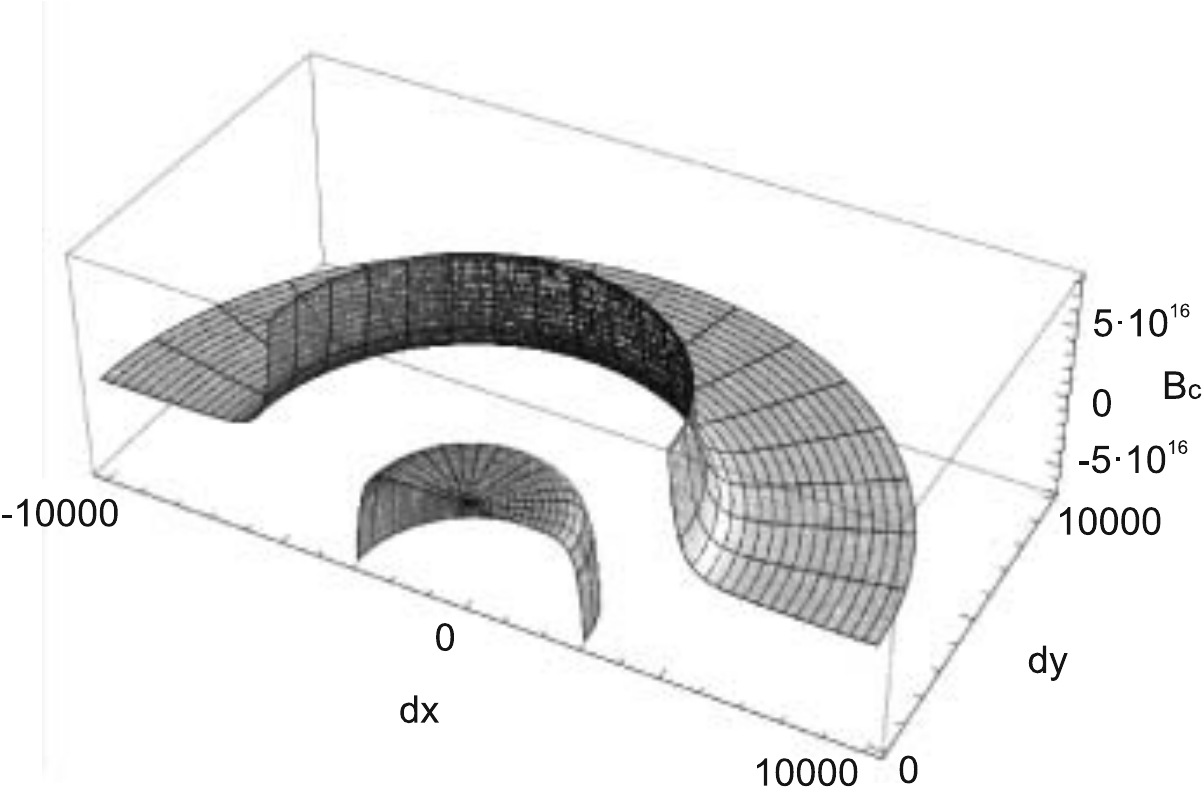}

\vspace*{-5cm}
B)\includegraphics[width=4.cm,height=3.5cm,angle=0]{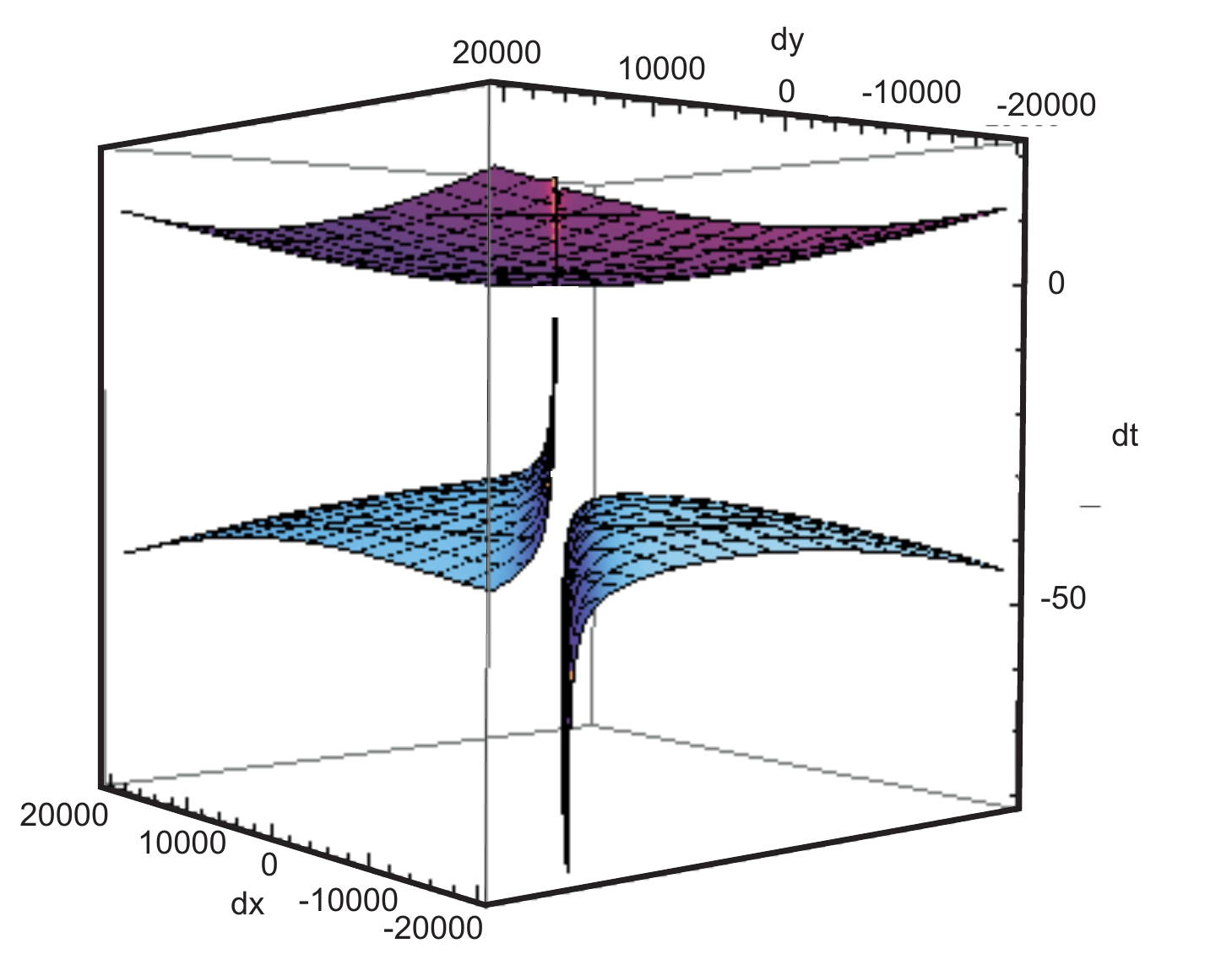}
\includegraphics[width=8.cm,height=8.5cm,angle=0]{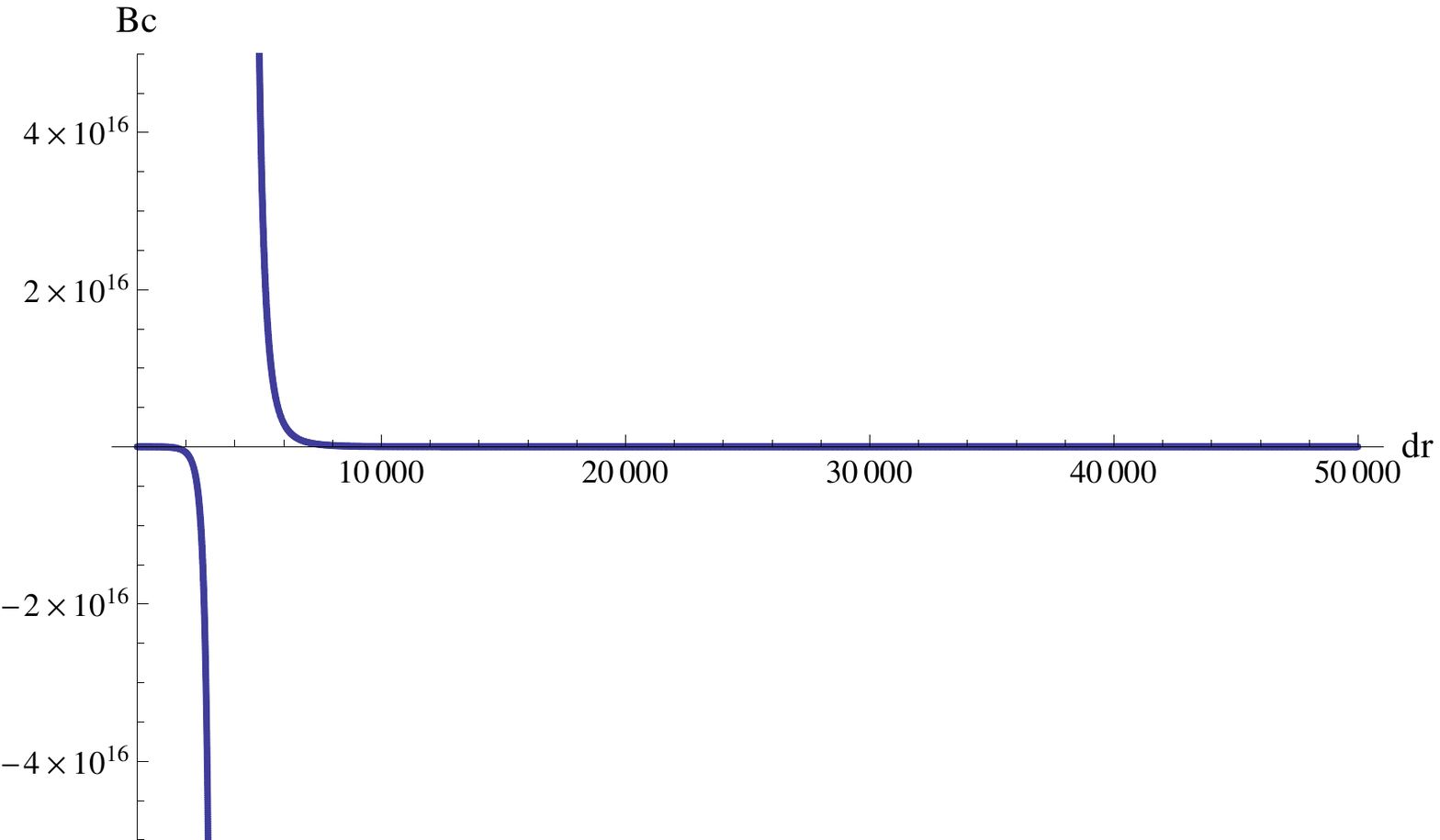}
\includegraphics[width=4.cm,height=3.5cm,angle=0]{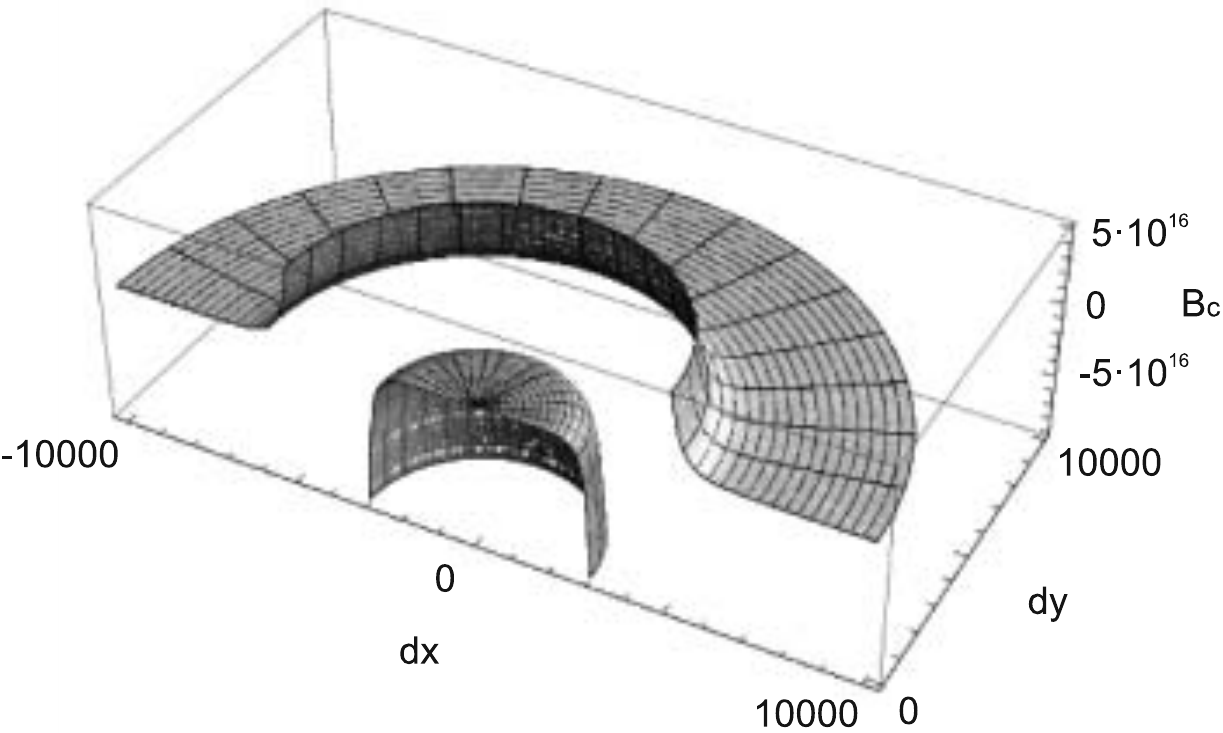}

\vspace*{-5cm}
C)\includegraphics[width=4.cm,height=3.5cm,angle=0]{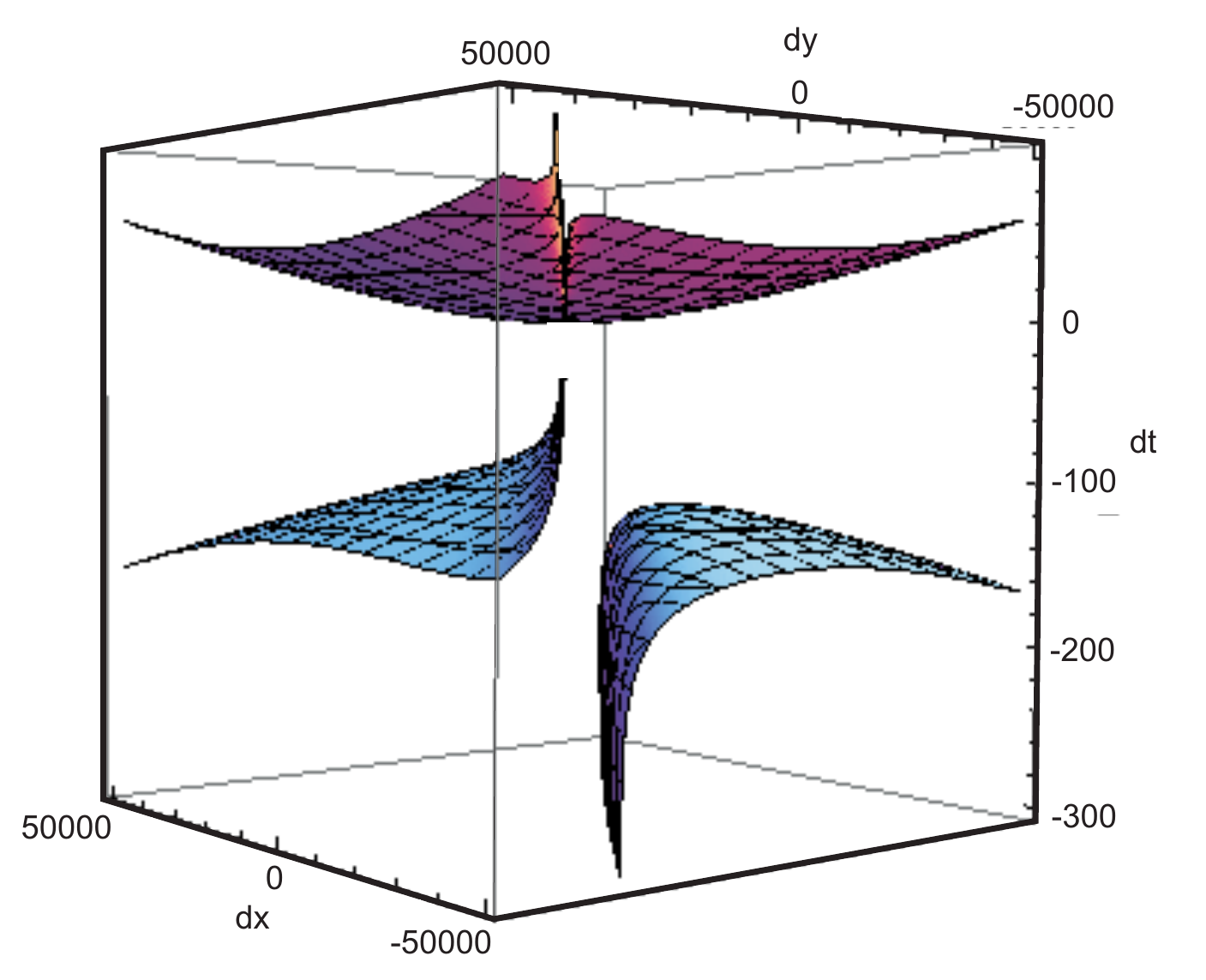}
\includegraphics[width=8.cm,height=8.5cm,angle=0]{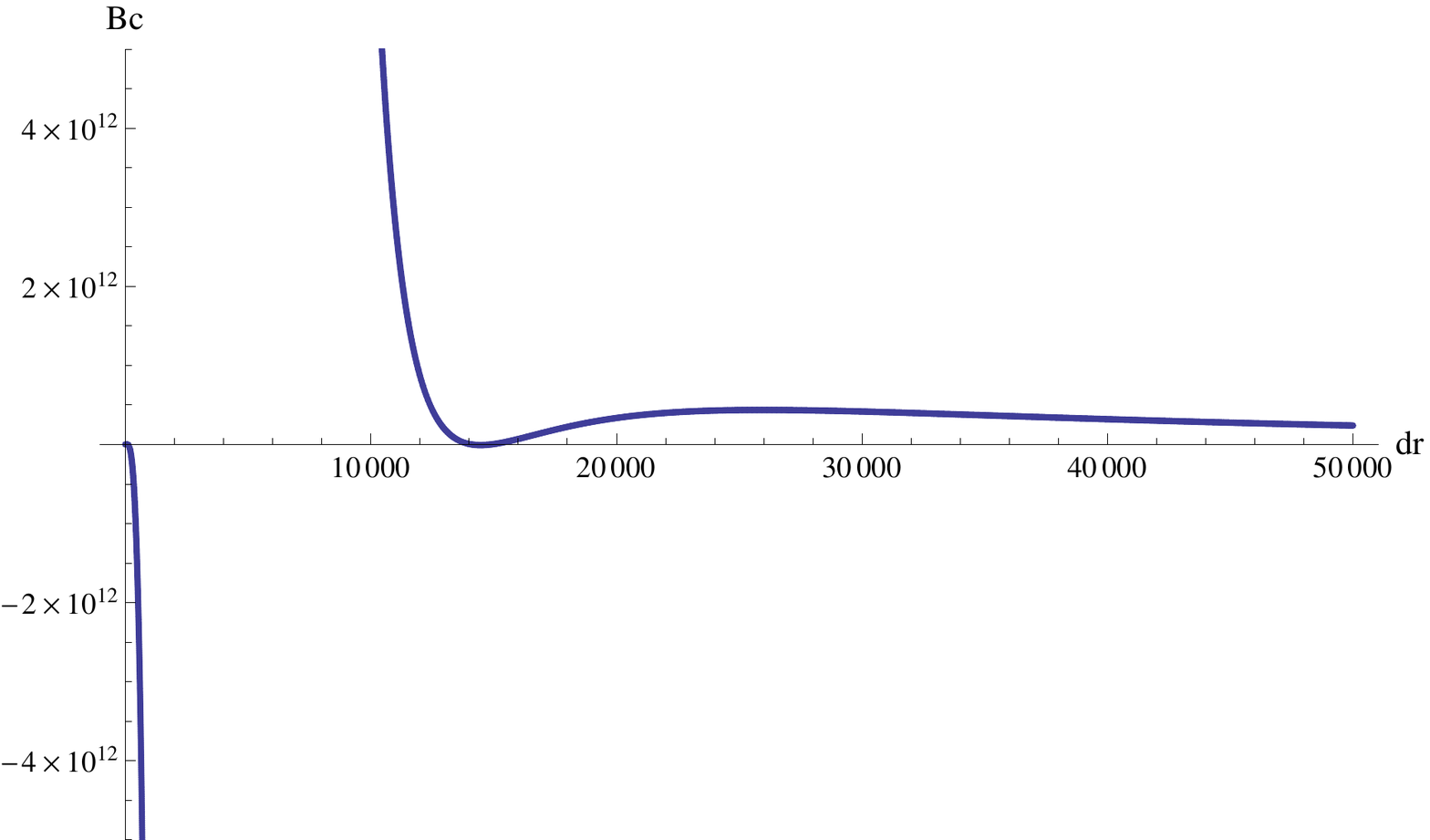}
\includegraphics[width=4.cm,height=3.5cm,angle=0]{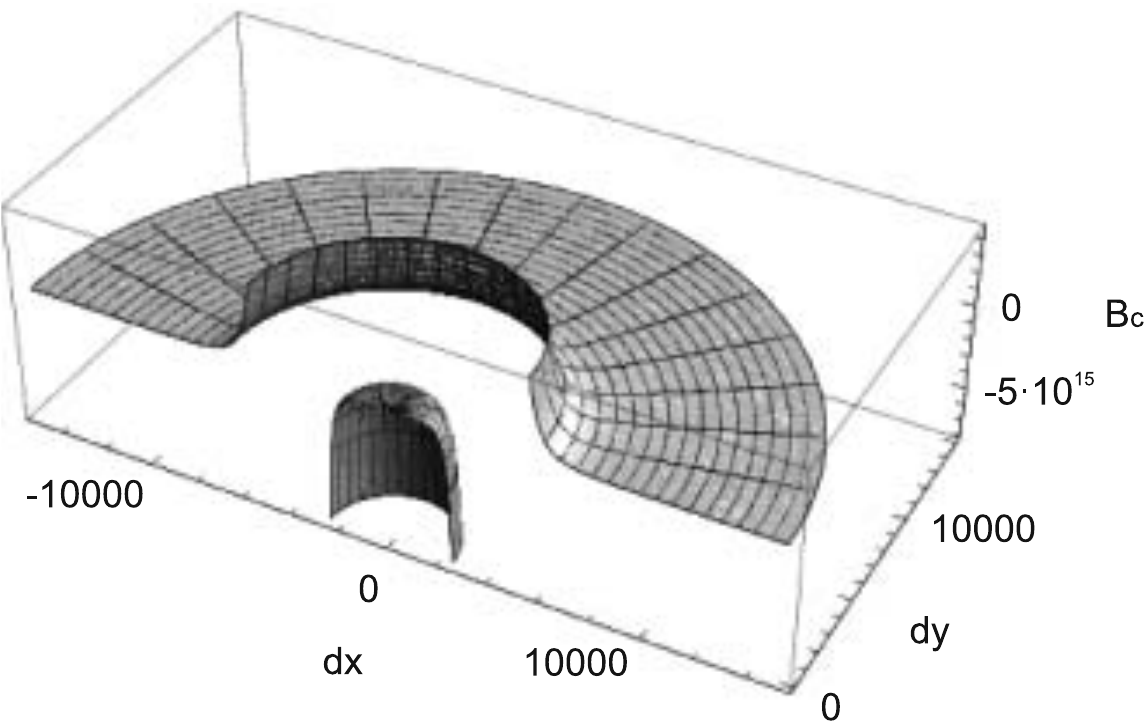}

\vspace*{-5cm}%
D)\includegraphics[width=4.cm,height=3.5cm,angle=0]{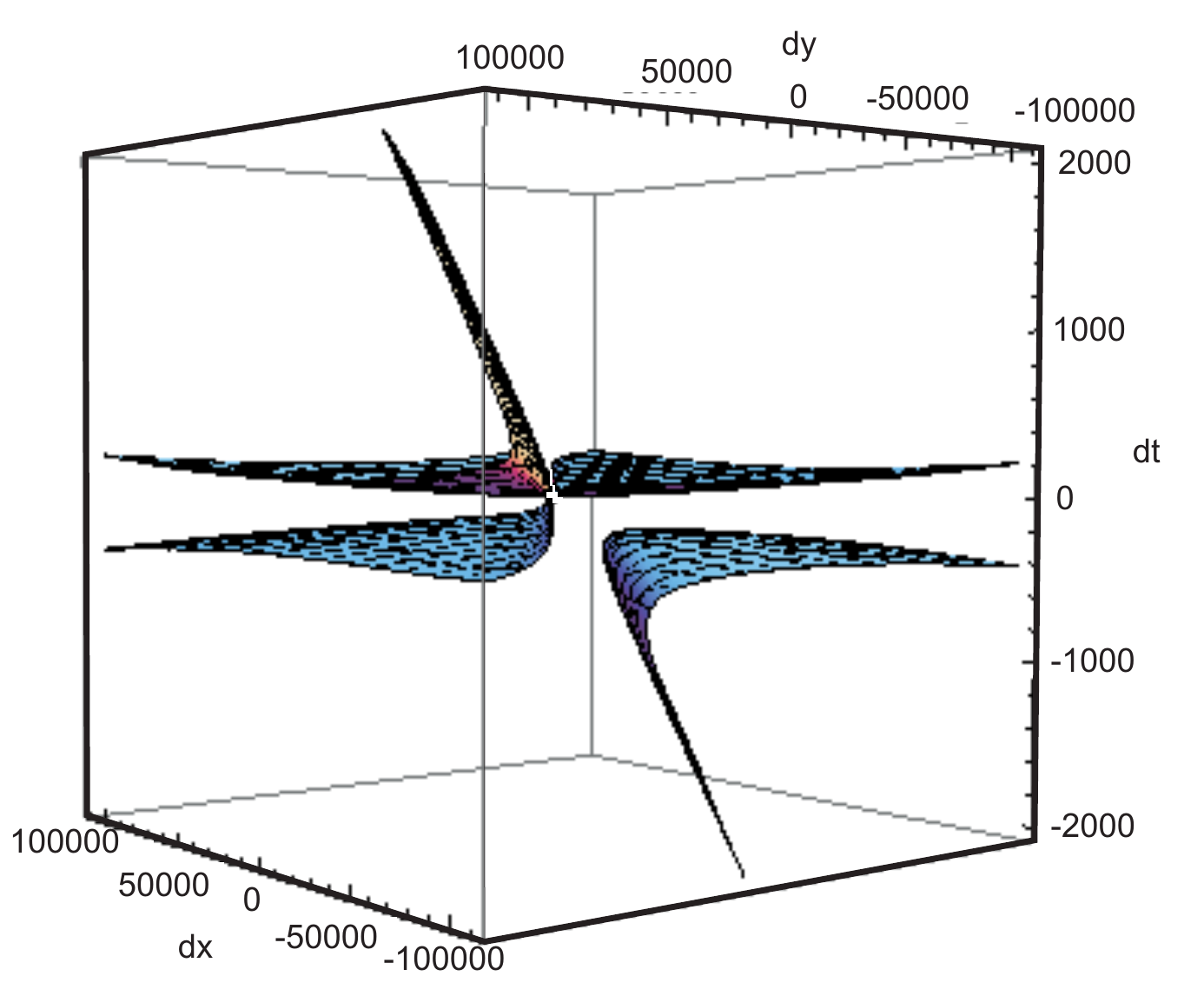}
\includegraphics[width=8.cm,height=8.5cm,angle=0]{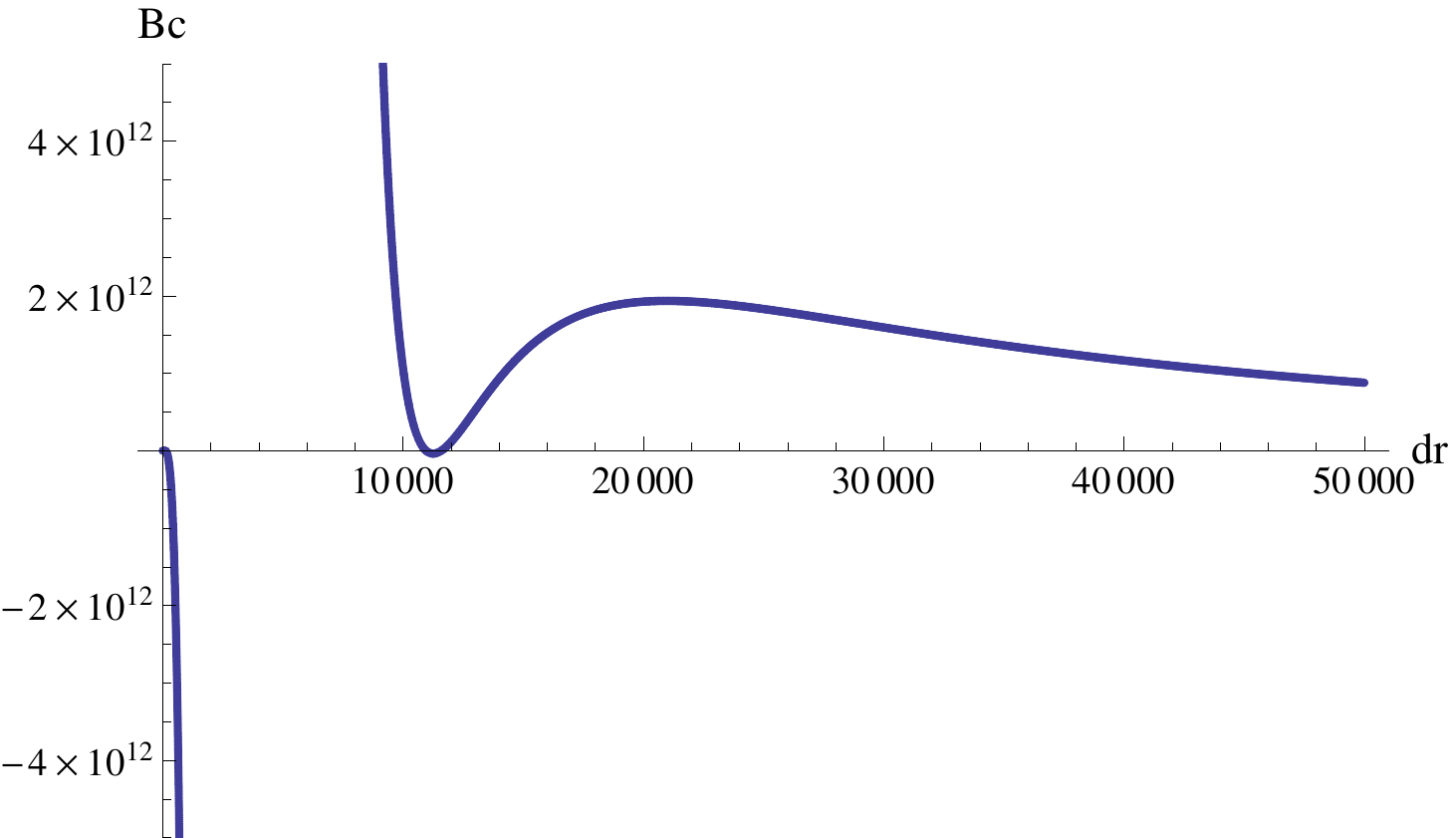}
\includegraphics[width=4.cm,height=3.5cm,angle=0]{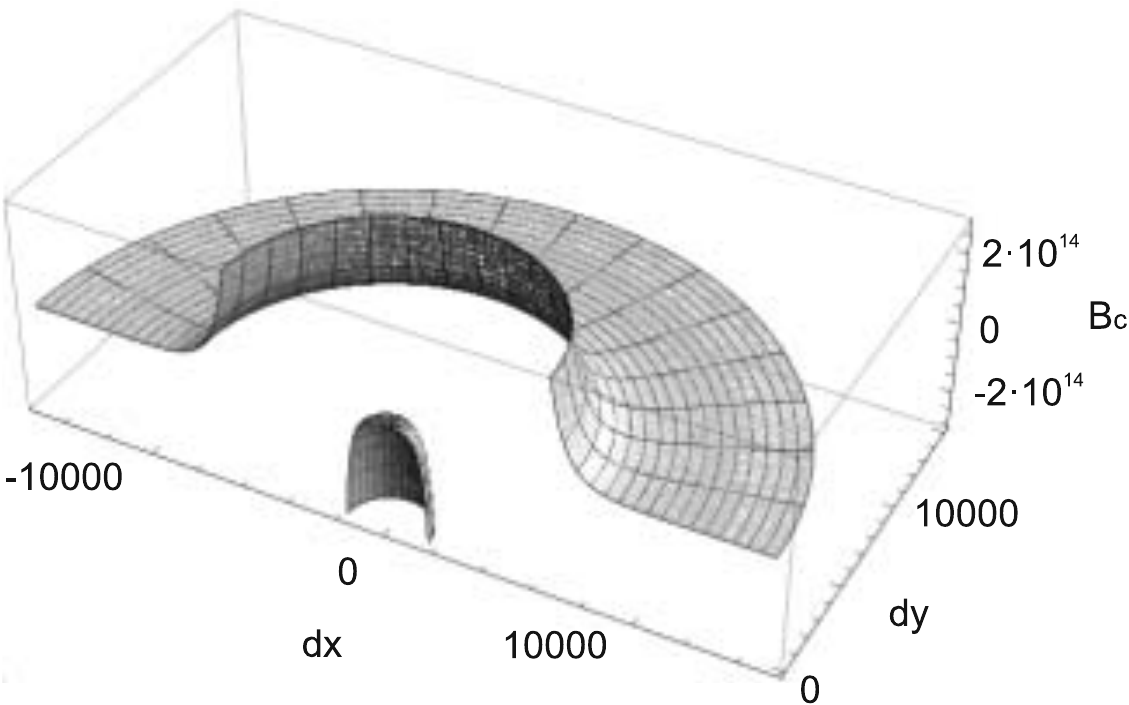}

\vspace*{-5cm}
E)\includegraphics[width=4.cm,height=3.5cm,angle=0]{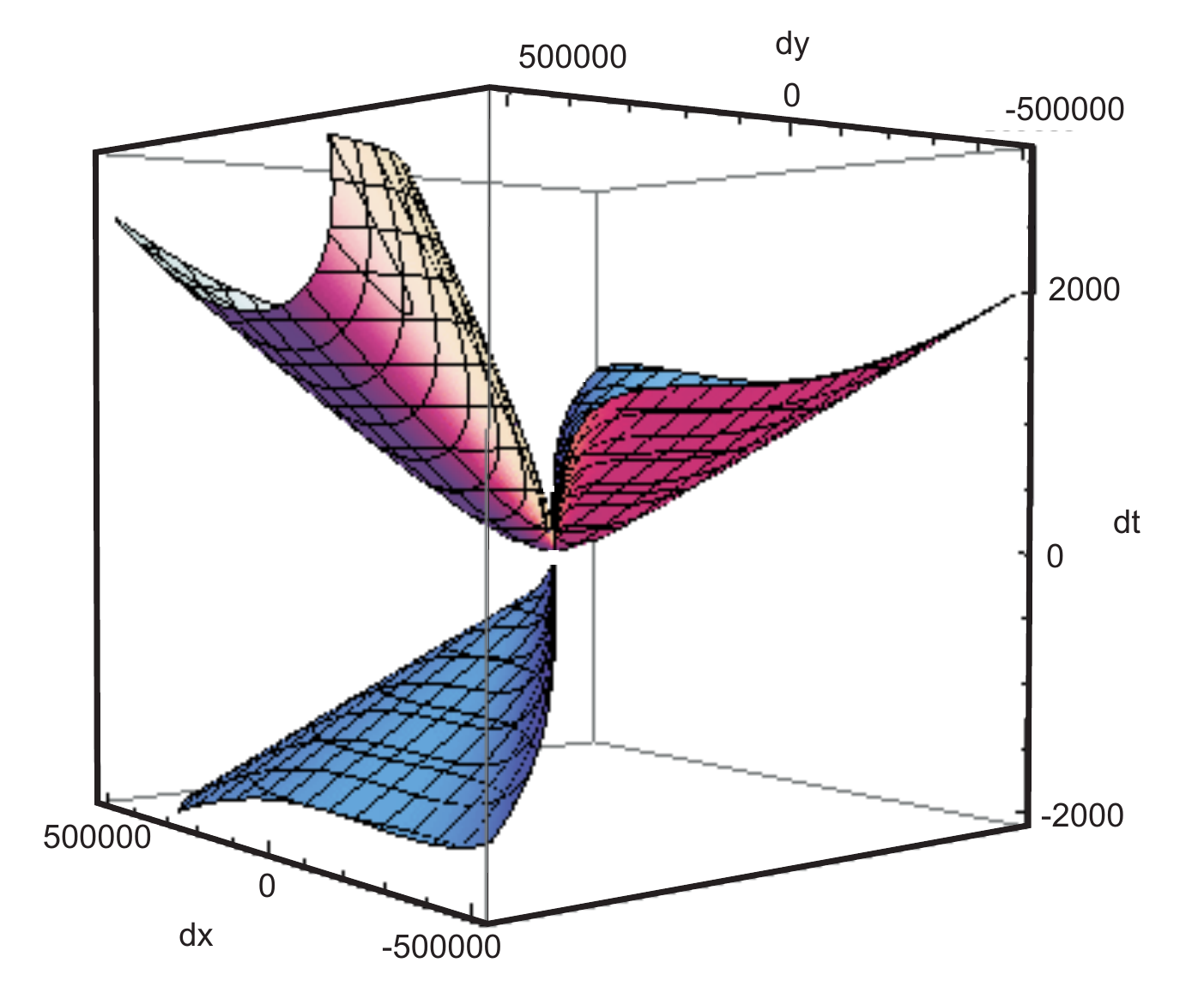}
\includegraphics[width=8.cm,height=8.5cm,angle=0]{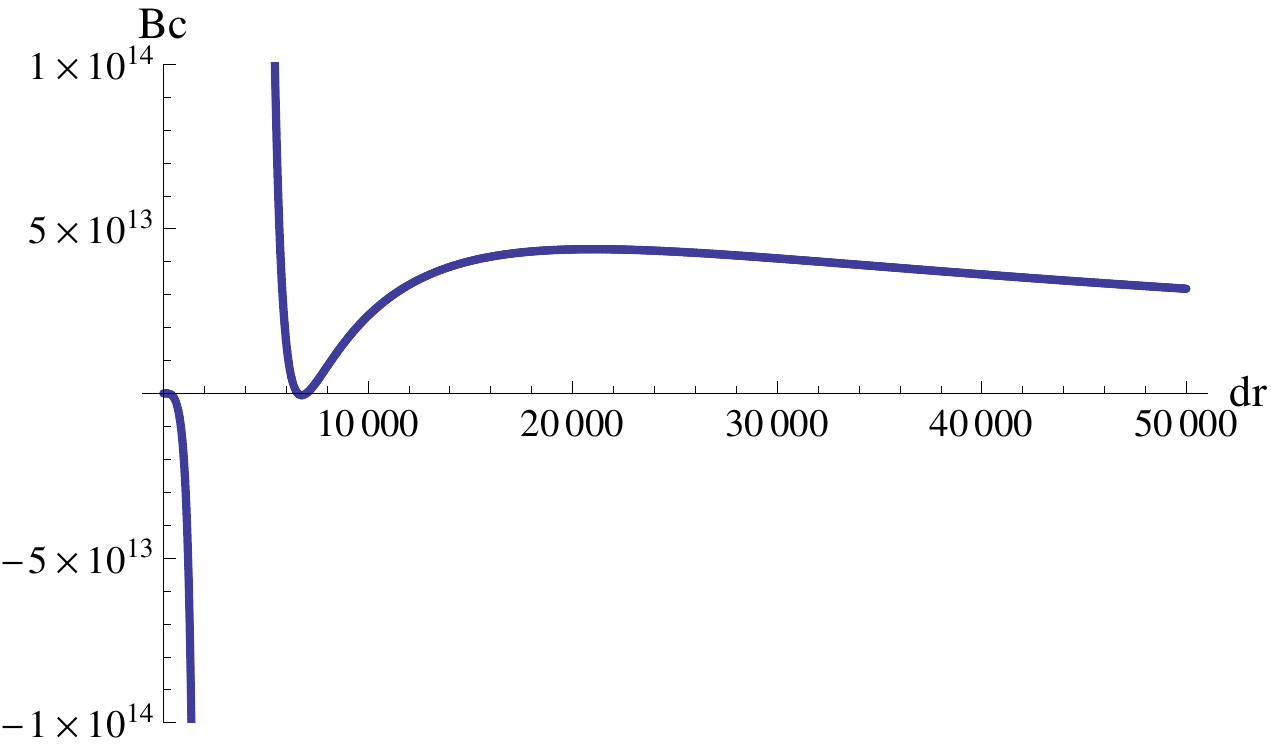}
\includegraphics[width=4.cm,height=3.5cm,angle=0]{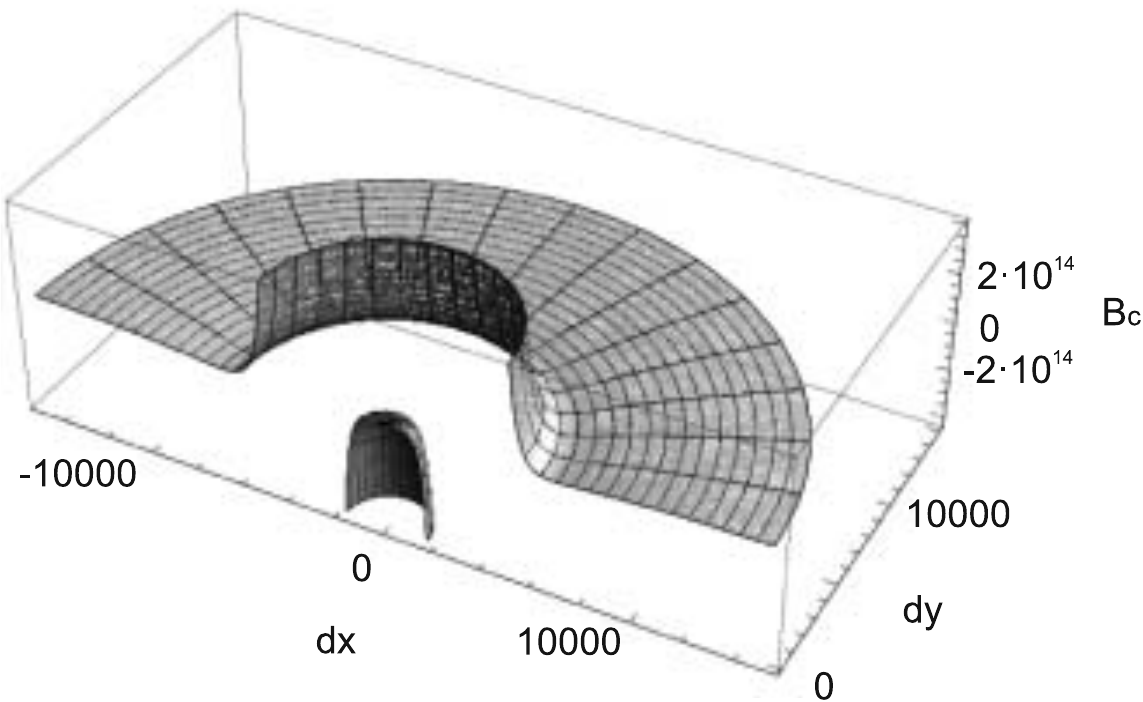}

\vspace*{-5cm}
F)\includegraphics[width=4.cm,height=3.5cm,angle=0]{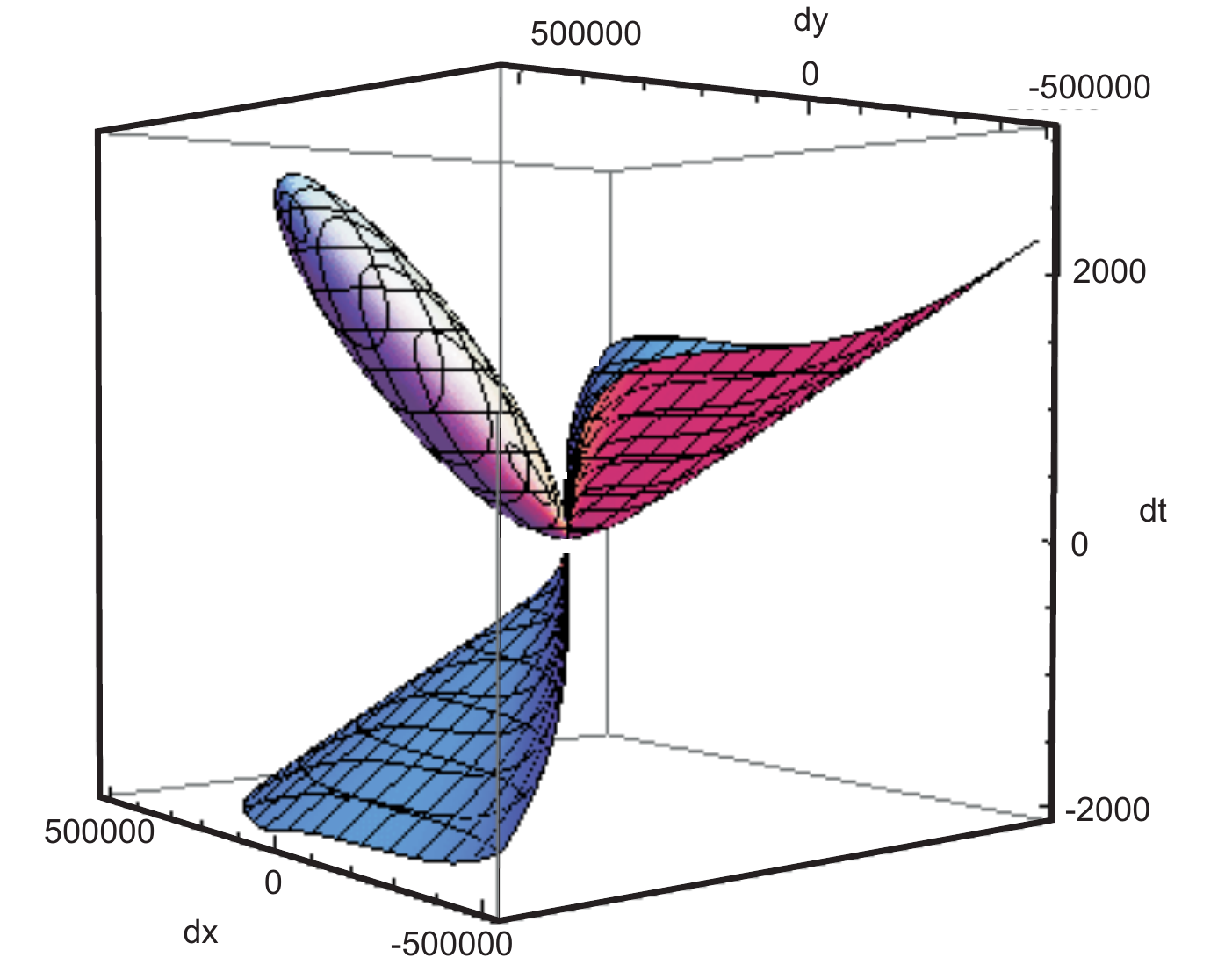}
\includegraphics[width=8.cm,height=8.5cm,angle=0]{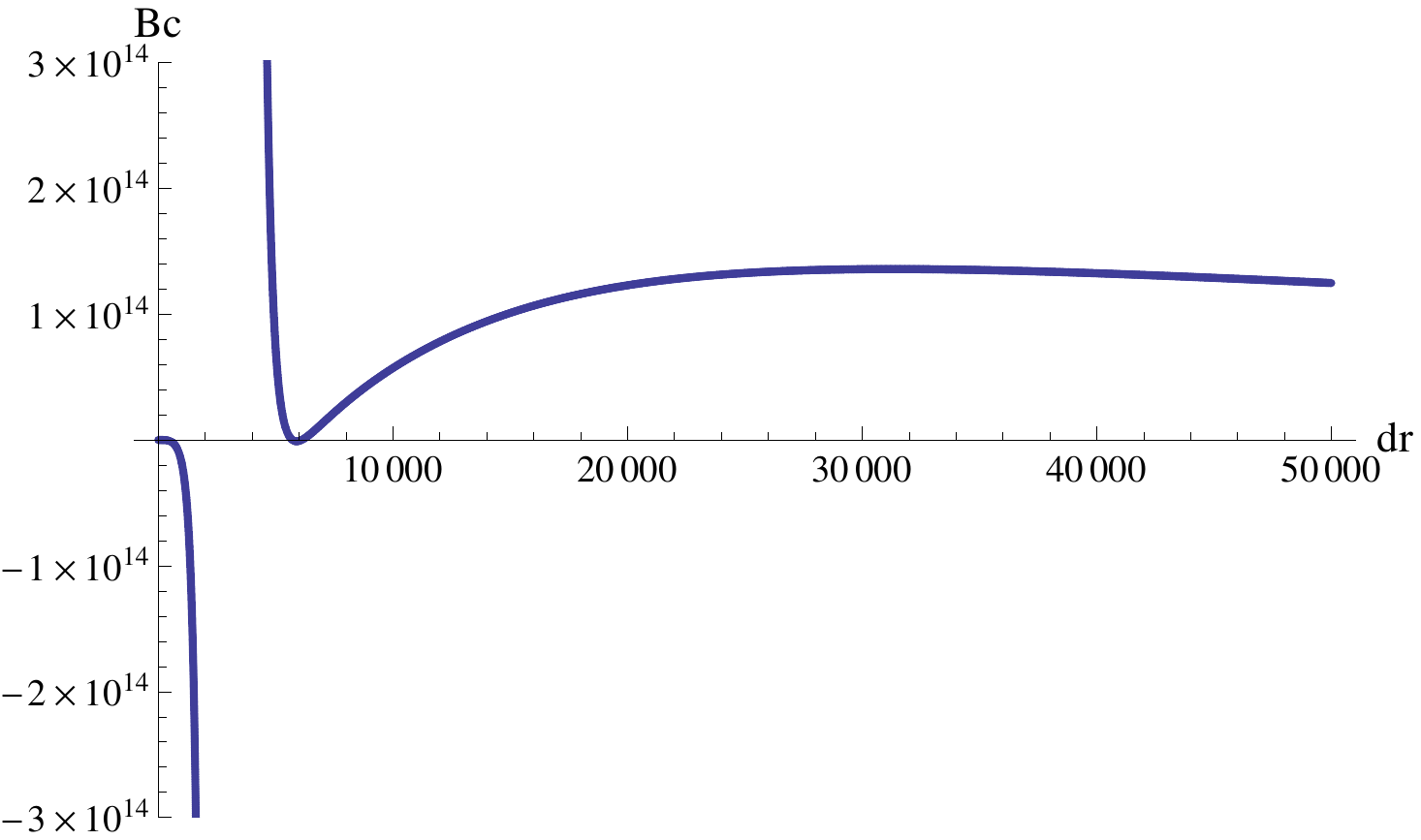}
\includegraphics[width=4.cm,height=3.5cm,angle=0]{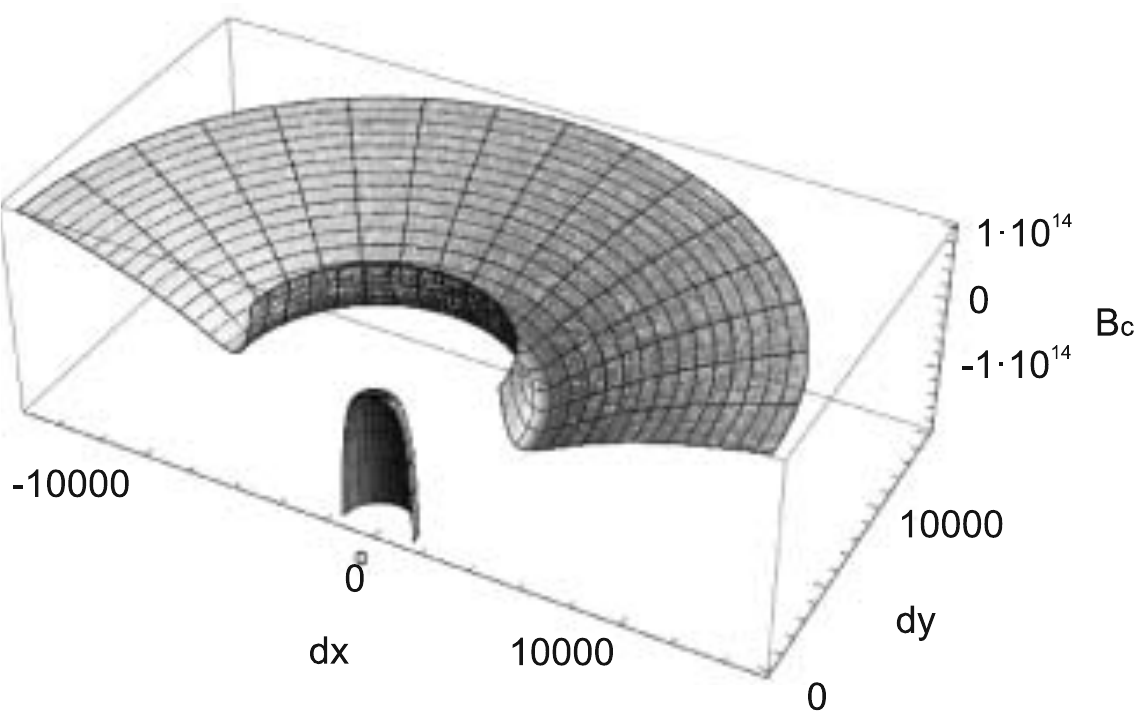}
\caption{\small Indicatrices (from the left) which relate to the
Berwald curvature $B_C$ (2d- and 3d-images in
    the middle and right respectively) having a single peculiarity  for $x_0=$0.0796186,\ 0.079618558,\
    0.079618557,\ 0.0796185568,\ 0.0796185566, \ 0.07961855657,\
    0.0796185565,  for $B=0.922721$ (A), 0.0315175 (B), 0.0102984 (C),
    0.00605459 (D), 0.00181076 (E), 0.00117419 (F) %, $-0.00031115$ (g)
    respectively. Here $C = 423\times 10^{-10}$, $A= 4.78079 \times 10^7$, $V = 15.0$.}
\label{figure5}
\end{figure}

\begin{figure}%[hbt]
\vspace*{-7cm}

A)\includegraphics[width=4.cm,height=3.5cm,angle=0]{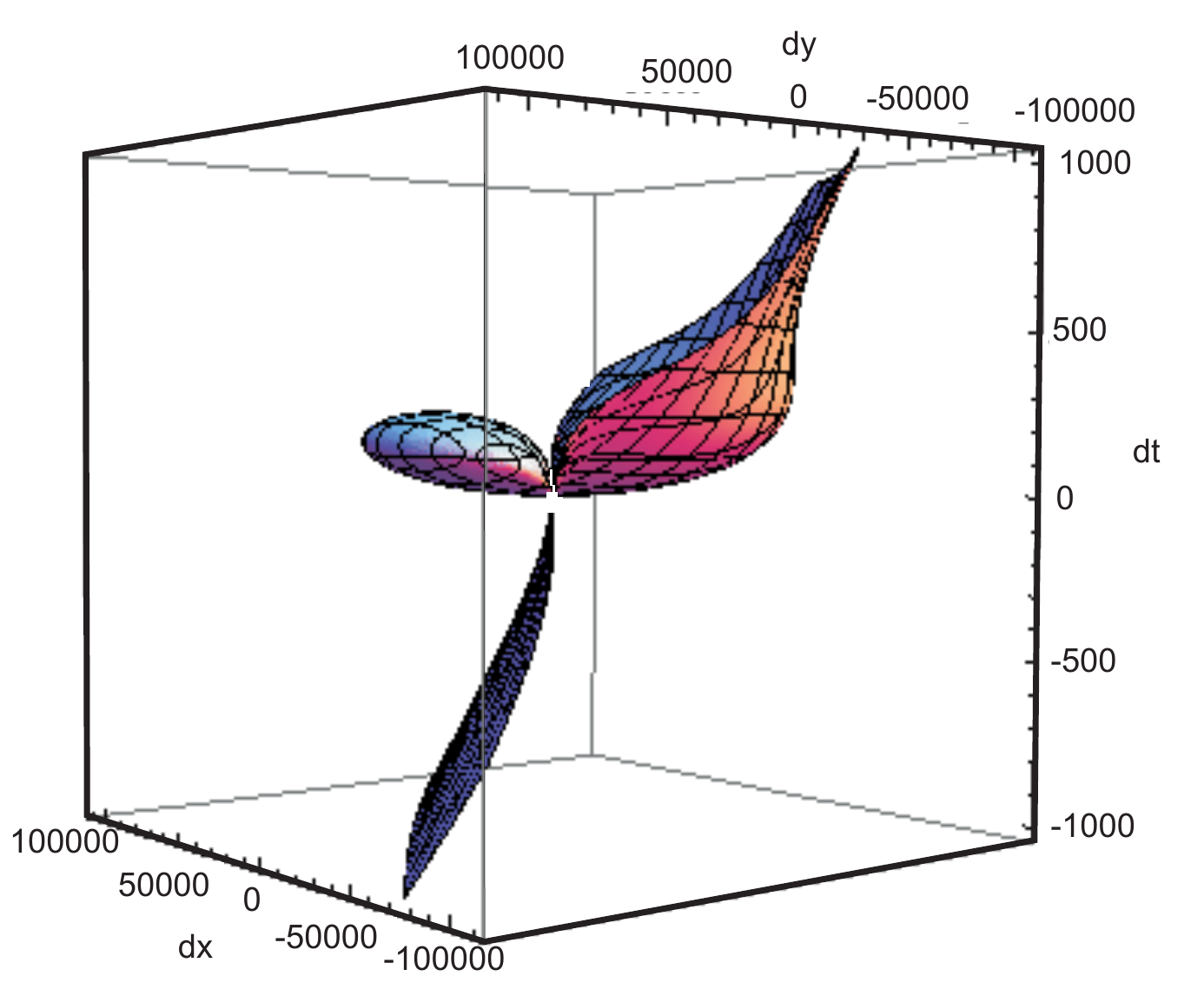}
\includegraphics[width=6.cm,height=8.5cm,angle=0]{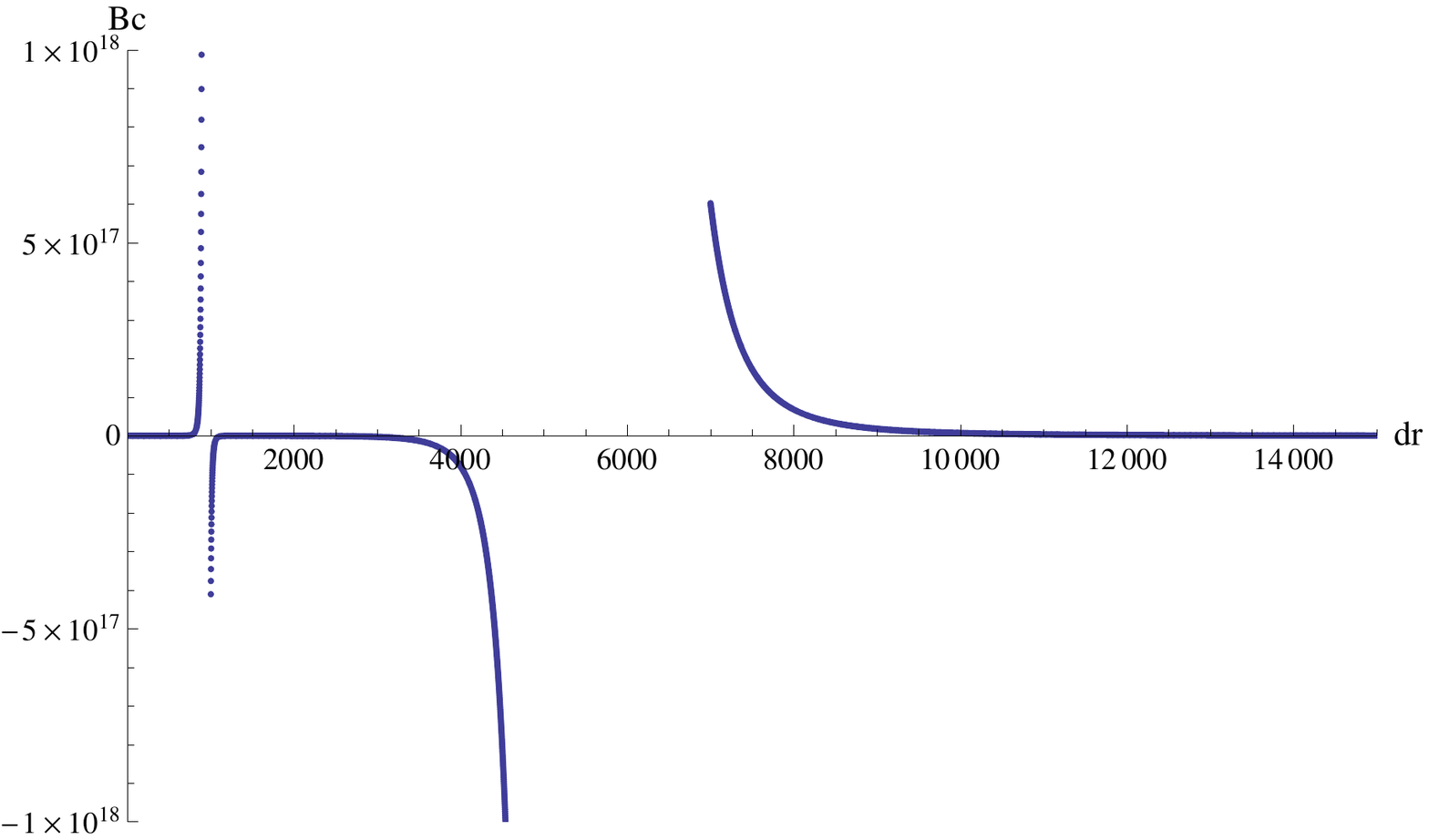}
\hspace*{-0.5cm}\includegraphics[width=8.cm,height=3.5cm,angle=0]{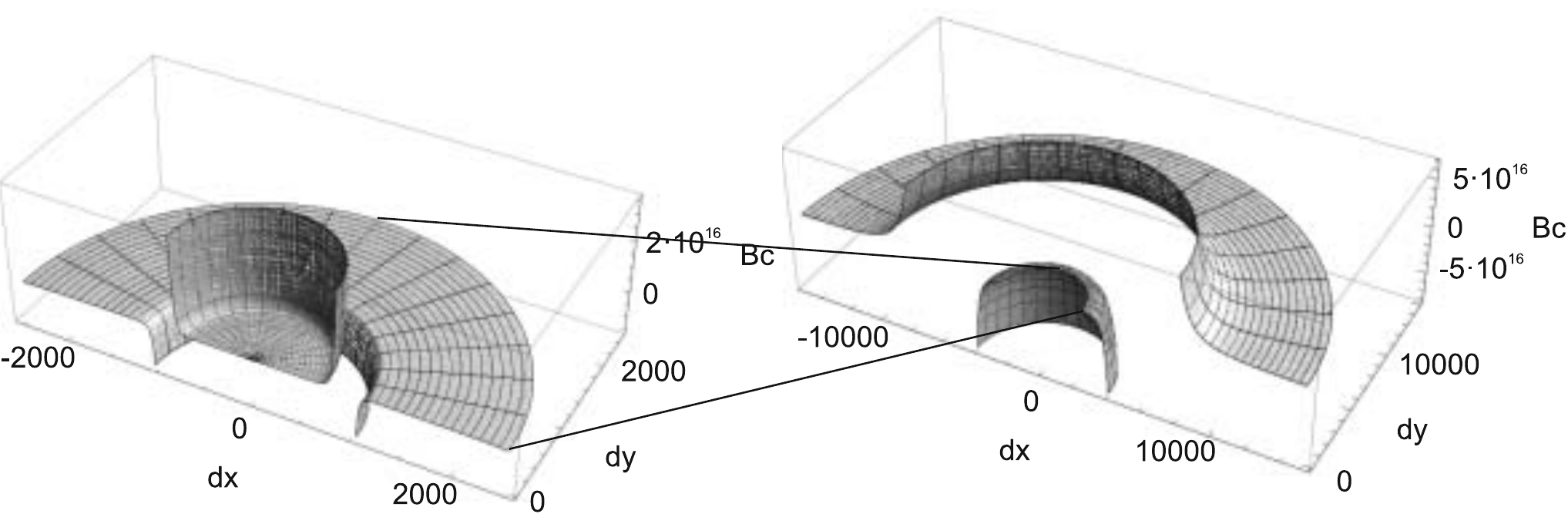}

\vspace*{-3cm}
B)\includegraphics[width=3.cm,height=3.5cm,angle=0]{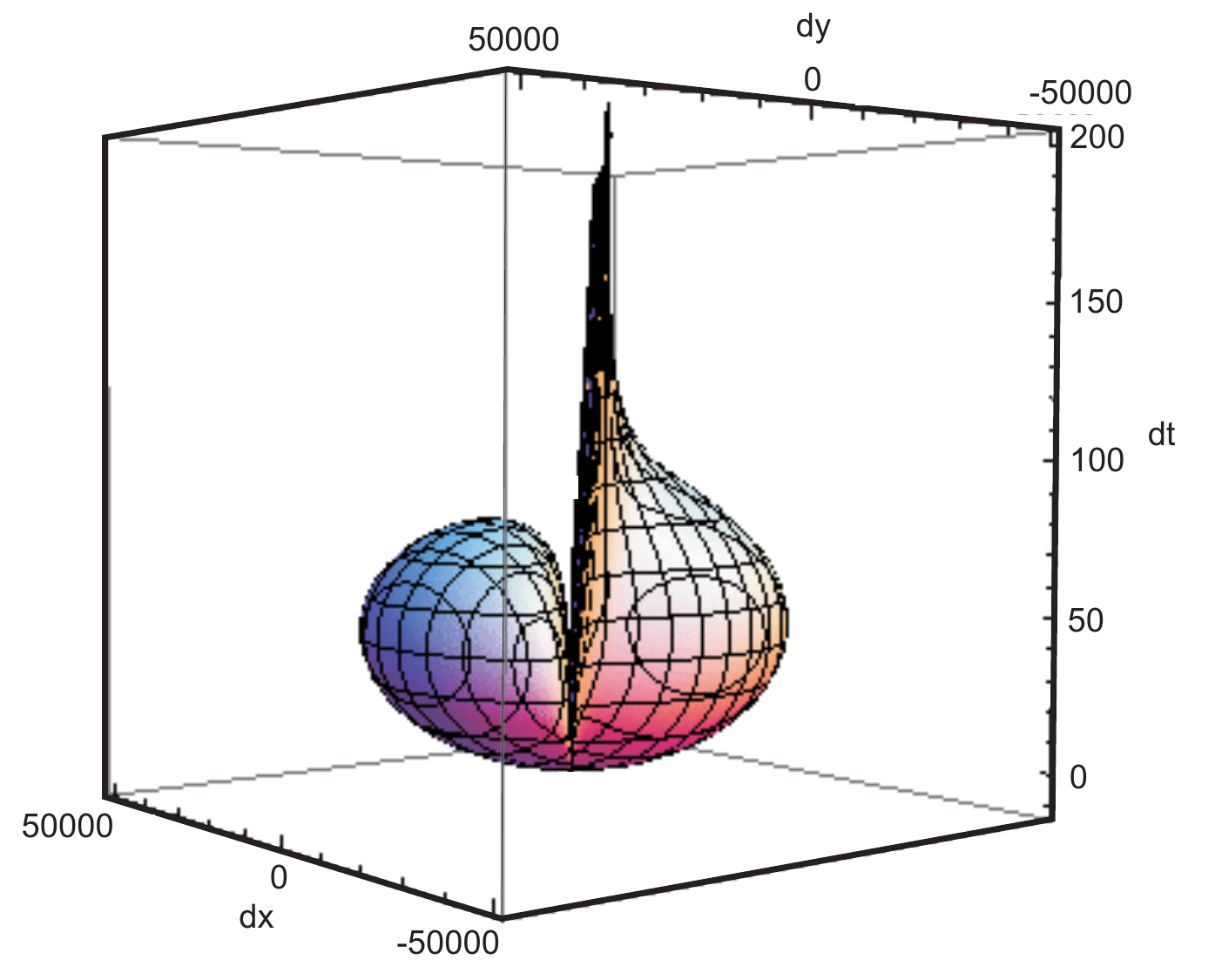}
\includegraphics[width=3.25cm,height=7.5cm,angle=0]{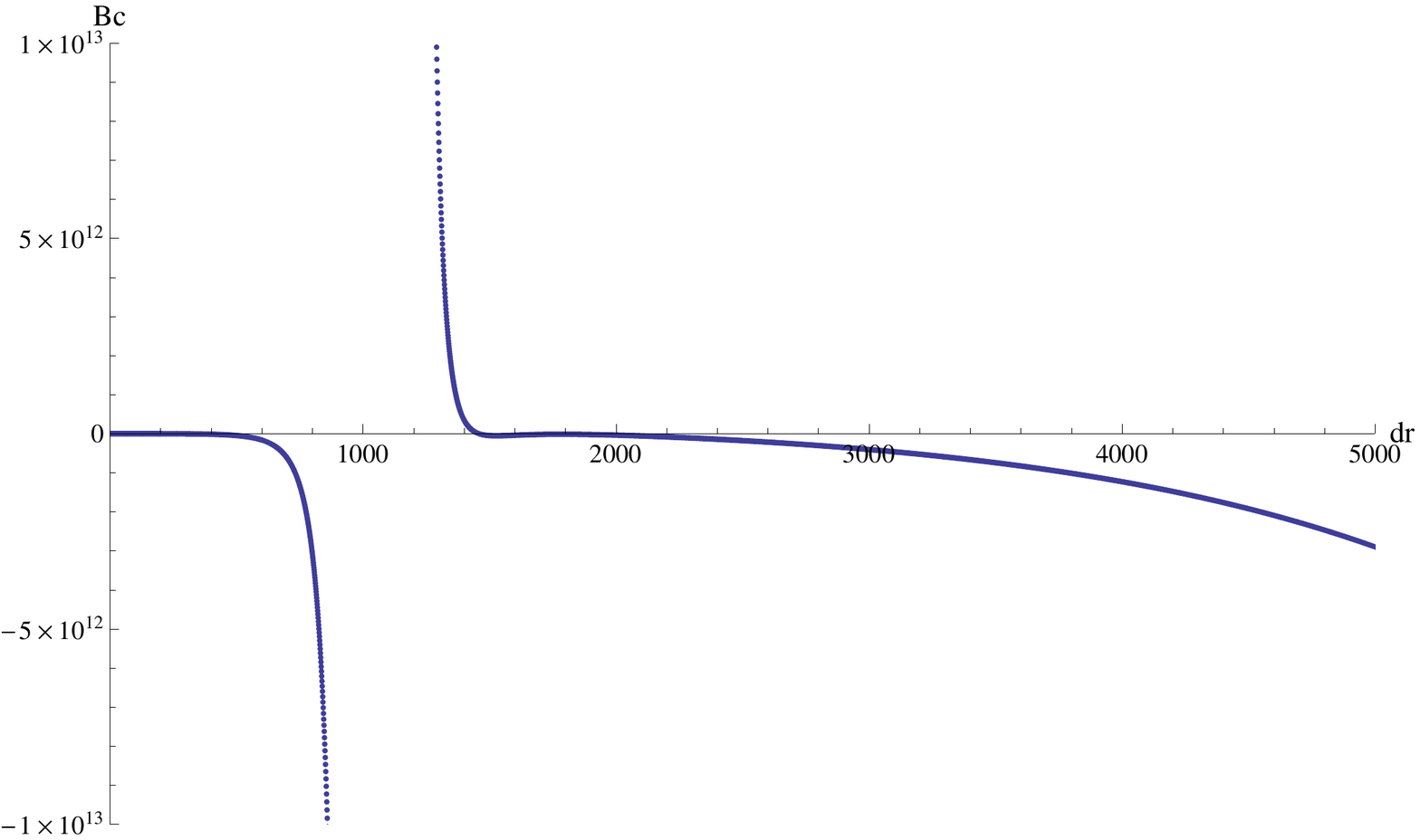}
\includegraphics[width=3.25cm,height=7.5cm,angle=0]{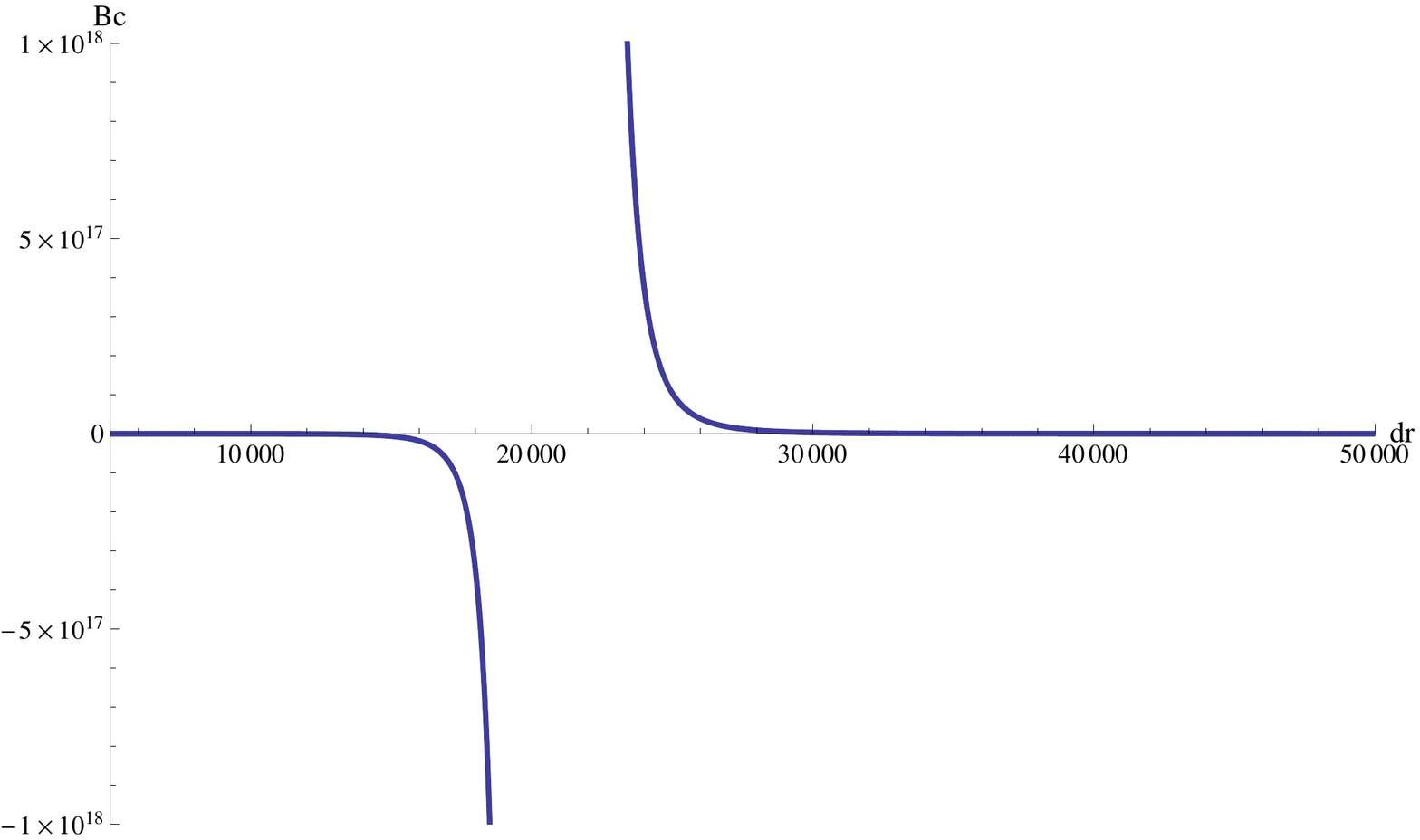}
\includegraphics[width=8.cm,height=3.5cm,angle=0]{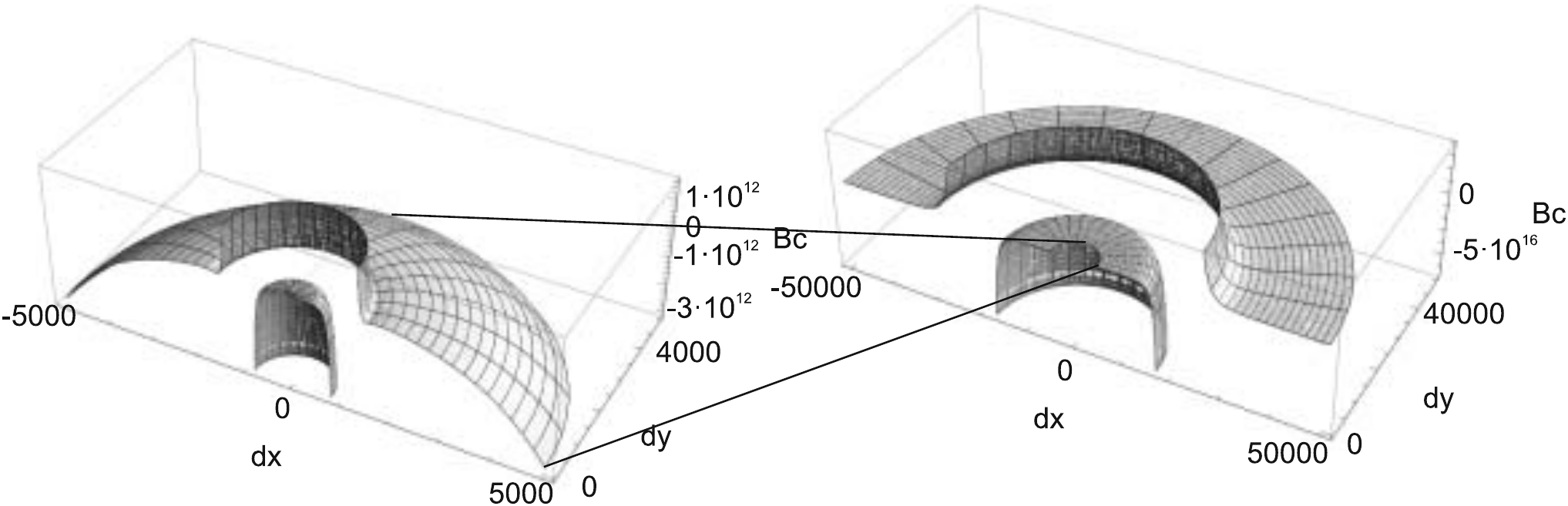}

\vspace*{-3cm}%
C)\includegraphics[width=3.cm,height=3.5cm,angle=0]{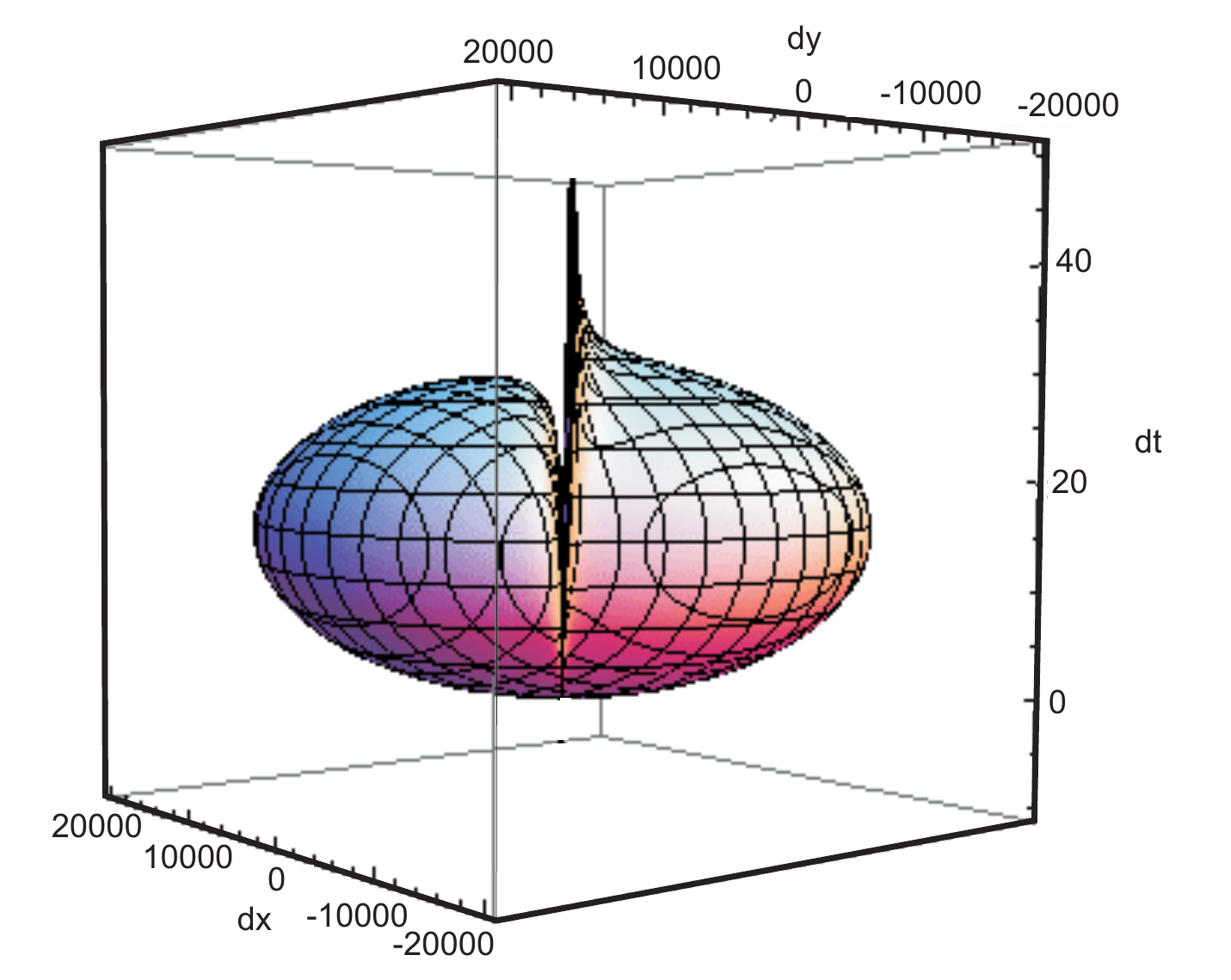}
\includegraphics[width=3.25cm,height=7.5cm,angle=0]{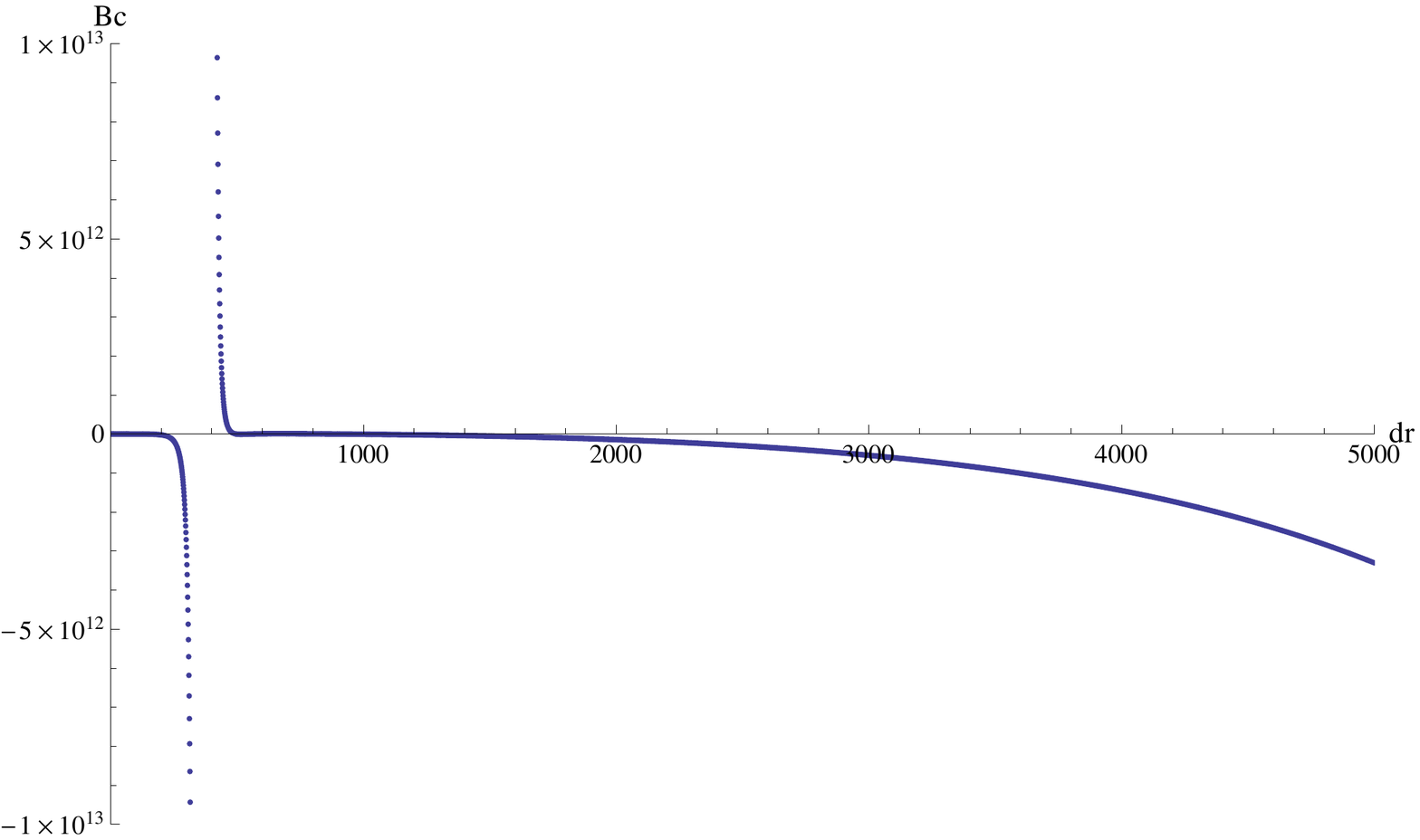}
\includegraphics[width=3.cm,height=7.5cm,angle=0]{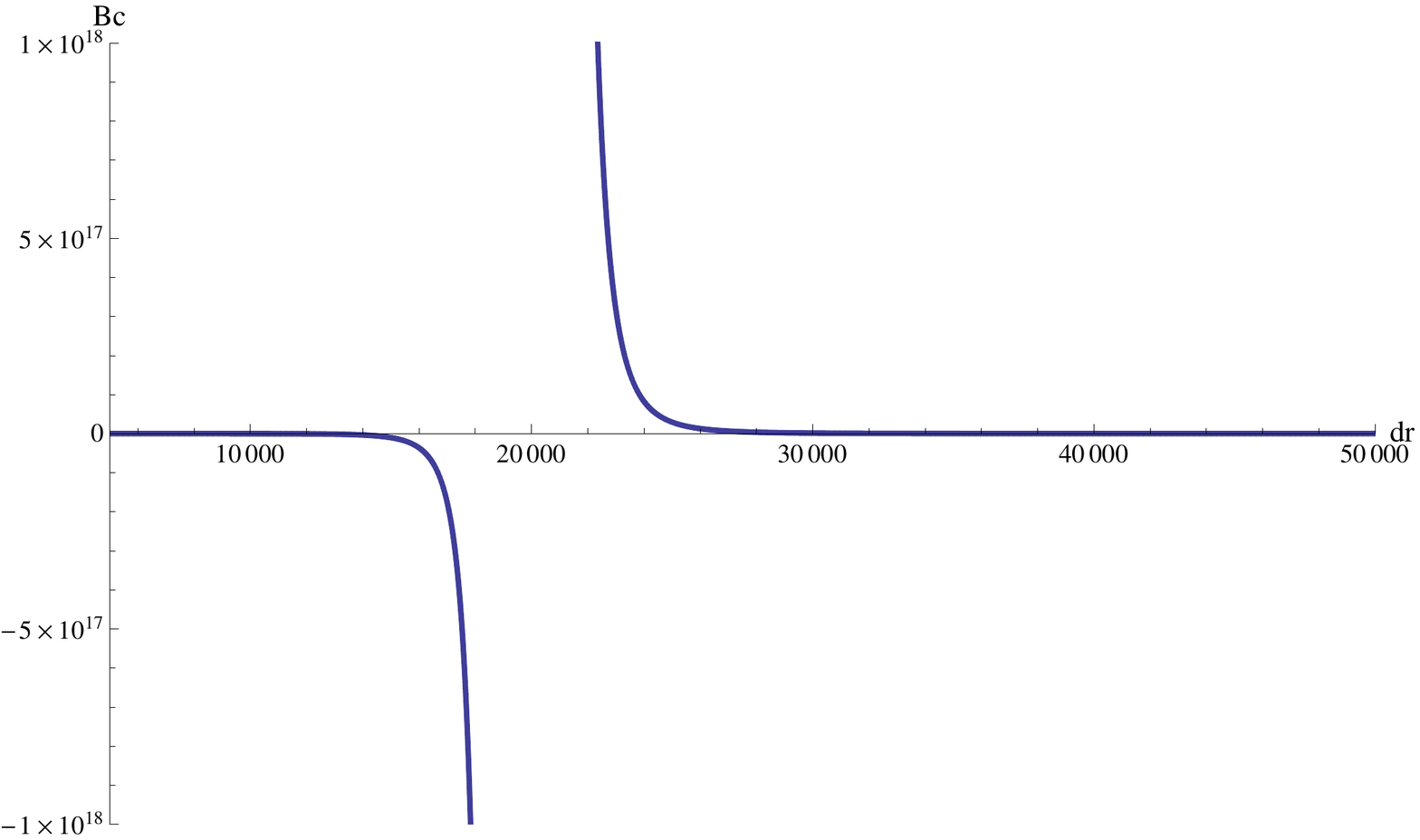}
\includegraphics[width=8.cm,height=3.5cm,angle=0]{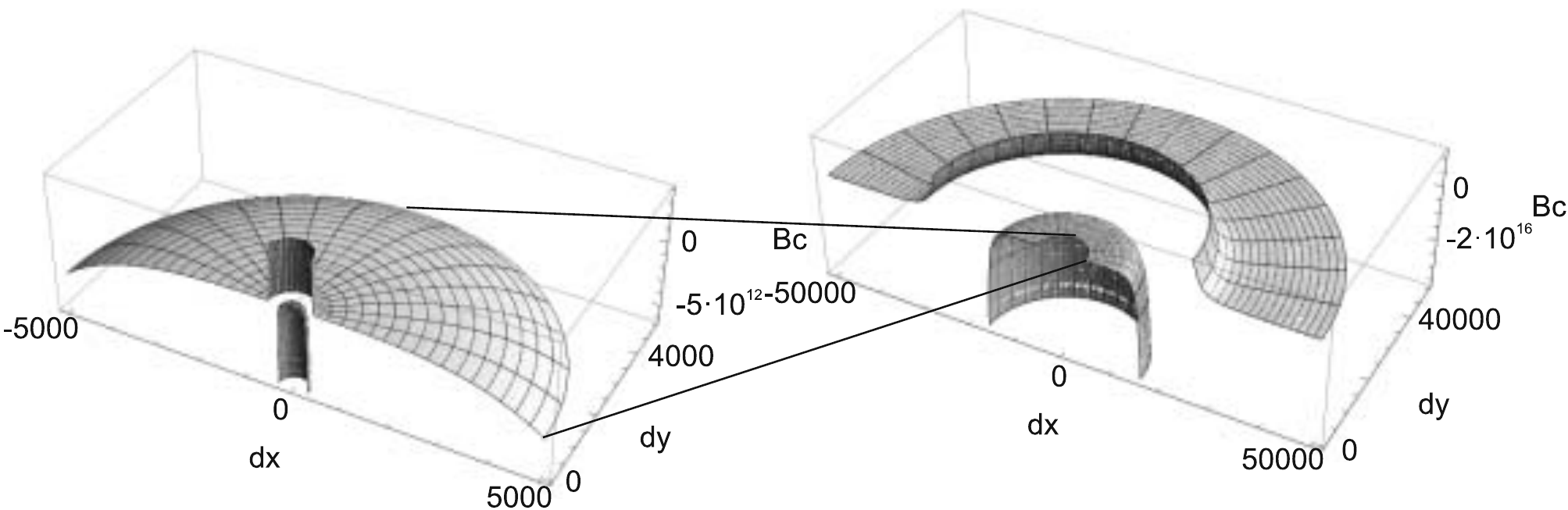}

\vspace*{-3cm}%
D)\includegraphics[width=3.cm,height=3.5cm,angle=0]{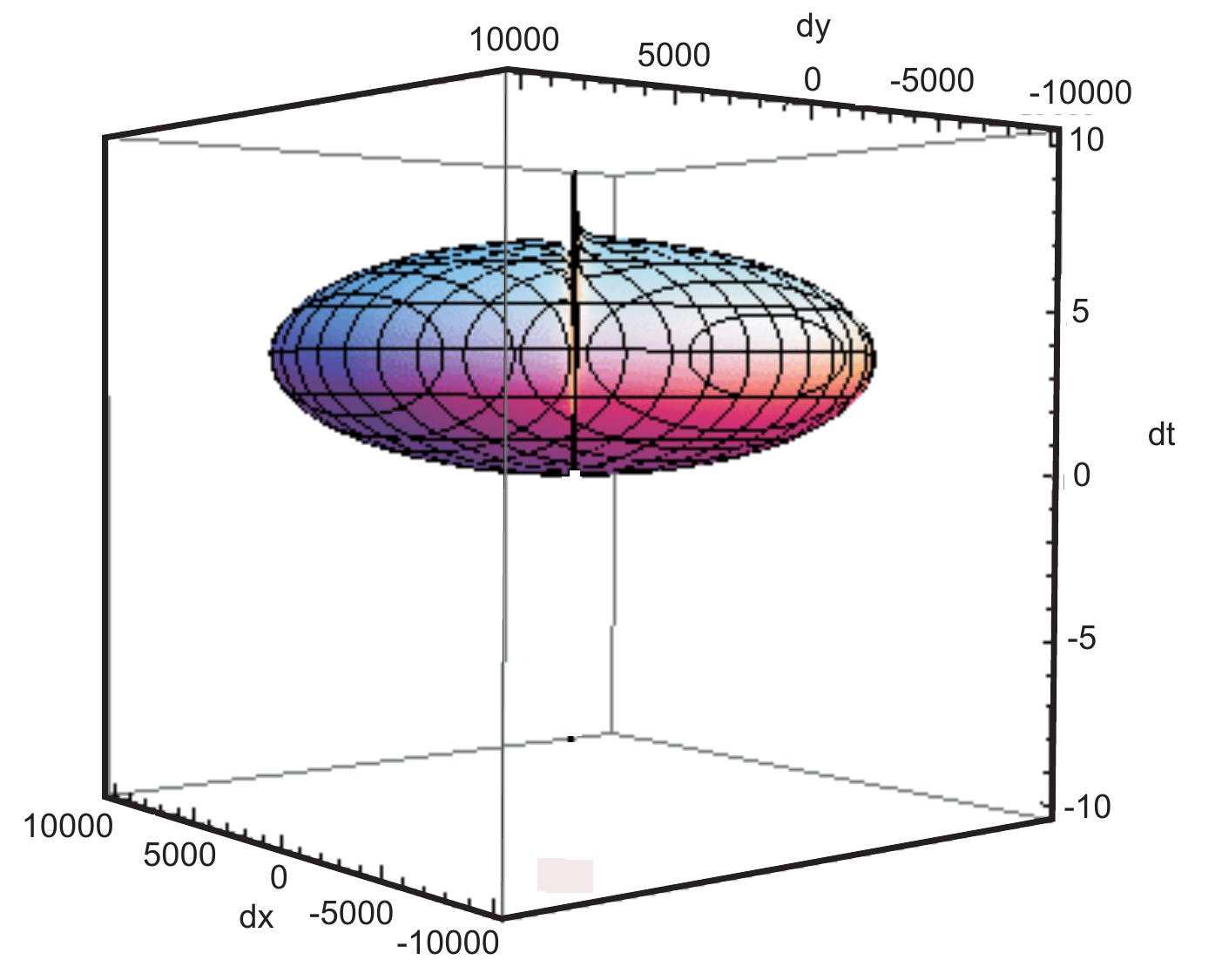}
\includegraphics[width=3.25cm,height=7.5cm,angle=0]{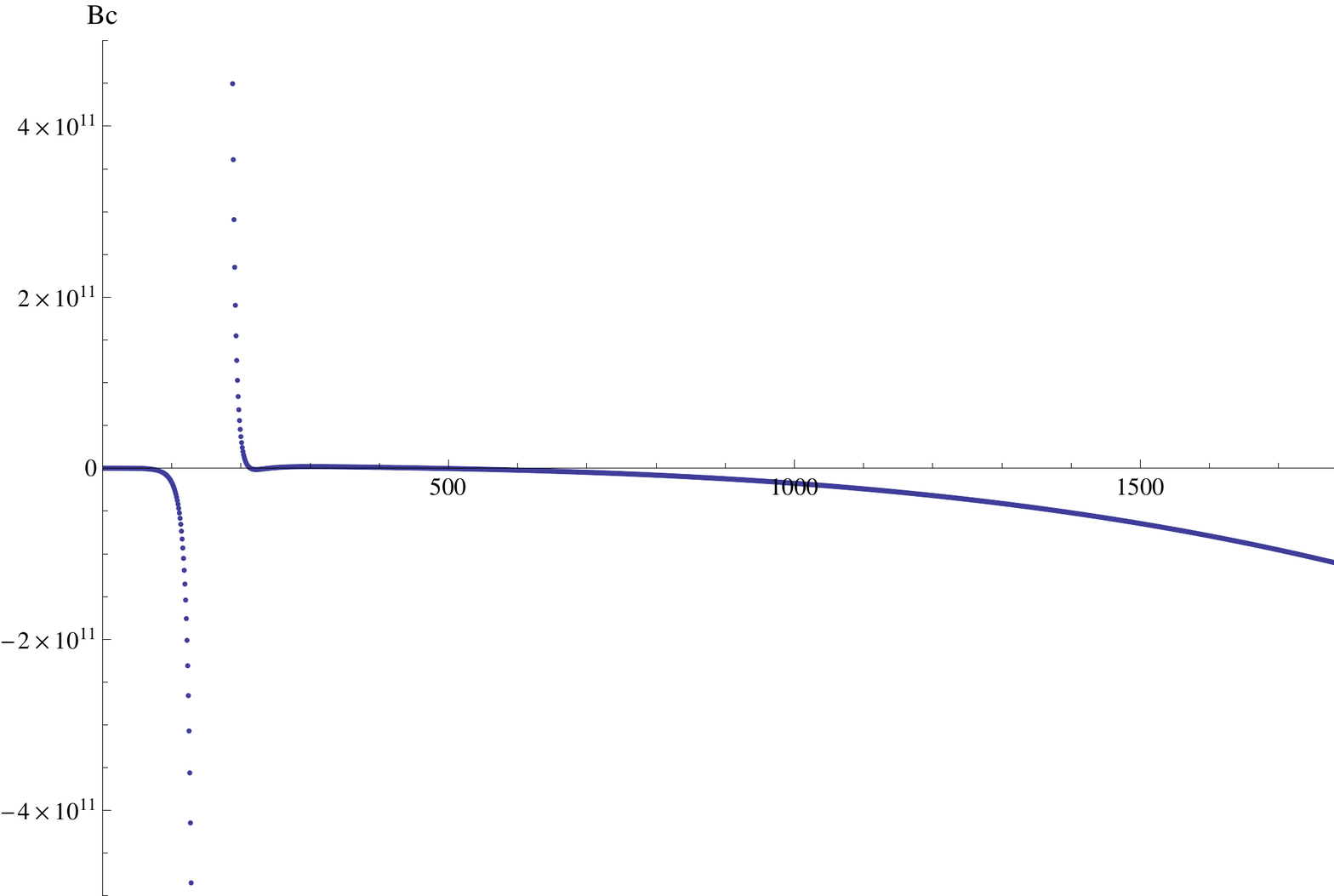}
\includegraphics[width=3.25cm,height=7.5cm,angle=0]{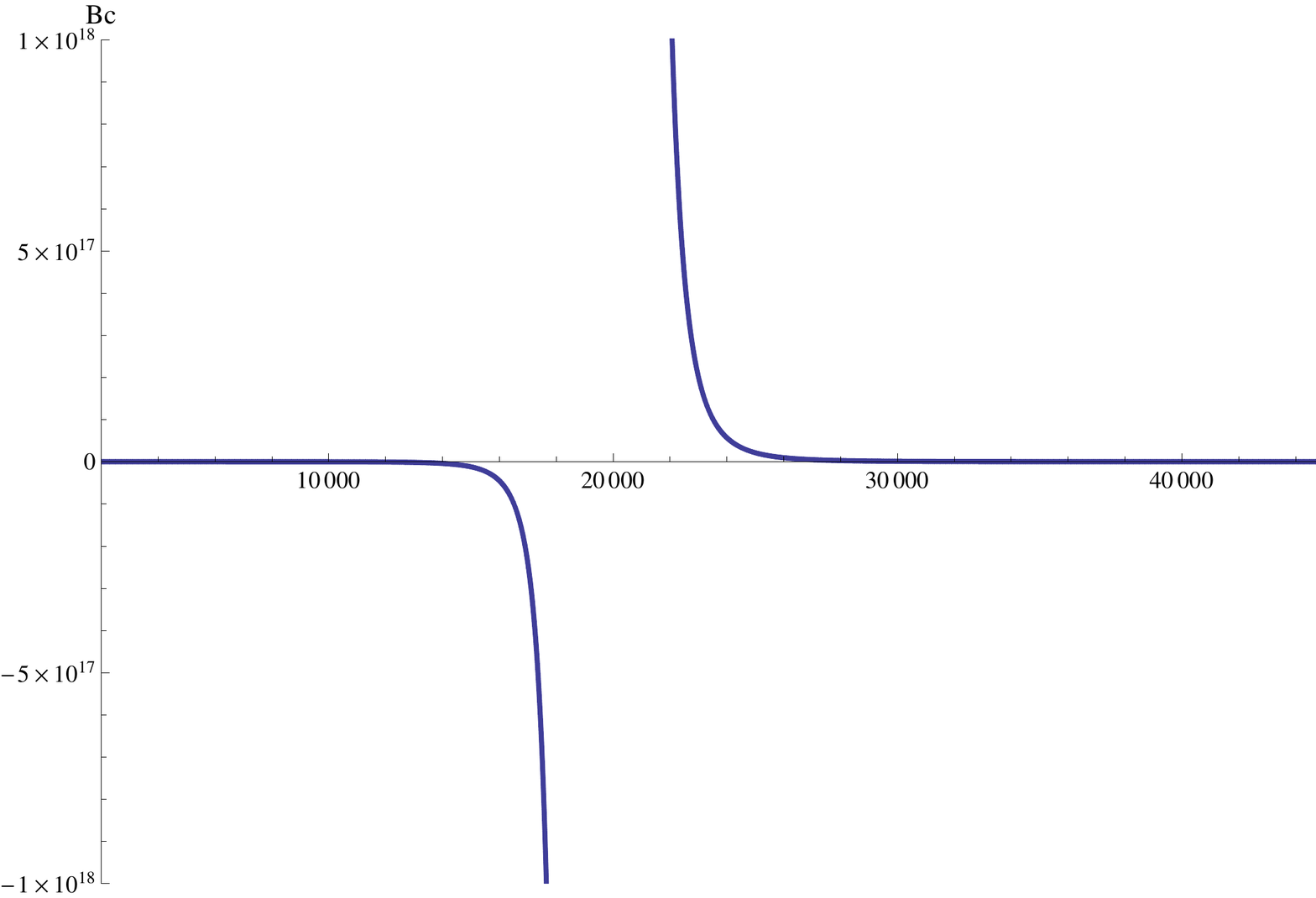}
\includegraphics[width=8.cm,height=3.5cm,angle=0]{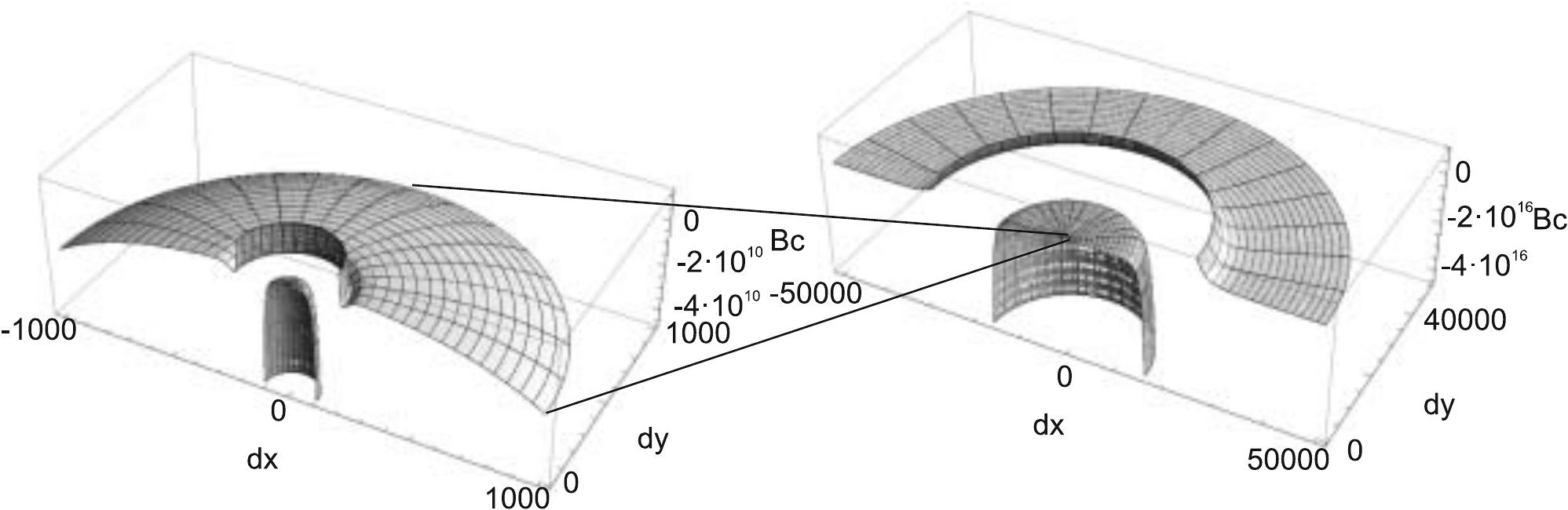}

\vspace*{-3cm}%
E)\includegraphics[width=3.cm,height=3.5cm,angle=0]{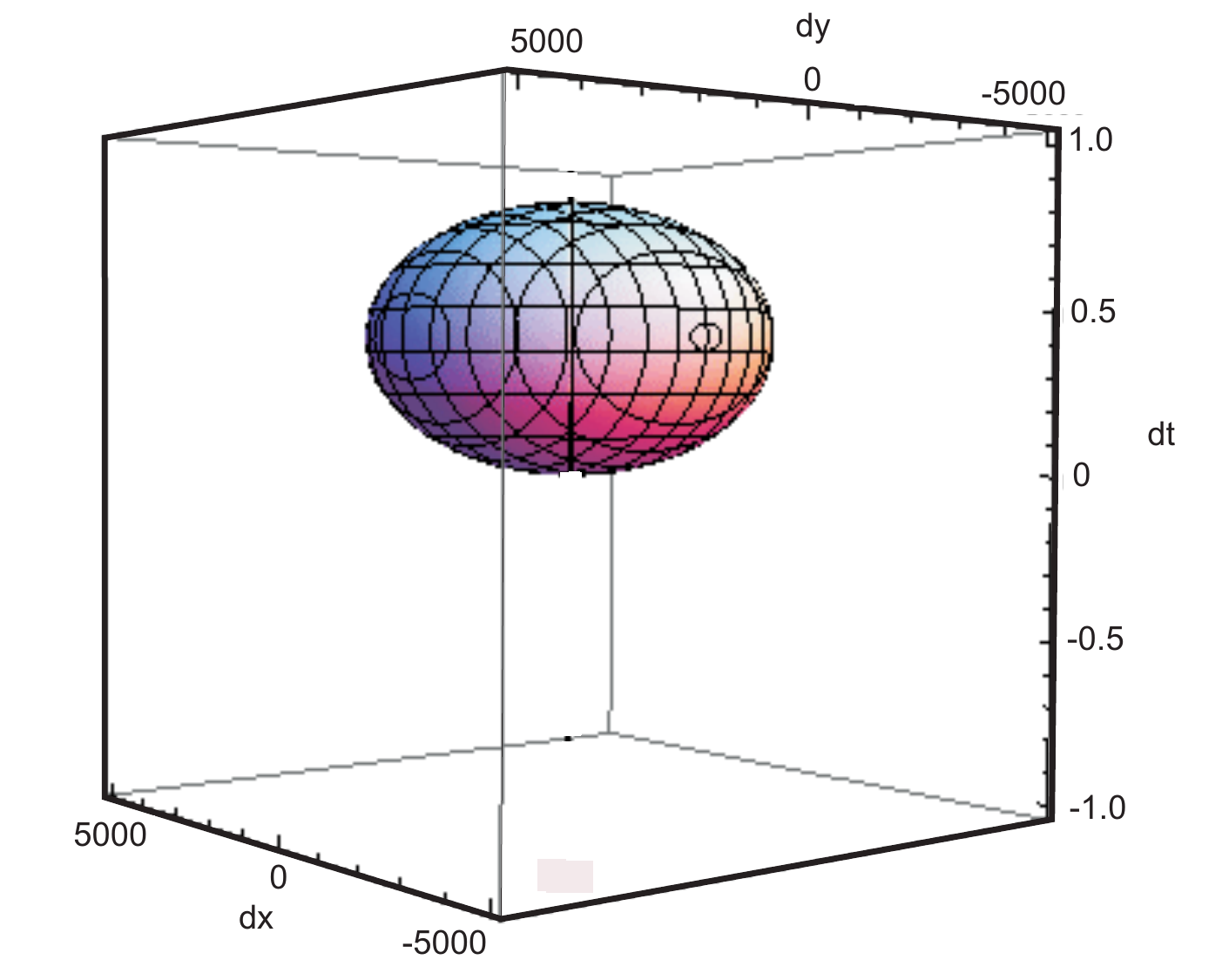}
\includegraphics[width=3.25cm,height=7.5cm,angle=0]{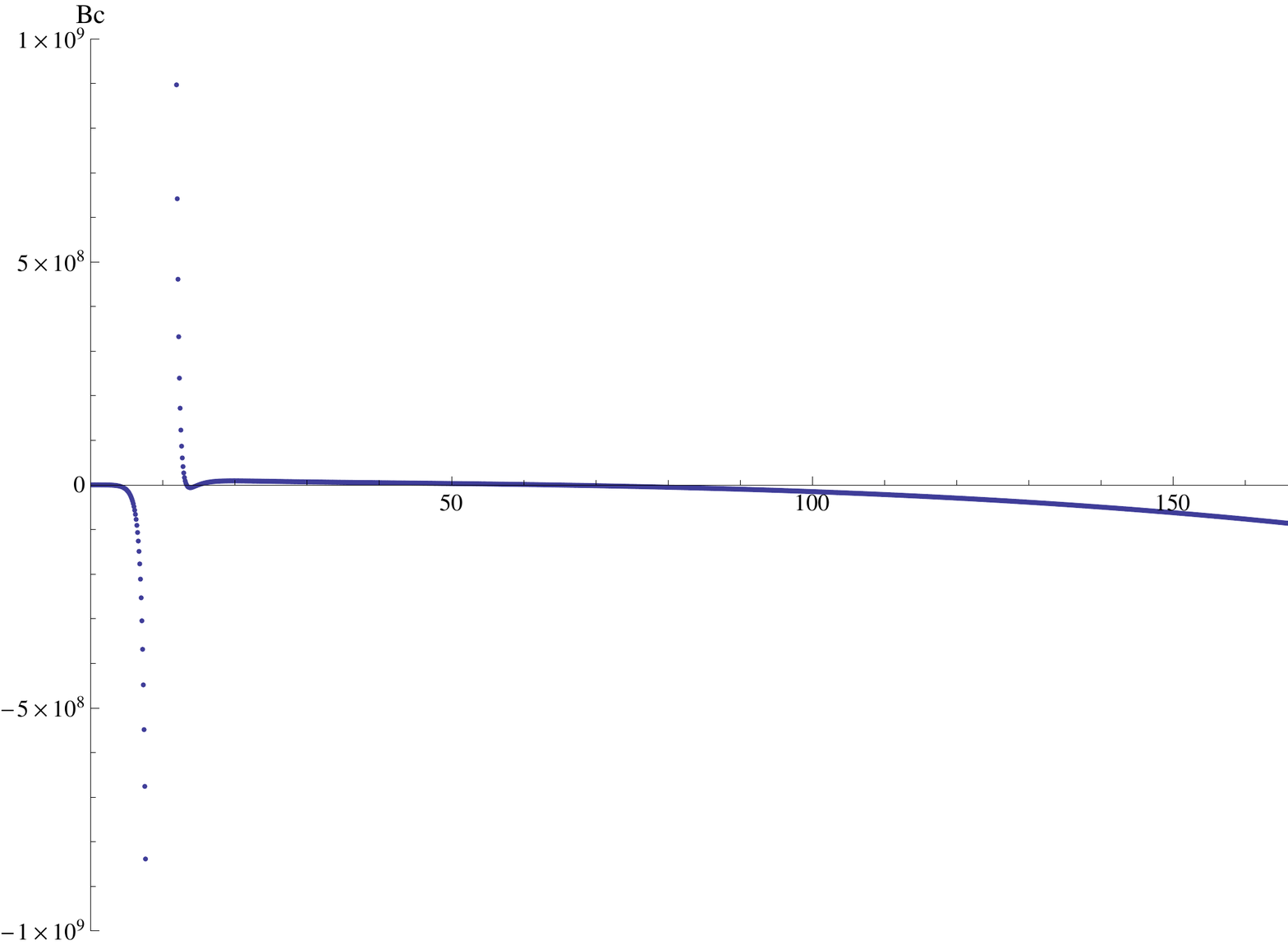}
\includegraphics[width=3.25cm,height=7.5cm,angle=0]{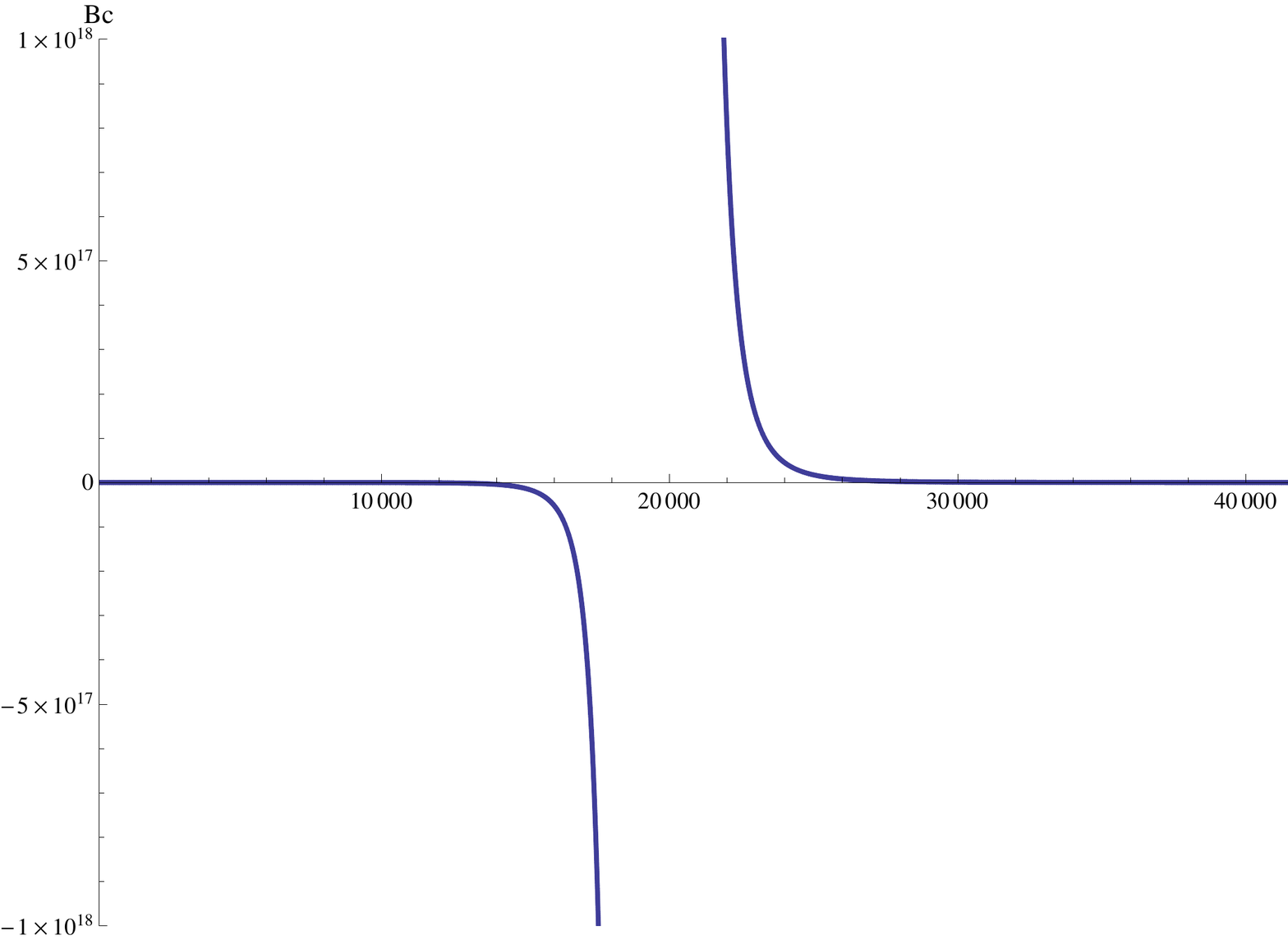}
\includegraphics[width=8.cm,height=3.5cm,angle=0]{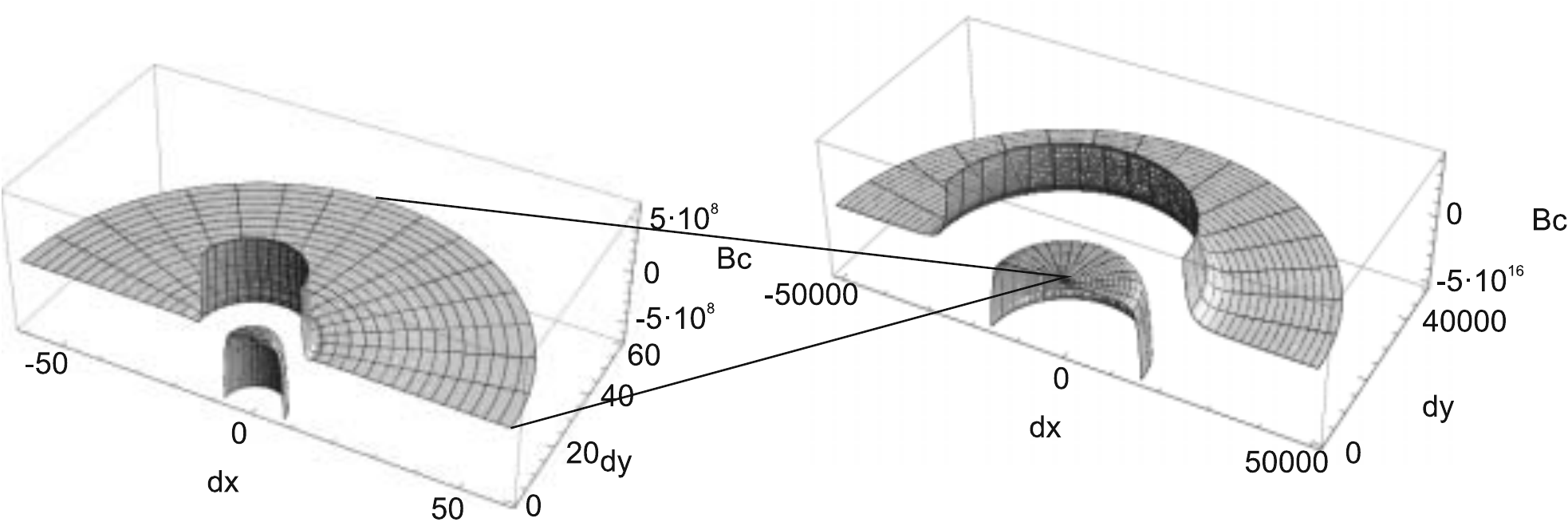}
\caption{\small Indicatrices (from the left) which relate to the
Berwald curvature $B_C$ (2d- and
    3d-images in the middle and right respectively) having two peculiarities for
    $x_0=$ 0.0796185564,\ 0.0796185562,\ 0.079618556,\ 0.079618555,\ 0.079618553,\ 0.0796185 for
    $B=-0.00243306$ (A), %(a),
    %$-0.00667689$ (b),
    $-0.0109207$ (B), %(c),
    $-0.0321398$ (C), %(d),
    $-0.0745781$ (D), %(e),
    $-1.19919$ (E) %(f)
    respectively. Here $C = 423\times 10^{-10}$, $A= 4.78079 \times 10^7$, $V = 15.0$.}
 \label{figure6}\end{figure}
\begin{figure}%[hbt]
\begin{center}
\vspace*{-2cm}
\includegraphics[width=10.cm,height=6.5cm,angle=0]{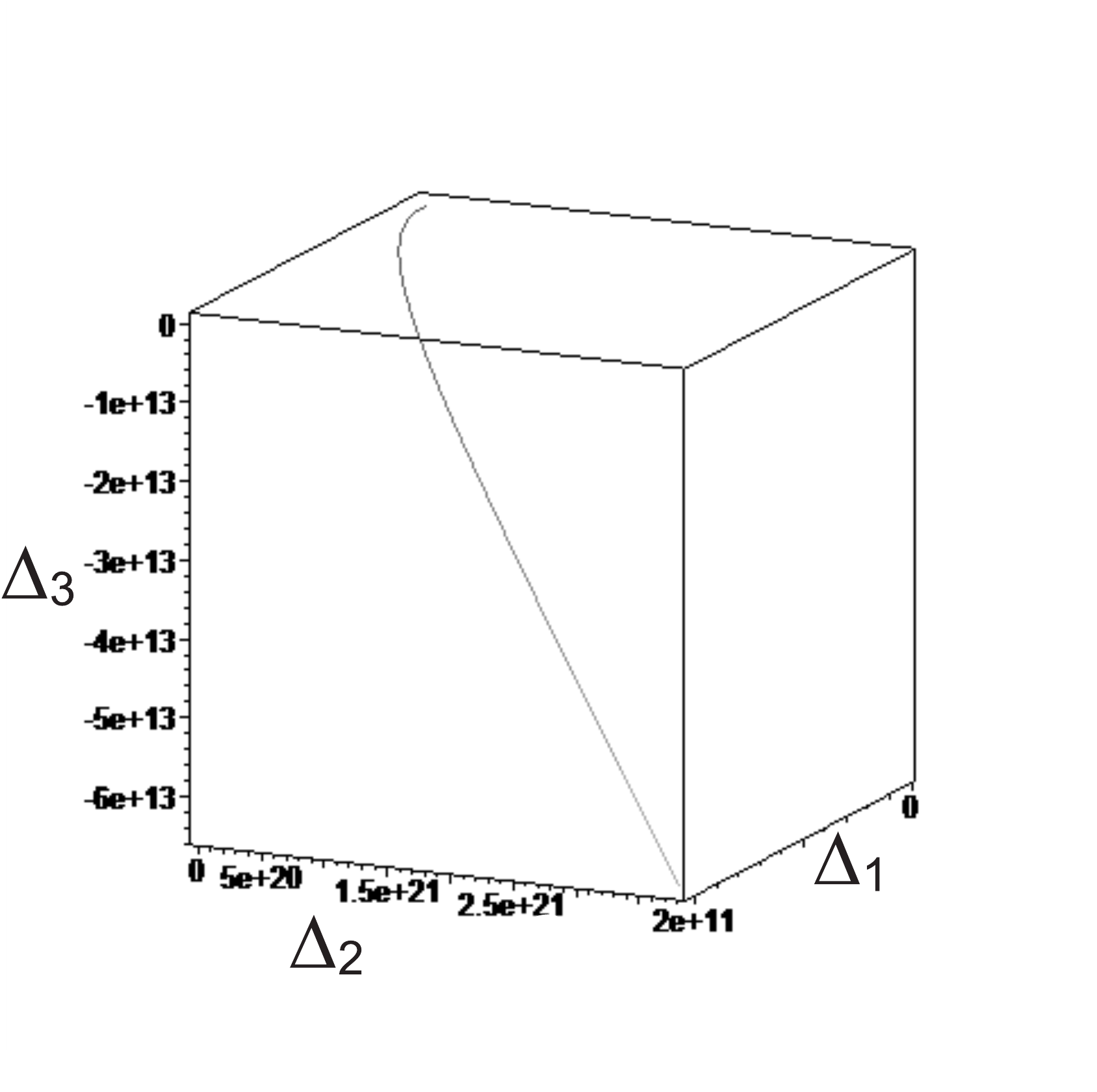}\end{center}
\caption{\small The 3d curve related to signature.}
\label{figure-signature}\end{figure}
\begin{figure*}%[hbt]
\vspace*{-2cm}\hspace{5cm} (a) \hspace{6cm} (b)  %\hspace{3cm} (c)
\begin{center}
\vspace*{-5cm}
\hspace*{-0cm}\includegraphics[width=9.cm,height=10.cm,angle=0]{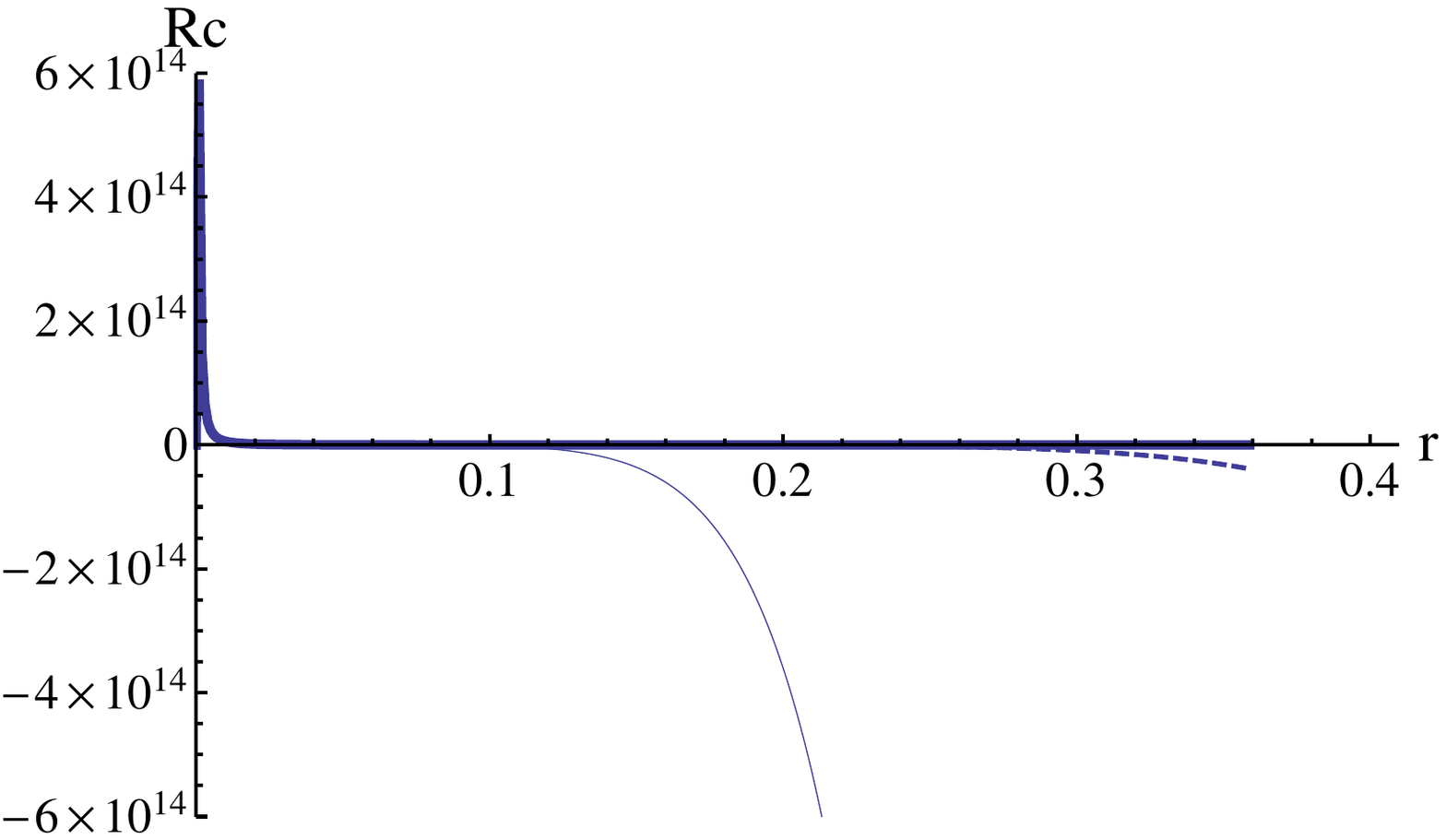}
\hspace*{-2cm}\includegraphics[width=10.cm,height=10.cm,angle=0]{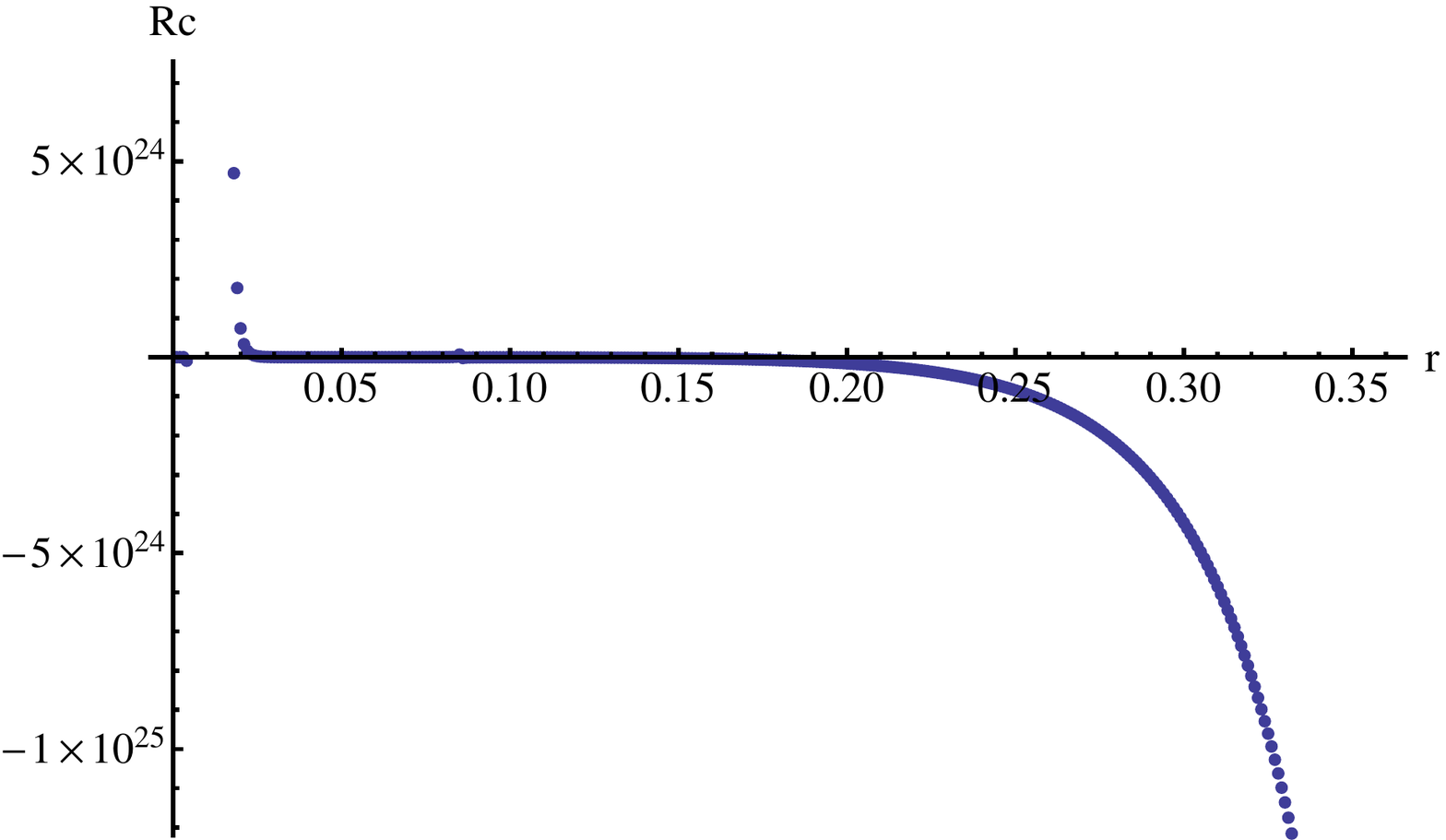}
\end{center}
\caption{\small The dependence of $R_C$ on the reference radius
$r$ at different monolayer compressing speeds (a)
$V=10^{-15}$~(thick solid line), $10^{-10}$~(dashed line),
   $10^{-8}$~(dotted line), $10^{-7}$~m/sec~(thin solid line); (b) $10$~m/sec for
    $t=0.01,\ \dot\xi =0.00001,\ \dot r =0.01,\ \dot\phi =0$.}
\label{figure-curvarure}
\end{figure*}
\begin{figure}%[hbt]
    \begin{center}
 \vspace*{-4cm}   \includegraphics[width=8.cm,height=4.5cm,angle=0]{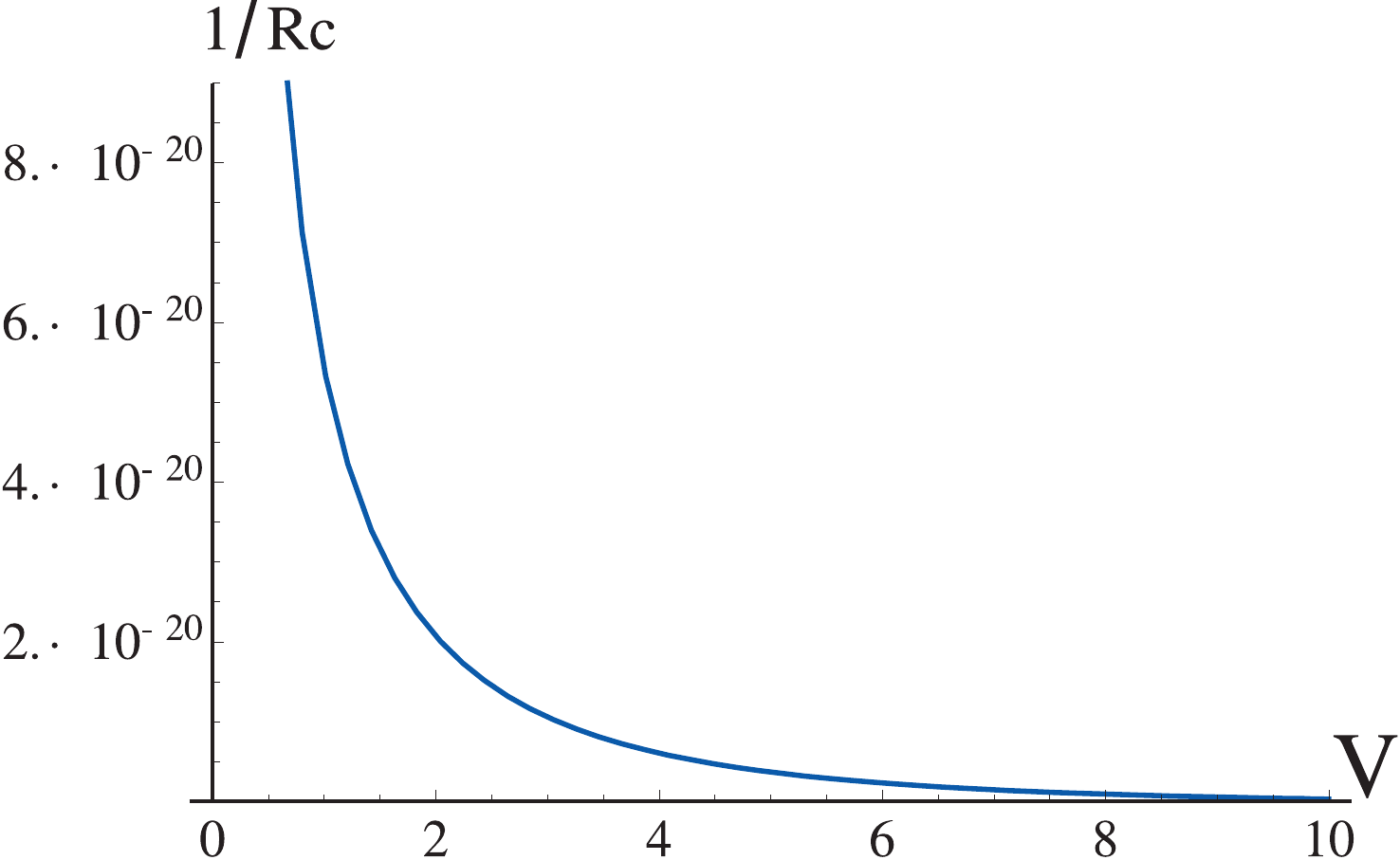}
    \end{center}
    \caption{ The dependence of the curvature radius $1/R_C$ on $V$.}
\label{figure2}\end{figure}

%\par\medskip
Using the following experimental values for the parameters
\begin{eqnarray}
m= 47 \times 10^{-26} \mbox{kg},\ p=8.93434 \times 10^9\mbox{ joule/m}^5,\
        c=3\times 10^8 \mbox{m/sec},
\end{eqnarray}
we shall perform an illustrative simulation. When the speed $V$ of the monolayer compression tends to zero,
    the coefficient $A$ tends to zero as well, and the indicatrix is a hyperboloid of two sheets -- which is
    characteristic for the Minkowski space. But at high speeds ($V$), there exists a region of Space--Time,
    in which the structural topological form of the indicatrix dramatically changes as Figs. \ref{figure5}
    and \ref{figure6} clearly show. We shall further analyze several invariants of the Finsler space endowed
    with the metric function $F$ given by (\ref{normal-quadratic-metric}).\par
Using the following steps, one can determine the horizontal $hh$-component $R^i_{jkl}$ of the curvature
    Ricci tensor $\Omega^i_{jkl}$:
\begin{eqnarray}
g_{ij}={1\over 2}{\partial^2 F^2\over\partial y^i \partial y^j},\\
G^i(y)=
{1\over 4}g^{il}(y)\left\{
{\partial^2 F^2(x,y)\over\partial x^k \partial y^l}y^k
-{\partial F^2(x,y) \over\partial x^l}
\right\}
= {1\over 4}g^{il}(y)\left\{
2{\partial g_{jl}\over\partial x^k}(y)
-{\partial g_{jk}\over\partial x^l}(y)
\right\}y^jy^k,\\
N^i_j(y)= {\partial G^{i}\over\partial y^j}(y),\\
{\delta \over\delta x^s}={\partial  \over\partial x^s}
-N^i_s{\partial  \over\partial y^i},\\
\Gamma^i_{jk}={g^{is}\over 2}\left(
{\delta g_{sj}\over\delta x^k}- {\delta g_{jk}\over\delta x^s}
+{\delta g_{ks}\over\delta x^j}
\right),\\
R^i_{jkl}=
{\delta \Gamma^i_{jl}\over\delta x^k}
-{\delta \Gamma^i_{jk}\over\delta x^l}
+\Gamma^i_{hk}\Gamma^h_{jl}-\Gamma^i_{hl}\Gamma^h_{jk}
\end{eqnarray}
where $x^k=\{t,r, \phi\},\ y^k=\{ \dot\xi, \dot r, \dot\phi \}$.

%\par\medskip
%=================================================================================
Let us investigate now the signature of the metric tensor $g_{ij}$.
The matrix associated to the metric tensor $g_{ij}$ is:
    \[g_{ij}=\left(\begin{array}{ccc}
        3A\dfrac{\dot{\xi}}{\dot{r}}+B & -\dfrac{3}{2}A\left( \dfrac{\dot{\xi}}{\dot{%
        r}}\right) ^{2} & 0 \\
        -\dfrac{3}{2}A\left( \dfrac{\dot{\xi}}{\dot{r}}\right) ^{2} & A\left( \dfrac{%
        \dot{\xi}}{\dot{r}}\right) ^{3}-\dfrac{C}{2} & 0 \\
        0 & 0 & -\dfrac{C}{2}r^{2}%
        \end{array}%
        \right) ,\]%
and hence the Jacobi minors are:
$$\left\{
\begin{array}{lll}
\Delta _{1} &=&3A\dfrac{\dot{\xi}}{\dot{r}}+B, %\medskip
\\
    \Delta _{2} &=&\frac{3A^{2}}{4}
    \left( \dfrac{\dot{\xi}}{\dot{r}}\right)^{4}+
        AB\left( \dfrac{\dot{\xi}}{\dot{r}}\right) ^{3}-\dfrac{3AC}{2}
        \dfrac{\dot{\xi}}{\dot{r}}-\dfrac{BC}{2}, %\medskip
        \\
    \Delta _{3} &=&-\frac{C}{2}r^{2} \Delta _{2}
%\left( \frac{3A^{2}}{4}\left(   \dfrac{\dot{\xi}}{\dot{r}}
%        \right) ^{4}+AB\left( \dfrac{\dot{\xi}}{\dot{r}}\right) ^{3}-
%        \dfrac{3AC}{2}\dfrac{\dot{\xi}}{\dot{r}}-\dfrac{BC}{2}\right)
.
\end{array}\right.
$$
Fig. \ref{figure-signature} displays the curve related to signature in 3D. It is easy to see that
    the metric is pseudo-Finsler. The signature of $g_{ij}$ is $(+,+,-)$. Hence, $\dot \xi$ and
    $\dot r$  have the same (temporal) character, and  $\dot \phi$ is spatial only. Since the components
    of the Cartan tensor are:
\[\begin{array}{lllll}
C_{111}=\dfrac{3A}{2}\dfrac{1}{\dot{r}}, &  &
C_{112}=-\dfrac{3A}{2}\dfrac{%
        \dot{\xi}}{\dot{r}} &  & C_{113}=0, \\
    &  &  &  &  \\
    C_{122}=\dfrac{3A}{2}\dfrac{\dot{\xi}^{2}}{\dot{r}^{3}}, &  &
    C_{222}=-\dfrac{%
    3A}{2}\dfrac{\dot{\xi}^{3}}{\dot{r}^{4}}, &  & C_{223}=0, \\
    &  &  &  &  \\
    C_{133}=0, &  & C_{233}=0, &  & C_{333}=0, \\
    &  &  &  &  \\
    C_{123}=0, &  &  &  &\end{array}\]
one gets due to the signature of the metric tensor that the Finsler space
under consideration is a pseudo-Finsler one with %flat 2d-Cartan tensor.
a Cartan tensor, whose effective components live on a 2-plane.

In order to determine the non-linear Barthel connection, we notice that:
    \[
    \begin{array}{lll}
G^{1} & = & \dfrac{A}{4\Delta _{2}%\delta
}\left[
\dfrac{\partial g_{11}}{\partial t}%
\dfrac{\dot{\xi}^{5}}{\dot{r}^{3}}+\left(
\dfrac{1}{2}\dfrac{\partial g_{11}%
}{\partial r}+3\dfrac{\partial g_{12}}{\partial t}\right)
\dfrac{\dot{\xi}%
^{4}}{\dot{r}^{2}}+\left( 2\dfrac{\partial g_{12}}{\partial r}+2\dfrac{%
\partial g_{22}}{\partial t}\right) \dfrac{\dot{\xi}^{3}}{\dot{r}}\right.
\\
&  &  \\
&  & \left. +\dfrac{3}{2}\dfrac{\partial g_{22}}{\partial r}\dot{\xi}^{2}
-%
\dfrac{3}{2}\dfrac{\partial g_{33}}{\partial r}\left( \dfrac{\dot{\xi}\dot{%
\varphi}}{\dot{r}}\right) ^{2}\right]  \\
&  &  \\
&  & -\dfrac{C}{8\Delta _{2} %\delta
}\left[
\dfrac{\partial g_{11}}{\partial t}\dot{\xi}%
^{2}+2
\dfrac{\partial g_{11}}{\partial r}\dot{\xi}\dot{r}+\left( 2\dfrac{%
\partial g_{12}}{\partial r}-\dfrac{\partial g_{22}}{\partial t}\right) \dot{%
r}^{2}\right],  \\
&  &  \\
G^{2} & = & \dfrac{3A}{4\Delta _{2} %\delta
}\left[
\dfrac{1}{2}\dfrac{\partial g_{11}}{%
\partial t}\dfrac{\dot{\xi}^{4}}{\dot{r}^{2}}+2\dfrac{\partial g_{12}}{%
\partial t}\dfrac{\dot{\xi}^{3}}{\dot{r}}\right.  \\
&  &  \\
&  & \left. +\left(
\dfrac{\partial g_{12}}{\partial r}+\dfrac{3}{2}\dfrac{%
\partial g_{22}}{\partial t}\right) \dot{\xi}^{2}+
\dfrac{\partial g_{22}}{%
\partial r}\dot{\xi}\dot{r}-
\dfrac{\partial g_{33}}{\partial r}\dfrac{\dot{%
\xi}\dot{\varphi}^{2}}{\dot{r}}\right]  \\
&  &  \\
&  & +\dfrac{B}{4 \Delta _{2} %\delta
}\left[ \left( 2\dfrac{\partial g_{12}}{\partial t}-%
\dfrac{\partial g_{11}}{\partial r}\right) \dot{\xi}^{2}+2\dfrac{\partial
g_{22}}{\partial t}\dot{\xi}\dot{r}+\dfrac{\partial g_{22}}{\partial r}\dot{%
\xi}^{2}-\dfrac{\partial g_{33}}{\partial r}\dot{\varphi}^{2}\right],  \\
&  &  \\
G^{3} & = & \dfrac{1}{r}\dot{r}\dot{\varphi}.%
\end{array}%
\]%
Considering that
\[
A_{t}=\frac{\partial A}{\partial t},A_{r}=\frac{\partial A}{\partial r}%
,B_{t}=\frac{\partial B}{\partial t},B_{r}=\frac{\partial B}{\partial r}
\]%
and
\[
\begin{array}{lcl}
\dfrac{\partial g_{11}}{\partial t}=
3A_{t}\dfrac{\dot{\xi}}{\dot{r}}+B_{t}, & & \dfrac{\partial
g_{11}}{\partial r}=3A_{r}\dfrac{\dot{\xi}}{\dot{r}}+B_{r},
\\
&  &  \\
\dfrac{\partial g_{12}}{\partial t}=-\dfrac{3}{2}A_{t}\left(
\dfrac{\dot{\xi}%
}{\dot{r}}\right) ^{2}, &  &
\dfrac{\partial g_{12}}{\partial r}=-\dfrac{3}{2}%
A_{r}\left( \dfrac{\dot{\xi}}{\dot{r}}\right) ^{2} , \\
&  &  \\
\dfrac{\partial g_{22}}{\partial t}=
A_{t}\left( \dfrac{\dot{\xi}}{\dot{r}}%
\right) ^{3}, &  & \dfrac{\partial g_{22}}{\partial r}=
A_{r}\left( \dfrac{%
\dot{\xi}}{\dot{r}}\right) ^{3}, \\
&  &  \\
\dfrac{\partial g_{33}}{\partial t}=0, &  &
\dfrac{\partial g_{33}}{\partial r%
}=-Cr%
\end{array}%
\]%
we get the components of the nonlinear connection
\begin{eqnarray*}
&&%
\begin{array}{ccl}
N_{1}^{1} & = & \dfrac{A}{4
\Delta _{2}%\delta
}\left[
-3A_{t}\dfrac{\dot{\xi}^{5}}{\dot{r%
}^{4}}+6A_{t}\dfrac{\dot{\xi}^{5}}{\dot{r}^{3}}+5B_{t}\dfrac{\dot{\xi}^{4}}{%
\dot{r}^{3}}+2B_{r}\dfrac{\dot{\xi}^{3}}{\dot{r}^{2}}+3Cr\dfrac{\dot{\xi}%
\dot{\varphi}^{2}}{\dot{r}^{2}}\right]  \\
&  &  \\
&  & -\dfrac{C}{4 \Delta _{2}%\delta
}\left[ 3A_{t}\dfrac{\dot{\xi}^{2}}{\dot{r}}+\left(
B_{t}+3A_{r}\right) \dot{\xi}+B_{r}\dot{r}\right],  \\
&  &  \\
N_{2}^{1} & = & \dfrac{A}{4\Delta _{2}%\delta
}\left[
-2A_{t}\dfrac{\dot{\xi}^{6}}{\dot{r%
}^{5}}-3B_{t}\dfrac{\dot{\xi}^{5}}{\dot{r}^{4}}-B_{r}\dfrac{\dot{\xi}^{4}}{%
\dot{r}^{3}}-3Cr\dfrac{\dot{\xi}^{2}\dot{\varphi}^{2}}{\dot{r}^{3}}\right]
\\
&  &  \\
&  & -\dfrac{C}{8 \Delta _{2}%\delta
}\left[ -2A_{t}\dfrac{\dot{\xi}^{3}}{\dot{r}^{2}}%
+B_{r}\dot{\xi}\right],  \\
&  &  \\
N_{3}^{1} & = & \dfrac{3AC}{4\Delta _{2}%\delta
}r
\dfrac{\dot{\xi}^{2}\dot{\varphi}}{%
\dot{r}^{2}},%
\end{array}
\\
&& \\
&&%
\begin{array}{ccl}
N_{1}^{2} & = & \dfrac{3A}{4\Delta _{2}%\delta
}\left( Cr+2B_{t}-2A_{r}\right) \dfrac{%
\dot{\xi}^{3}}{\dot{r}^{2}}-\dfrac{B}{2\Delta _{2}%\delta
}\left[ 2A_{t}\dfrac{\dot{\xi}%
^{3}}{\dot{r}^{2}}+3A_{r}\dfrac{\dot{\xi}^{2}}{\dot{r}}+B_{t}\dot{\xi}
\right],
\\
&  &  \\
N_{2}^{2} & = & \dfrac{3A}{4\Delta _{2}%\delta
}\left[ \left( A_{r}-B_{t}\right) \dfrac{%
\dot{\xi}^{4}}{\dot{r}^{3}}-Cr\dfrac{\dot{\xi}\dot{\varphi}^{2}}{\dot{r}^{2}}%
\right] +\dfrac{B}{2 \Delta _{2}%\delta
}\left[ A_{t}\dfrac{\dot{\xi}^{4}}{\dot{r}^{3}}%
+A_{r}\dfrac{\dot{\xi}^{3}}{\dot{r}^{2}}\right],  \\
&  &  \\
N_{3}^{2} & = & \dfrac{3AC}{2\Delta _{2}%\delta
}r
\dfrac{\dot{\xi}\dot{\varphi}}{\dot{r}}%
+\dfrac{BC}{2 \Delta _{2} %\delta
}r\dot{\varphi},%
\end{array}
\\
&& \\
&&%
\begin{array}{ccc}
N_{1}^{3} & = & 0, \\
&  &  \\
N_{2}^{3} & = & \dfrac{\dot{\varphi}}{r}, \\
&  &  \\
N_{3}^{3} & = & \dfrac{\dot{r}}{r}.%
\end{array}%
\end{eqnarray*}%

We shall further study the dependence of the Ricci scalar curvature
\begin{equation}
R_C(x,y,V) = g^{ij}R^k_{ikj}
\end{equation}
on the speed $V$ and on the reference point in the monolayer.

%\par\smallskip
%
The following results from below were obtained by numerical
simulation, using constants taken    from the experimental setup.
The curvature $R_C$ is represented by a function of $x_i,\ y_i$
which is nonzero one for any values of parameter $V$ as one can
see in Fig.~\ref{figure-curvarure}. Therefore, $R_C$ is not an
invariant of the flat monolayer    space, and hence it cannot be
associated with a quantity describing the studied physical
process. But, $R_C$ trends to zero in a limit of very small speeds
$V$ ($V=10^{-15}$ and less) for the large membrane radius $r$,
corresponding to a compression beginning. One  notes that the
limit $V\to 0$ leads to
    $$A\to 0,\ B \to mc^2 +{4\over 3}pr^3.$$
In this case the curvature  radius $1/R_C$ tends to infinity (see
Fig.~\ref{figure2}).
Hence, the Finsler structure tends to a pseudo-Riemannian.

\begin{figure}%[hbt]
    \begin{center}
\vspace*{-6cm}\includegraphics[width=9.cm,height=10.5cm,angle=0]{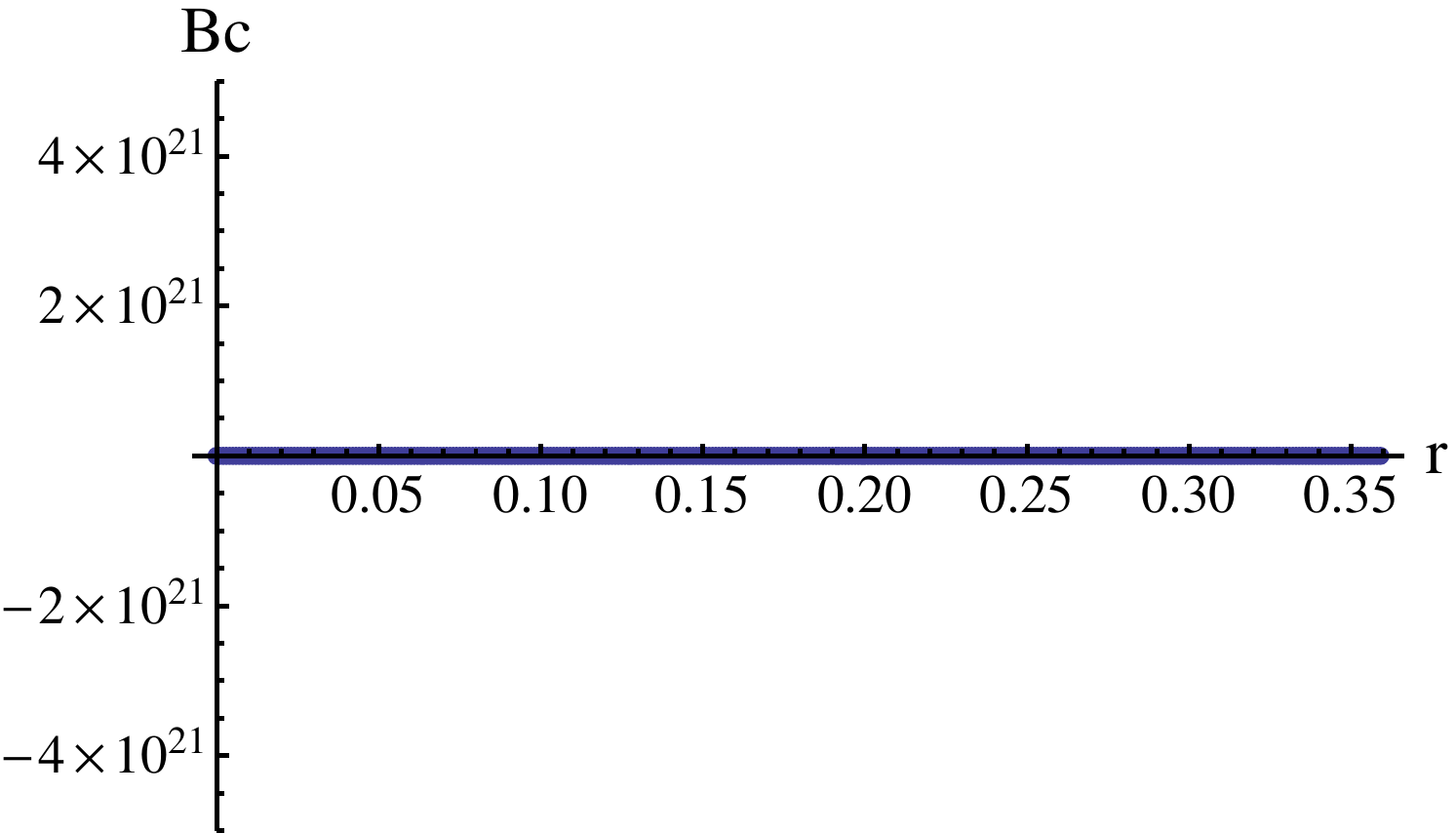}
\hspace*{-1.5cm}\includegraphics[width=9.cm,height=10.5cm,angle=0]{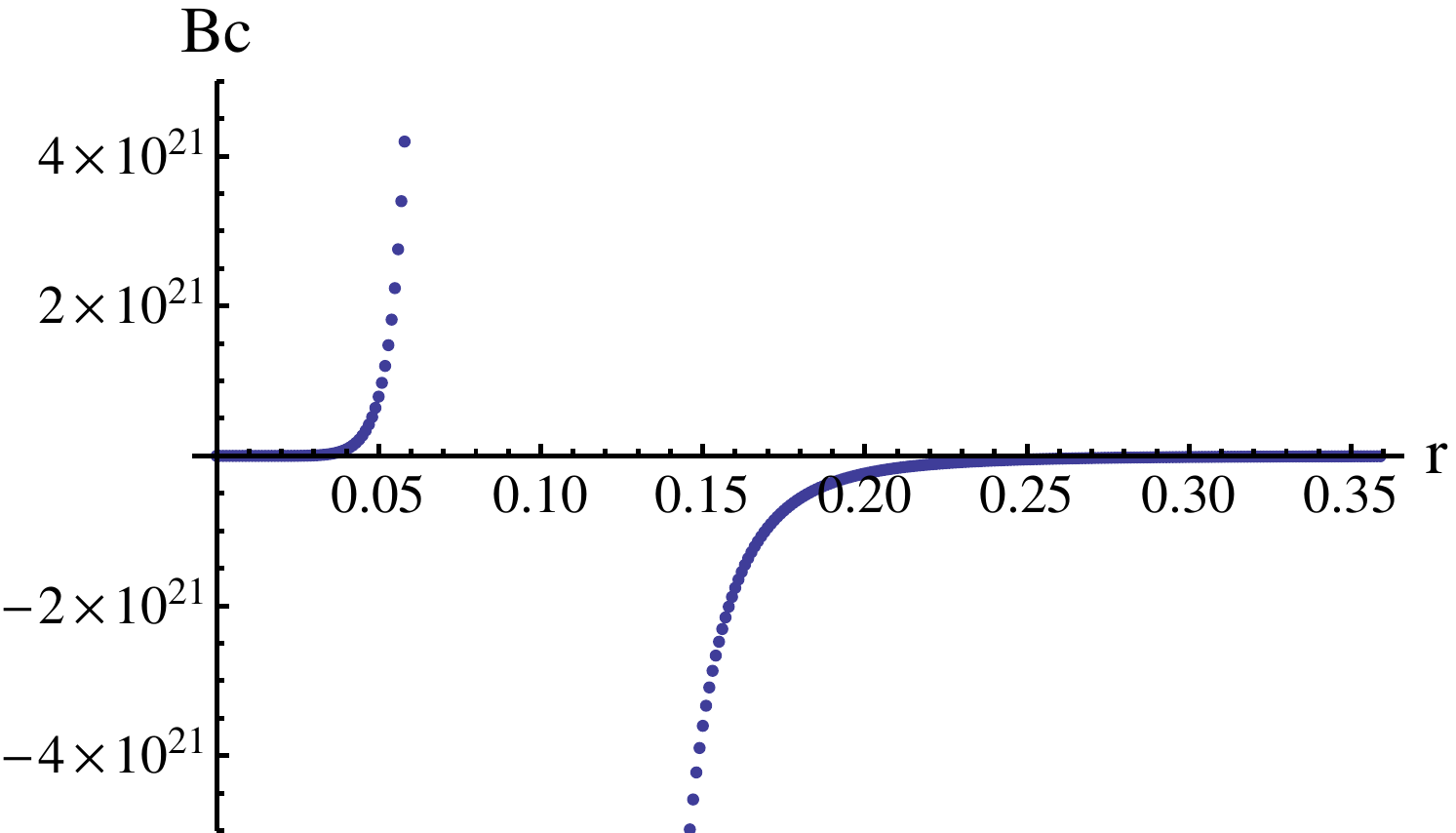}\end{center}
\caption{\small The dependence of $B_C$ on the reference point $r$ at different monolayer compressing
    speeds $V=10^{-15}$ (left), $10^{-1} $ (right)~m/sec of the barrier for $t=0.01,\  \dot\xi=0.001,\ \dot r =0.01$
    and $\dot\phi =0$.}
    \label{figure3}
\end{figure}

The Berwald curvature tensor and its total trace are respectively given by
   $$B^i_{jkl}= {\partial^3 G^i\over \partial y^j\partial y^k \partial y^l},\qquad
    B_C(x,y,V) = g^{ij}B^k_{ikj}.$$
The dependence of the Berwald scalar curvature $B_C$ on the reference
radius $r$ at different monolayer
    compressing speeds $V$ is represented in Fig.~\ref{figure3}.
It is easy to see that $B_C$ is equal to zero as $V$ tends to
zero. This means that the  space under consideration is a flat one
in Finsler geometry. At large speeds $V$ the scalar function $B_C$
is zero everywhere, excluding anomalous areas.

\section{Curvature and first order phase transitions}
We note that an integral of the  $B_C$ at large values of $V$
behaves like the compressibility $\kappa$ being the first order
derivative ${\partial  s\over \partial \tilde\pi }$ of the $s -
\tilde\pi$--isotherm with respect to $\tilde\pi$ in an region of
the phase transition. This allows us to assume the following
relationship between $B_C$ and $\tilde\pi$ in a neighborhood of
the phase transition:
$$
\kappa \stackrel{def}{=\!\!= }{\partial  s\over \partial
\tilde\pi} \propto
  \int  B_C\, d^2 r.
$$
This elucidates the physical sense of $B_C$,
    whose anomalous behavior testifies the possibility of the first order
    phase transition.

\begin{figure}%[hbt]
 \vspace*{-6cm}
    \begin{center}
\vspace*{-14cm} \hspace*{3cm}
\includegraphics[width=22.cm,height=24.5cm,angle=0]{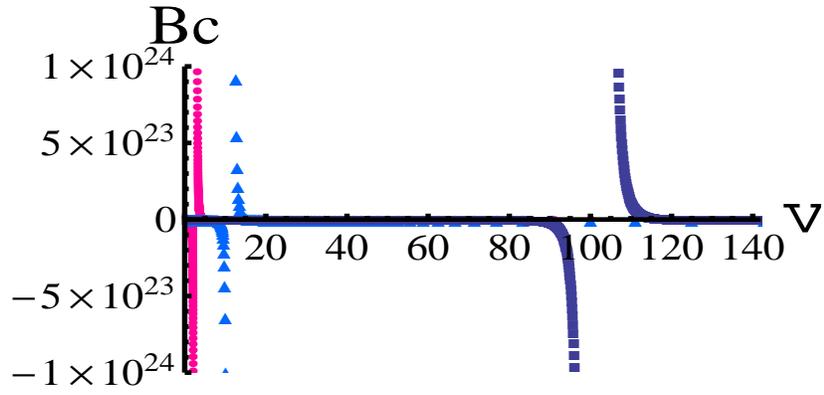}
\end{center}
\caption{\small Dependence of $B_C$ on the particle velocity
$v=\dot{r}/\dot{\xi}$ at different monolayer compressing speeds:
$V = 0.001$ (red circles), $V = 0.1$    (light blue triangles) and
$V = 10.0$ (dark blue squares)  for  $\dot r =1,\dot \phi =0,
t=0.01, r=0.1$.} \label{figure4}\end{figure}

The dependencies of $B_C$ on the particle velocity $v =\dot
r/\dot\xi $ at the following monolayer compressing    speeds $V =
0.001,\ 0.1,\ 10.0$ for $\dot r =1.0,\  \dot\phi=0,\ t=0.01,\
r=0.1$, are shown in    Fig.~\ref{figure4}. One can see here
 (Fig.~\ref{figure4}) %\xx{??4}
that the increment of the compressing speed $V$ leads to the
disappearing of the singularity of    the function $B_C(v)$, while
this singularity still reappears at some another location. The
shift    of the singularity towards the direction of larger values
$v$ is a consequence of the disordering    influence of the
compressing barrier. Therefore the phase transition becomes
enabled at some    increment of the molecule velocity $v$.\par
The last one is in accordance with the experimental data presented
in \cite{Hrushevsky}, and can be    easily understood from the
following explanation: while the speed $v$ of the particle
increases,    the bigger becomes the area occupied by the
trajectory which is bent by the electro-capillary    interaction,
and there increases the probability of collisions between the
particles of the monolayer,    leading to the formation of
structures.

\begin{figure}%[hbt]
\vspace*{-0cm} A)
\includegraphics[width=4.cm,height=3.5cm,angle=0]{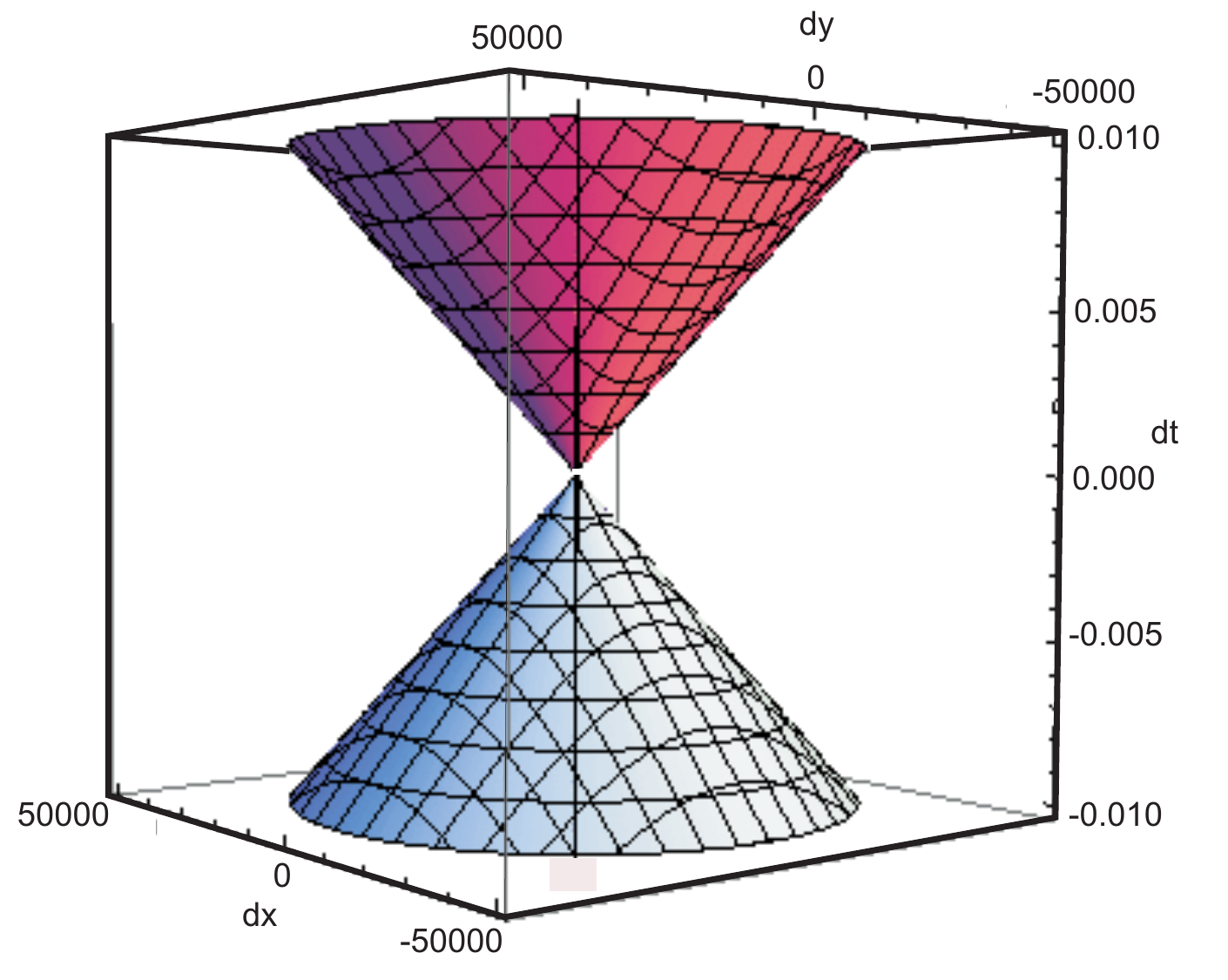}
\includegraphics[width=4.cm,height=3.5cm,angle=0]{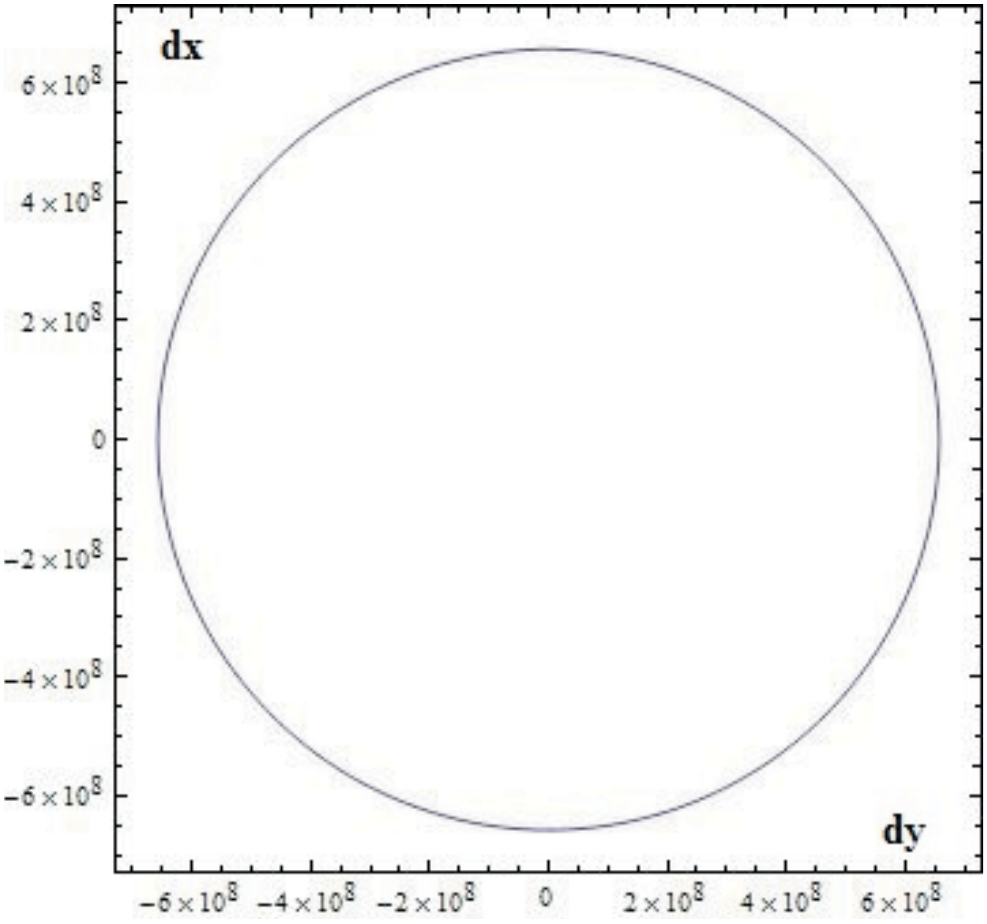}
\includegraphics[width=4.cm,height=3.5cm,angle=0]{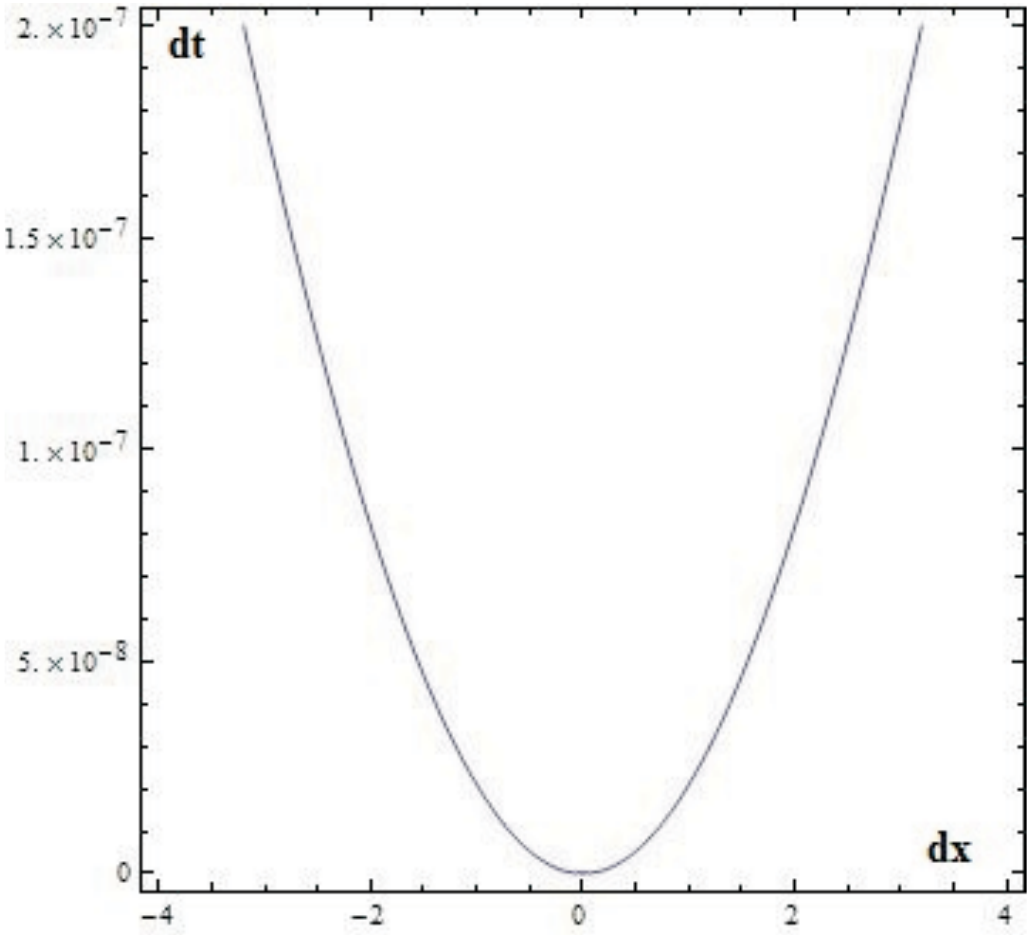}

B)
\includegraphics[width=4.cm,height=3.5cm,angle=0]{Graphic3}
\includegraphics[width=5.7cm,height=3.5cm,angle=0]{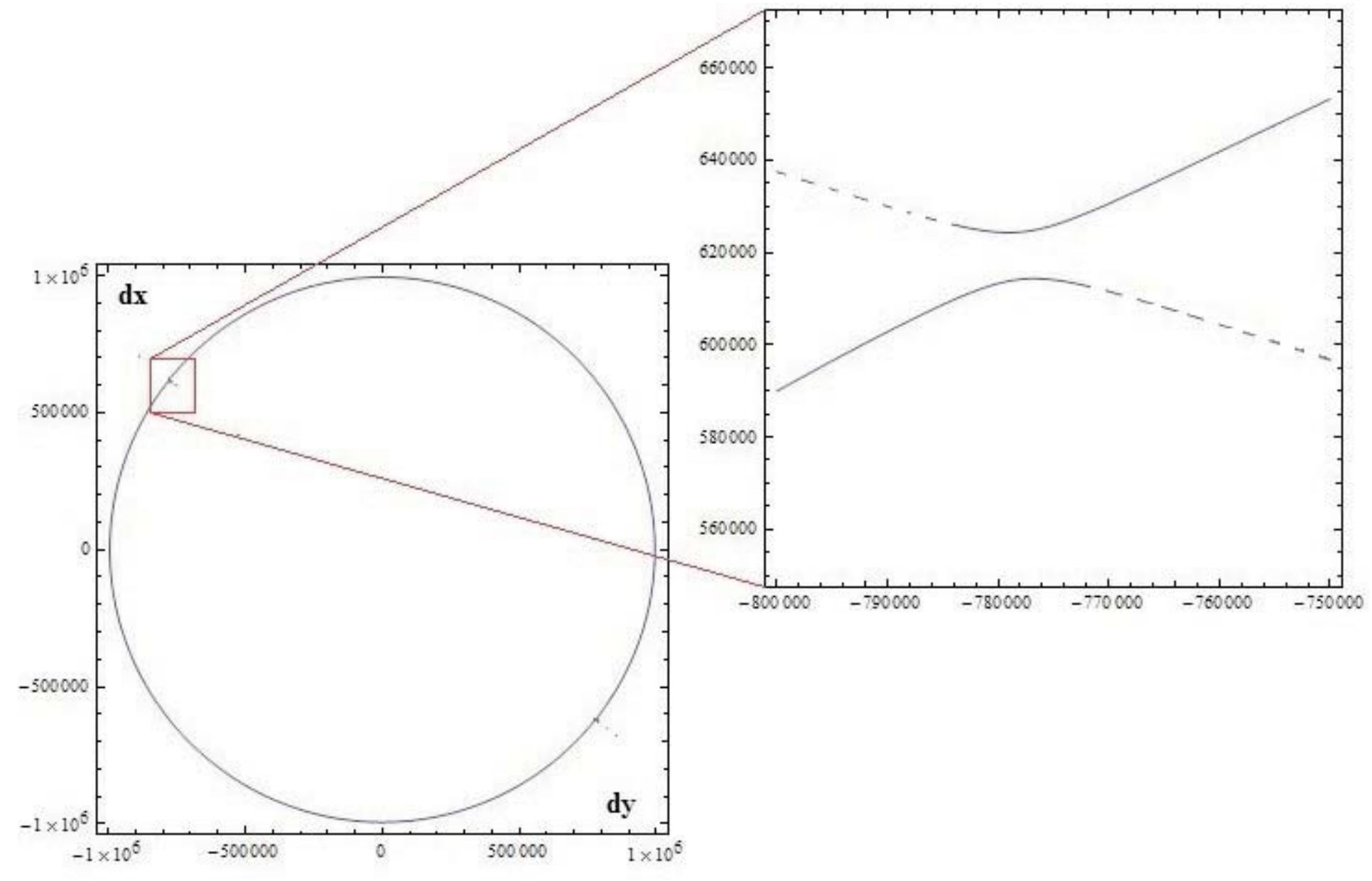}
\includegraphics[width=4.cm,height=3.5cm,angle=0]{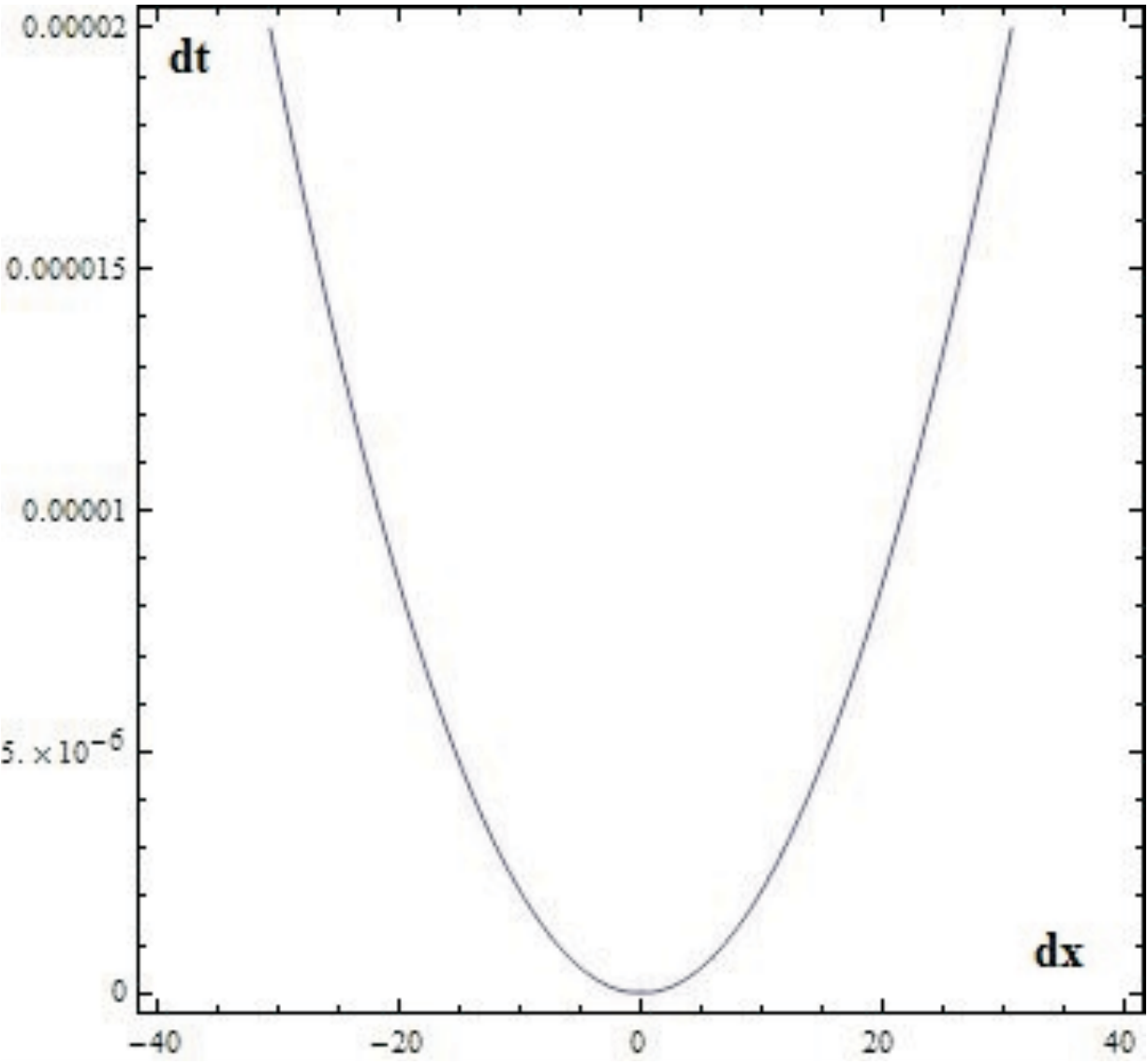}

C)
\includegraphics[width=4.cm,height=3.5cm,angle=0]{Graphic12}
\includegraphics[width=4.cm,height=3.5cm,angle=0]{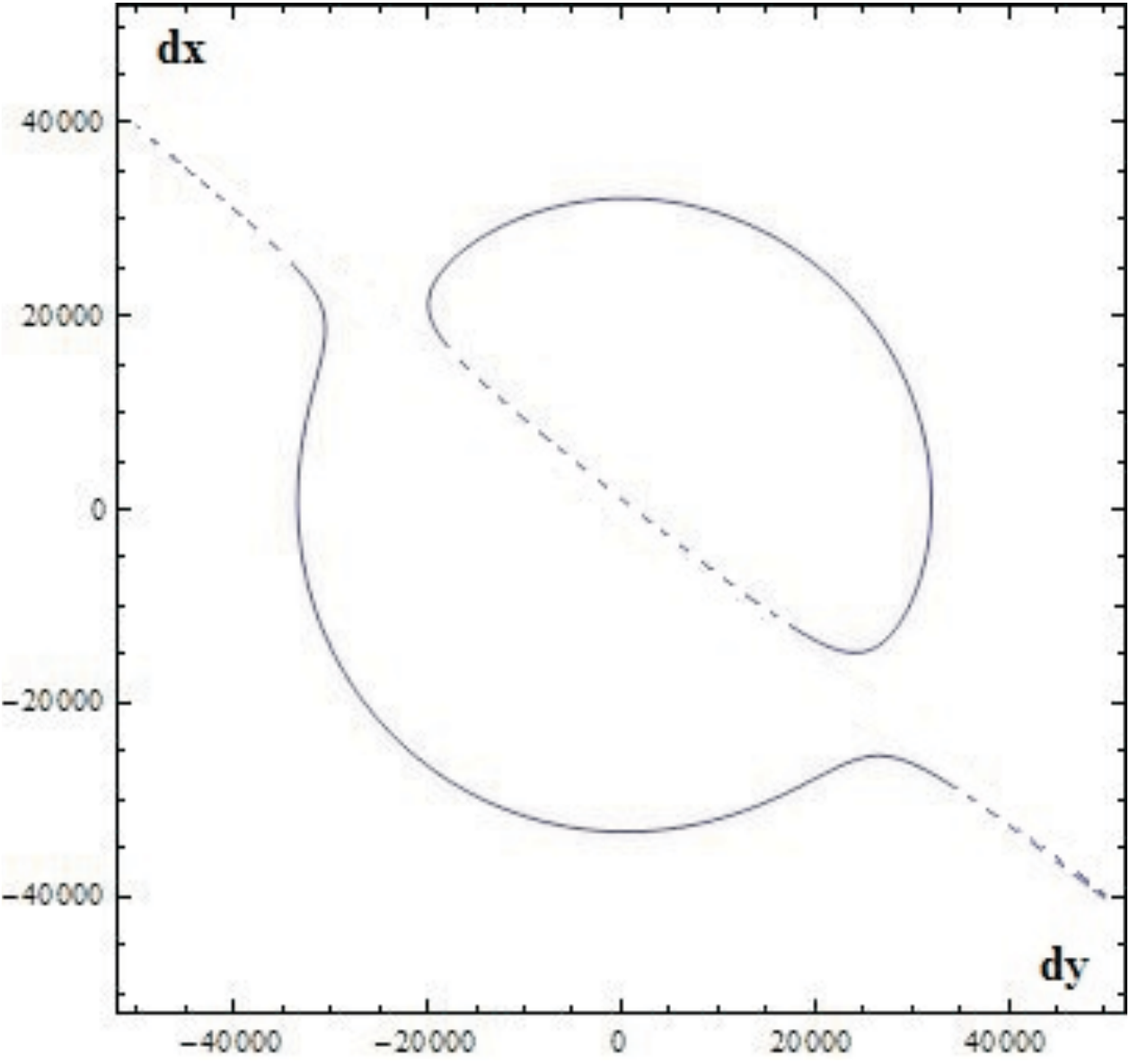}
\includegraphics[width=4.cm,height=3.5cm,angle=0]{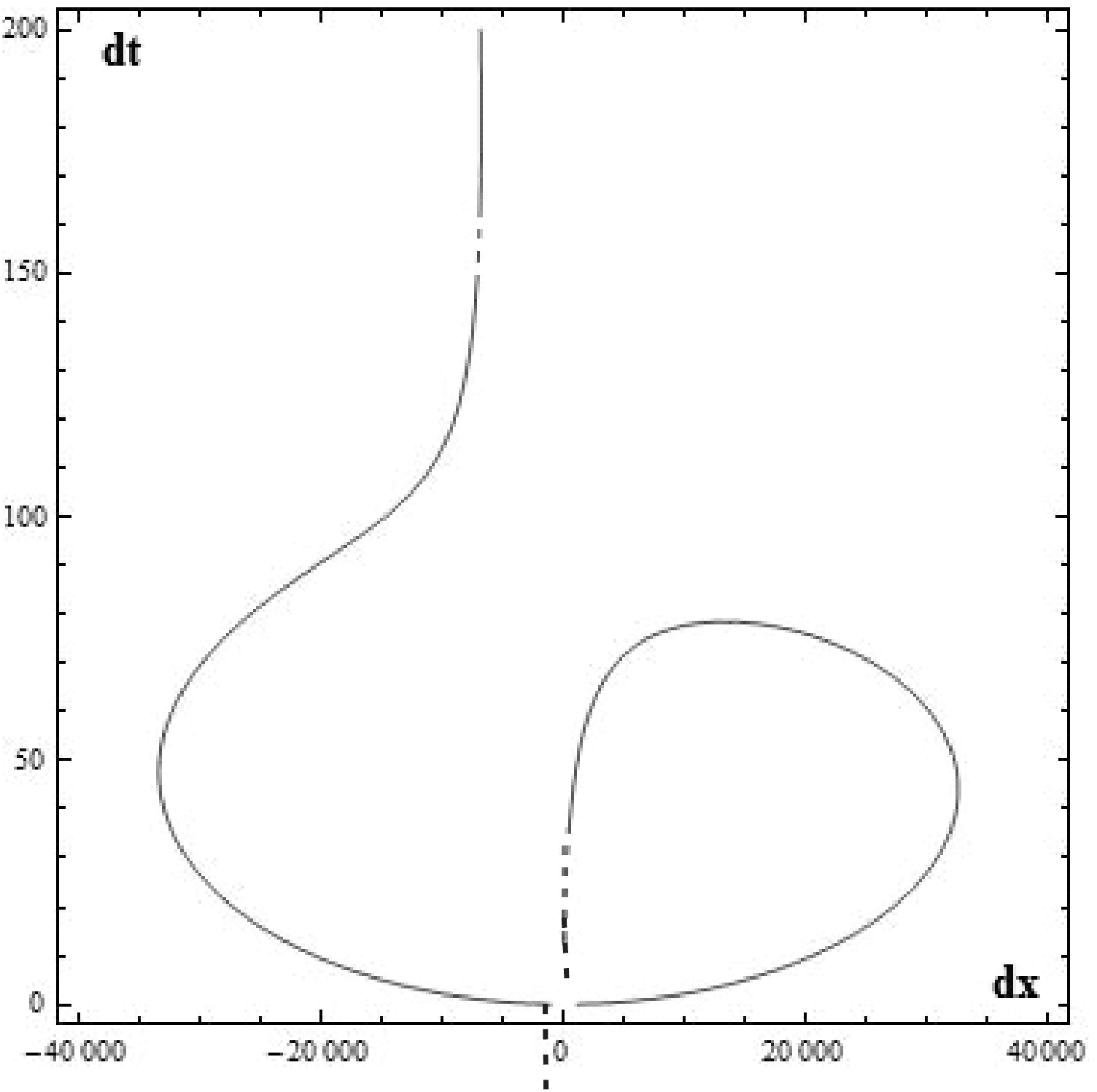}

D)\includegraphics[width=4.cm,height=3.5cm,angle=0]{Graphic15}
\includegraphics[width=4.cm,height=3.5cm,angle=0]{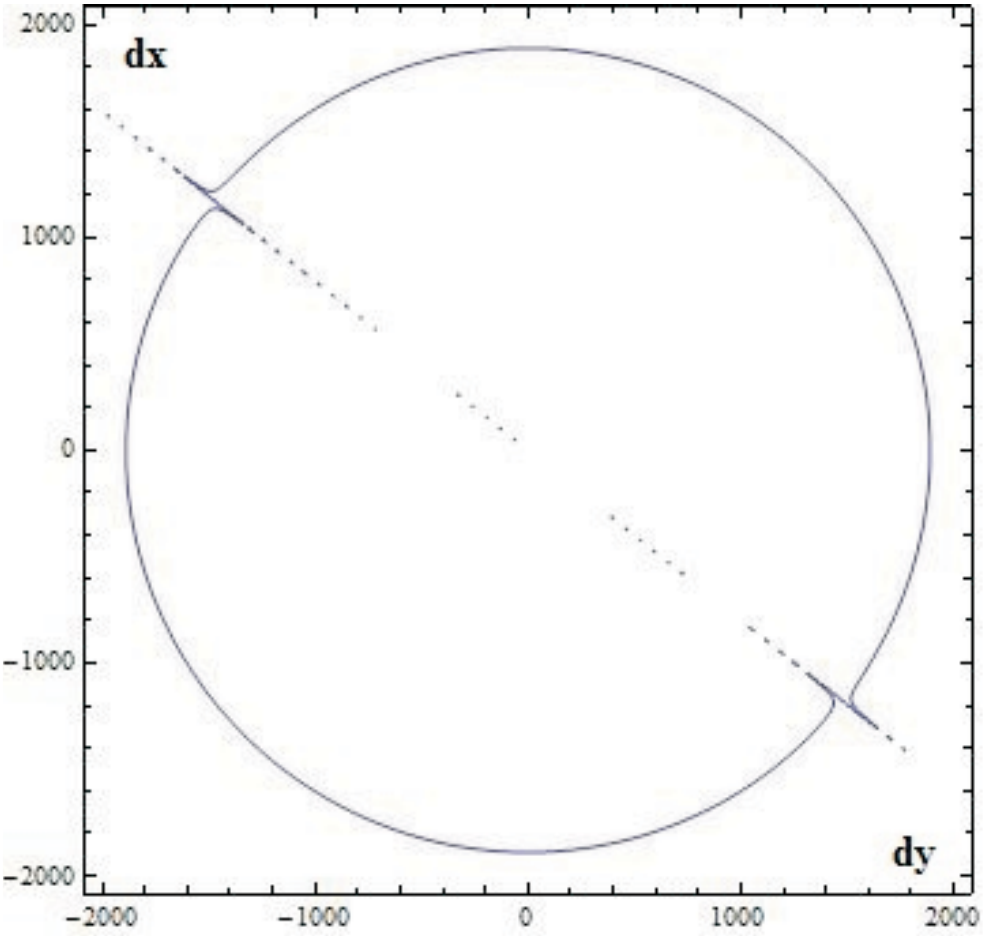}
\includegraphics[width=4.8cm,height=3.5cm,angle=0]{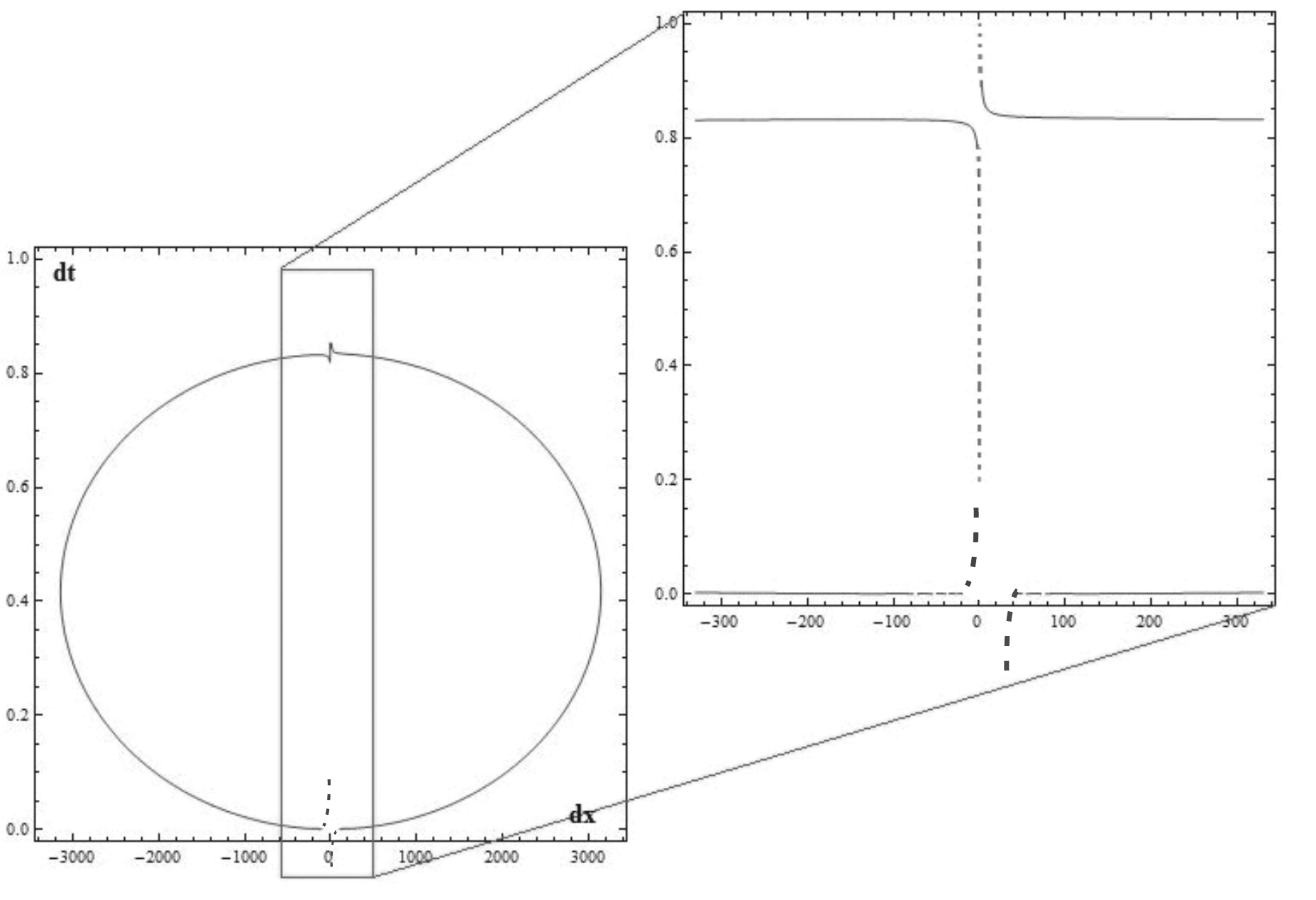}

\caption{\small Projections of indicatrices (from the left) into
the spatial
    (in the middle) and time--spatial  (right) planes.}
    \label{indicatries-projection}\end{figure}

% which relate to the Berwald curvature $B_C$ (2D-
% and 3D-images
%
One can conclude according to the examination performed above that
$B_C$ is an invariant for    the monolayer structurization.
Therefore, the Berwald scalar curvature $B_C$ -- due to its
invariance -- allows us to classify indicatrices according to the
behavior of $B_C$  in the following way. Figs.~\ref{figure5}
and~\ref{figure6} illustrate eleven
    %thirteen
types of indicatrices -- for large compression speed $V$ equal to
15.0. But the Berwald scalar    curvature $B_C$  for corresponding
indicatrices reveal the following descriptive classifying picture.
It is easy to see observing the behavior of the scalar Berwald
curvatures $B_C$ and the indicatrices    corresponding to them in
Figs.~\ref{figure5} and~\ref{figure6}, that for  $V=15.0$, $B_C$
may have    one or two anomalies only. The second singularity
occurs in a form of a stick-slip inversion of the first anomalous
dependence $B_C=f(\dot r)$ (see Fig.~\ref{figure6}A). At the start of the compression
process, the scalar    curvature $B_C$ for the corresponding
indicatrix has just one singularity    (see Fig.~\ref{figure5}),
and afterwards the second singularity and another indicatrix which
corresponds to a curvature $B_C$ having two singularities, appear
(see Fig.~\ref{figure6}). We note that in contrast to the
numerically derived large diversity of indicatrices, their
curvatures $B_C$    behave uniformly. One can  distinguish two
topological classes of indicatrices:    (A -- F) in
Fig.~\ref{figure5} and (A -- E) in Fig.~\ref{figure6}, whose
curvatures $B_C$ behave    structurally different. The curvature
$B_C$ everywhere vanishes, excluding anomalous areas similar
    to the phase transition areas.
Therefore, one can conclude that the behavior of $B_C$ reflects several features of physically
    feasible processes, namely, the monolayer compression.\par
We further show that though there seemingly exist a large
diversity of indicatrices attached to the    Finsler structure,
these basically classify into only three types of indicatrices,
which    topologically differ from each other. This can be
established by the the peculiarities of the orthogonal projections
of indicatrices to the    spatial $\{x,\ y\}$ and to the
time-spatial $\{x,\ t\}$ or $\{y,\ t\}$ planes. At very small
speed $ V$, $V\to 0$, $B_C$ vanishes identically, and the
indicatrix  tends to the one    of the pseudo-Riemannian space for
sufficiently large values of the reference radius $r$ (see
Fig.~\ref{indicatries-projection}A). The type of such indicatrix
is referred as {\em first type}.\par
The indicatrices (A -- F) in Fig.~\ref{figure5} are referred as
indicatrices of {\em second type}. These indicatrices hold
anomalous areas in the spatial plane (see
Fig.~\ref{indicatries-projection}B),
    and the relevant function $B_C$  exhibits one singularity.\par
The indicatrices (A -- E) in Fig.~\ref{figure6} will be referred
as indicatrices of {\em third type}. They hold anomalous areas
both in spatial and time-spatial planes (see
Fig.~\ref{indicatries-projection}C and
\ref{indicatries-projection}D), and the relevant function $B_C$
has two singularities.\par
We finally conclude that the monolayer is flat in the framework of
Finsler geometry, and therefore the
    particles which move within it exhibit acceleration.
\section{ Conclusions}
We proved that the curvature scalar $R_c$ is not an invariant of
the monolayer, and this leads to its complex dependence on the
monolayer parameters. On the contrary, $B_c$ vanishes for very
small speeds, and is null as well outside certain domains of
anomaly, for larger speeds of compression. Hence we conclude that
the monolayer is flat  in the framework of Finsler geometry,
because the particles
    which move within it, exhibit acceleration.
It is assumed that the domains of anomaly correspond to the phase
transition and that the integral of $B_c$ represents the
compressibility $\kappa$. It is shown that, in spite of the fact
that
    the manifold of Finsler indicatrices seems to be large, still there exist only three types of indicatrices
    for the three types of dependence of the Berwald scalar curvature in terms of the tangent vector.
These types can be identified by the three different projections on the spatial plane and on space-time plane.
    For small speeds -- when $B_c$ is identically zero - the indicatrix tends to the one of the
    pseudo-Euclidean Minkowski space for rather large components of the reference point.
The indicatrices which exhibit anomalies only in the spatial plane are characterized by having a
    singularity of $B_c$. The third type of indicatrices have two singularities for $B_c$, which
    corresponds to an anomalous behavior of the indicatrix both in the space and in the space-time planes.\par
As well, there are determined the Cartan tensor and the nonlinear Barthel connection, which provide means
    to estimate how far is the Finsler model from the pseudo-Euclidean one, and to construct Finsler
    adapted frames for the module of sections of the tangent bundle.
Simulations illustrate this inter-relation for several classes of structure formation - which depend on compression
    speed and characteristics of the double electrical layer.
Vladimir Balan\\
University Politehnica of Bucharest, Splaiul Independentei 313, \\
060042 Bucharest, Romania, E-mail: vladimir.balan@upb.ro\\\\
Halina V. Grushevskaya and Nina G. Krylova\\
Faculty of Physics, Belarusan State University, 4 Nezavisimosti Ave.,\\
220030 Minsk, The Republic of Belarus, E-mail: grushevskaja@bsu.by , nina-kr@tut.by\\\\
Alexandru Oana\\
University Transilvania of Brasov, 50 Iuliu Maniu Str.,\\
500091 Brasov, Romania, E-mail: alexandruo@gmail.com

\end{document}